\documentclass{aa}

\usepackage{graphicx}
\usepackage{txfonts}
\usepackage{epsfig}
\usepackage{natbib}
\defcitealias{Sarzi2018}{S18}
\defcitealias{Pinna2019a}{P19a}
\defcitealias{Pinna2019b}{P19b}

\newcommand{\placetabone}{
\renewcommand{\tabcolsep}{4pt}
\begin{sidewaystable*}
\caption{Main morphological, photometric, and kinematic properties of the F3D galaxies.}
\centering
\begin{tabular}{lccccccccccccccc}
\hline\hline
\noalign{\smallskip}
Object &  Morph. &  $m_r$ & $R_{\rm e}$ & $M_\ast$ & $R_{\rm proj}/R_{\rm vir}$ & $cz$  & $R_{\rm max}/R_{\rm e}$ & PA$_{\rm kin}$ & $\Delta$PA & Kinem. & $\sigma_{\rm e}$ & $\lambda_{R_{\rm e}}$ & $\epsilon_{\rm e}$  & Kinem. & Altern.\\
    & Type & & & & & & &  &  & Structure & & & & Type & Name\\
       &       &    (mag) & (arcsec) & ($10^{10}$ M$_{\odot}$) & & (km~s$^{-1}$) &  & (deg) & (deg) & & (km~s$^{-1}$) & & & \\
(1)    & (2)        & (3)       & (4)        & (5)  & (6)   & (7) & (8) & (9)   & (10) & (11) & (12) & (13) & (14) & (15) & (16)\\
\noalign{\smallskip}
\hline
\noalign{\smallskip}
FCC~083 & E5         & 10.84 &  35.7 & ...  & 0.83 & $1532\pm1$     & 2.0 & $142\pm1$    & $2\pm1$    & RR  & 103 & 0.51 & 0.35 & FR & NGC~1351\\
FCC~090 & E4 pec     & 13.62 &  12.1 & 0.08 & 0.82 & $1832\pm1$     & 2.2 & $188\pm12$   & $72\pm13$  & NR  & 33  & 0.17 & 0.20 & FR & ...\\
FCC~113 & ScdIII pec & 14.44 &  20.6 & 0.02 & 0.62 & $1395\pm1$     & 1.9 & $2\pm5$    & $-11\pm8$  & RR  & ... & 0.36 & 0.40 & FR & ESO~358-G015\\
FCC~119 & S0 pec     & ...   &  17.4 & 0.1  & 1.39 & $1409.7\pm0.4$ & 1.5 & $43\pm6$     & $1\pm7$  & RR  & 17  & 0.33 & 0.07 & FR & ...\\
FCC~143 & E3         & 12.66 &  11.0 & 0.28 & 0.37 & $1379\pm1$     & 2.4 & $188\pm3$    & $0\pm3$    & DC  & 62  & 0.14 & 0.04 & FR & NGC~1373\\
FCC~147 & E0         & 10.50 &  24.8 & 2.40 & 0.32 & $1350\pm1$     & 2.6 & $123\pm2$    & $2\pm3$    & RR  & 131 & 0.22 & 0.04 & FR & NGC~1374\\
FCC~148 & S0         & 11.70 &  28.3 & 0.58 & 0.32 & $777\pm1$      & 2.1 & $264\pm2$    & $-3\pm4$   & RR  & 43  & 0.65 & 0.60 & FR & NGC~1375\\
FCC~153 & S0         & 11.70 &  19.8 & 0.76 & 0.57 & $1642\pm1$     & 3.2 & $82\pm2$     & $-1\pm3$   & RR  & 55  & 0.73 & 0.65 & FR & IC~1963\\
FCC~161 & E0         & 10.47 &  28.6 & 2.63 & 0.24 & $1369\pm1$     & 2.0 & $179\pm2$    & $-4\pm5$   & RR  & 96  & 0.19 & 0.02 & FR & NGC~1379\\
FCC~167 & S0/a       &  9.27 &  56.4 & 9.85 & 0.30 & $1878\pm2$     & 1.9 & $186\pm1$    & $1\pm1$    & RR  & 143 & 0.53 & 0.49 & FR & NGC~1380\\
FCC~170 & S0         & 10.99 &  15.9 & 2.25 & 0.20 & $1743\pm2$     & 4.3 & $138\pm1$    & $-3\pm2$   & RR  & 113 & 0.45 & 0.50 & FR & NGC~1381\\
FCC~176 & SBa        & 11.74 &  53.7 & 0.68 & 0.43 & $1428\pm1$     & 0.7 & $170\pm5$    & $-15\pm7$  & RR  & 46 & 0.46 & 0.15 & FR & NGC~1369\\
FCC~177 & S0         & 11.80 &  35.9 & 0.85 & 0.38 & $1591\pm1$     & 1.6 & $180\pm3$    & $5\pm4$    & RR  & 42  & 0.69 & 0.70 & FR & NGC~1380A \\
FCC~179 & Sa         & 10.68 &  30.0 & 1.58 & 0.32 & $915\pm4$      & 3.0 & $24.7\pm0.4$ & $7\pm1$  & KDC & 70 & 0.77 & 0.50 & FR & NGC~1386\\
FCC~182 & SB0 pec    & 13.58 &   9.9 & 0.15 & 0.15 & $1705.1\pm0.4$ & 2.2 & $151\pm13$   & $-25\pm15$ & DC  & 39  & 0.17 & 0.05 & FR & ...\\
FCC~184 & SB0        & 10.00 &  35.5 & 4.70 & 0.15 & $1302.6\pm0.6$ & 2.5 & $54\pm3$    & $36\pm4$   & KDC & 143 & 0.19 & 0.10 & FR & NGC~1387\\
FCC~190 & SB0        & 12.26 &  18.3 & 0.54 & 0.18 & $1794\pm1$     & 2.1 & $128\pm3$    & $-83\pm4$  & RR  & 75  & 0.32 & 0.20 & FR & NGC~1380B\\
FCC~193 & SB0        & 10.69 &  28.2 & 3.32 & 0.19 & $976\pm1$      & 2.0 & $221\pm1$    & $9\pm1$    & KDC & 95  & 0.55 & 0.40 & FR & NGC~1389\\
FCC~219 & E2         &  8.57 & 161.0 & 12.7 & 0.08 & $1933\pm2$     & 0.5 & $153\pm1$    & $10\pm1$   & RR  & 154 & 0.22 & 0.20 & FR & NGC~1404\\
FCC~249 & E0         & ...   &   9.6 & 0.5  & 1.02 & $1579\pm1$     & 3.3 & $156\pm5$     & $2\pm12$  & DC  & 104 & 0.08 & 0.04 & FR & NGC~1419\\
FCC~255 & S0         & ...   &  13.8 & 0.5  & 0.85 & $1328\pm3$     & 2.7 & $-7\pm3$    & $0\pm3$    & RR  & 38  & 0.41 & 0.77 & FR & ESO~358-G050\\
FCC~263 & SBcdIII    & 12.70 &  27.2 & 0.04 & 0.41 & $1746\pm2$     & 1.4 & $165\pm10$   & $-33\pm12$   & RR  & 28 & 0.60 & 0.50 & FR & ESO~358-G051\\
FCC~276 & E4         & 10.15 &  44.7 & 1.81 & 0.38 & $1434\pm1$     & 1.7 & $73\pm3$     & $-5\pm3$   & DC  & 123 & 0.09 & 0.31 & SR & NGC~1427\\
FCC~277 & E5         & 12.34 &  12.8 & 0.34 & 0.42 & $1657\pm2$     & 2.6 & $116\pm4$    & $3\pm2$    & RR  & 80  & 0.77 & 0.36 & FR & NGC~1428\\
FCC~285 & SdIII      & 13.02 &  49.9 & 0.02 & 0.60 & $901\pm1$      & 0.7 & $124\pm7$    & $40\pm10$  & RR  & ... & 0.41 & 0.20 & FR & NGC~1437A\\
FCC~290 & ScII       & 11.08 &  48.5 & 0.64 & 0.54 & $1398\pm4$     & 1.5 & $152\pm3$    & $-3\pm4$   & RR  & 39 & 0.60 & 0.30 & FR & NGC~1436\\
FCC~301 & E4         & 12.65 &  11.7 & 0.20 & 0.70 & $1043\pm2$     & 2.3 & $153\pm3$    & $-2\pm3$   & DC  & 49  & 0.36 & 0.20 & FR & ESO~358-G059\\
FCC~306 & SBmIII     & 15.18 &   9.7 & 0.01 & 0.83 & $902\pm34$     & 1.7 & $2\pm7$      & $-47\pm9$    & PR  & 69 & 0.12 & 0.40 & FR & ...\\
FCC~308 & Sd         & 12.54 &  37.1 & 0.04 & 0.85 & $1519\pm2$     & 1.3 & $10\pm2$     & $5\pm3$   & RR  & 43 & 0.33 & 0.60 & FR & NGC~1437B\\
FCC~310 & SB0        & 11.81 &  35.6 & 0.54 & 0.97 & $1383\pm1$     & 1.0 & $219\pm5$    & $-35\pm5$  & RR  & 48  & 0.46 & 0.27 & FR & NGC~1460\\
FCC~312 & Scd        & 10.89 & 109.5 & 1.48 & 0.82 & $1937\pm4$     & 1.0 & $-47\pm1$    & $3\pm1$  & RR  & 63 & 0.74 & 0.50 & FR & ESO~358-G063\\
\noalign{\smallskip}
\hline
\end{tabular}
\tablefoot{
(1) and (2) Galaxy name and morphological type from \citet{Ferguson1989}. 
(3)--(5) Total $r$-band magnitude, $r$-band effective radius, and total stellar mass from \citet{Iodice2018} for ETGs and \citet{Raj2019} for LTGs. 
(6) Ratio between the projected distance from the cluster centre and 
virial radius of the cluster ($R_{\rm vir}=0.7$ Mpc, \citealt{Drinkwater2001}).
(7) Heliocentric systemic velocity from the stellar velocity field.
(8) Ratio between the maximum radial extension of the MUSE data along the galaxy major axis and the $r$-band effective radius. 
(9) and (10) Kinematic position angle and difference with respect to the photometric position angle from \citet{Iodice2018}.
(11) Kinematic structure following the classification by \citet{Krajnovic2011}: 
RR = regularly rotating galaxy, NR = non-rotating galaxy, DC = galaxy with a distinct core, KDC = galaxy with a kinematically decoupled core, PR = prolate rotating galaxy.
(12) Average stellar velocity dispersion inside the $r$-band effective radius.
(13) Effective stellar angular momentum.  
(14) Ellipticity at the effective radius.
(15) Kinematic type following the classification by \citet{Emsellem2011}: FR = fast-rotating galaxy, SR = slowly-rotating galaxy.
(16) Galaxy alternative name.
}
\label{tab:sample}
\end{sidewaystable*}
}

\newcommand{\placetabtwo}{
\renewcommand{\tabcolsep}{4pt}
\begin{table*}
\caption{Star-formation rate and H$\alpha$ equivalent width for the F3D galaxies.}
\centering
\begin{tabular}{lccccccc}
\hline\hline
\noalign{\smallskip}
Object & $cz_{{\rm gas}}$ & SFR & SFR$_{\rm up}$ & EW(H$\alpha$) & EW(H$\alpha$)$_{\rm up}$ & Notes on star formation\\
    &    (km~s$^{-1}$) & (M$_{\odot}$ yr$^{-1}$)   & (M$_{\odot}$ yr$^{-1}$) &  (\AA) & (\AA)  &   \\
    (1) & (2)        & (3)       & (4)        & (5) & (6) & (7)   \\
\noalign{\smallskip}
\hline
\noalign{\smallskip}
FCC~090 & 1756.5$\pm$0.2 & {0.035} & {0.038} &  {21.46} &  {23.22} &	pervasive\\
FCC~113 & 1323.4$\pm$0.3 & {0.042} & {0.042} &  {29.63} &  {30.05} &	pervasive\\     
FCC~119 & 1340.1$\pm$0.2 & {0.001} & 0.002 &   {0.87} & {1.20} &	central\\
FCC~167 & 1850$\pm$16 & 0.000 & {0.003} &   0.00 &   {0.06} &	traces in the centre\\
FCC~179 & 828$\pm$9 & {0.155} & {0.228} &   {4.95} & {7.26} &	extended star-forming ring\\
FCC~184 & 1230$\pm$2 & {0.008} & {0.082} &   {0.26} & {2.54} &	circumnuclear star-forming ring\\	
FCC~219 & 1948$\pm$1 & 0.000 & 0.000 &   0.00 &   0.00 &	no star formation\\
FCC~263 & 1664$\pm$1 & {0.282} & {0.285} &  {50.65} &  {51.07} &	pervasive\\
FCC~285 & 831.6$\pm$0.2 & {0.145} & {0.149} &  {75.47} &  {77.31} &	pervasive \\
FCC~290 & 1334.1$\pm$0.3 & {0.127} & {0.148} &  {12.52} & {14.62} &	pervasive	\\
FCC~306 & 823.6$\pm$0.4 & {0.016} & {0.016} &  {15.29} &  {15.49} &	pervasive \\
FCC~308 & 1464.0$\pm$0.4 & {0.167} & {0.170} &  {36.22} &  {36.81} & pervasive \\
FCC~312 & 1871$\pm$1 & {0.751} & {0.783} & {54.33} & {56.65} &	pervasive \\
\noalign{\smallskip}
\hline
\end{tabular}
\tablefoot{
(1) Galaxy name from \citet{Ferguson1989}. 
(2) Heliocentric systemic velocity from the ionised-gas velocity field.
(3) Star formation rate.
(4) Upper limit of the star formation rate derived by assuming that also all the H$\alpha$ emission from transition regions is powered by O-type stars.
(5) and (6) H$\alpha$ equivalent width and its upper limit.
(7) Notes about the star formation activity.}
\label{tab:emission}
\end{table*}
}

\newcommand{\placetabthree}{
\renewcommand{\tabcolsep}{4pt}
\begin{table*}
\caption{Mean stellar population properties of the F3D ETGs.}
\centering
\begin{tabular}{lccccccc}
\hline\hline
\noalign{\smallskip}
Object & age$_{0.5 R_{\rm e}}$ & [$M$/H]$_{0.5 R_{\rm e}}$ & [Mg/Fe]$_{0.5 R_{\rm e}}$ & $R_{\rm tr}$ & age$_{R>R_{\rm tr}}$ & [$M$/H]$_{R>R_{\rm tr}}$ & [Mg/Fe]$_{R>R_{\rm tr}}$\\
        & (Gyr) & (dex) & (dex) & (arcsec)  & (Gyr)  & (dex) & (dex)  \\
    (1) & (2)   & (3)   & (4)   & (5)       & (6)    & (7) & (8) \\
\noalign{\smallskip}
\hline
\noalign{\smallskip}
FCC~083 & 13.2 & $-0.20$ & 0.26 & 43.0 & 12.0 & $-0.34$ & 0.22 \\
FCC~090 & 1.4  & $-0.57$ & 0.14 & 15.0 & 5.4  & $-1.01$ & 0.10 \\
FCC~119 & 6.7  & $-0.51$ & 0.11 & 10.0 & 10.0 & $-0.57$ & 0.13 \\
FCC~143 & 12.6 & $-0.18$ & 0.18 &  6.0 & 11.5 & $-0.49$ & 0.21 \\
FCC~147 & 13.5 & $ 0.04$ & 0.23 &  4.0 & 12.6 & $-0.34$ & 0.31 \\
FCC~148 & 9.8  & $-0.22$ & 0.09 & 42.0 & 10.7 & $-0.57$ & 0.11 \\
FCC~153 & 10.7 & $-0.05$ & 0.11 & 42.5 & 10.0 & $-0.08$ & 0.13 \\
FCC~161 & 12.9 & $-0.13$ & 0.20 & 14.3 & 12.9 & $-0.33$ & 0.22 \\
FCC~167 & 13.5 & $ 0.09$ & 0.20 & 43.5 & 12.6 & $-0.15$ & 0.17 \\
FCC~170 & 13.2 & $-0.05$ & 0.17 &  8.1 & 13.5 & $ 0.00$ & 0.16 \\
FCC~177 & 9.8  & $-0.14$ & 0.11 & 18.0 & 11.5 & $-0.17$ & 0.14 \\
FCC~182 & 12.6 & $-0.22$ & 0.11 &  5.0 & 10.7 & $-0.32$ & 0.22 \\
FCC~184 & 13.2 & $ 0.21$ & 0.19 & 18.0 & 10.0 & $-0.07$ & 0.24 \\
FCC~190 & 12.9 & $-0.13$ & 0.16 &  9.0 & 11.7 & $-0.36$ & 0.24 \\
FCC~193 & 11.7 & $-0.09$ & 0.13 & 27.4 & 10.5 & $-0.31$ & 0.13 \\
FCC~219 & 11.7 & $ 0.14$ & 0.18 & 38.7 & 12.6 & $-0.27$ & 0.26 \\
FCC~249 & 13.5 & $-0.26$ & 0.24 &  9.6 & 12.6 & $-0.60$ & 0.26 \\
FCC~255 & 4.6  & $-0.17$ & 0.10 & 13.8 & 4.5  & $-0.27$ & 0.10 \\
FCC~276 & 13.8 & $-0.25$ & 0.20 & 23.5 & 12.3 & $-0.44$ & 0.20 \\
FCC~277 & 11.7 & $-0.34$ & 0.11 & 10.4 & 11.5 & $-0.45$ & 0.15 \\
FCC~301 & 10.2 & $-0.38$ & 0.09 & 21.3 & 10.5 & $-0.61$ & 0.14 \\
FCC~310 & 12.0 & $-0.30$ & 0.14 & 21.8 & 8.9  & $-0.20$ & 0.16 \\
\noalign{\smallskip}
\hline
\end{tabular}
\tablefoot{
(1) Galaxy name from \citet{Ferguson1989}. 
(2)--(4) Mean values of age, total metallicity, and [Mg/Fe] abundance ratio in the central parts of the galaxy ($R \leq 0.5 R_{\rm e})$.
(5) Transition radius from \citet{Spavone2019}.
(6)--(8) Mean values of age, total metallicity, and [Mg/Fe] abundance ratio in the galaxy outskirts ($R \geq R_{\rm tr})$.}
\label{tab:SP_analysis}
\end{table*}
}

\begin{document} 

\title{The Fornax3D project: Tracing the assembly history of the cluster from the kinematic and line-strength maps} 
\titlerunning{Structure of the Fornax cluster}
\subtitle{}

\author{E.~Iodice\inst{1}\thanks{enrichetta.iodice@inaf.it}
\and 
M.~Sarzi\inst{2}
\and 
A.~Bittner\inst{3,4}
\and 
L.~Coccato\inst{3}
\and 
L.~Costantin\inst{5}
\and 
E.~M.~Corsini\inst{6,7}
\and 
G.~van~de~Ven\inst{3,8}
\and 
P.~T.~de~Zeeuw\inst{9,10}
\and 
J.~Falc\'on-Barroso\inst{11,12}
\and 
D.~A.~Gadotti\inst{3}
\and 
M.~Lyubenova\inst{3}
\and 
I.~Mart\'in-Navarro\inst{13,14}
\and 
R.~M.~McDermid\inst{15,16}
\and 
B.~Nedelchev\inst{17}
\and 
F.~Pinna\inst{11}
\and 
A.~Pizzella\inst{6,7}
\and 
M.~Spavone\inst{1}
\and 
S.~Viaene\inst{18}
}
     
\institute{INAF--Osservatorio Astronomico di Capodimonte, 
via Moiariello 16, I-80131 Napoli, Italy
\and Armagh Observatory and Planetarium, 
College Hill, Armagh, BT61 9DG, Northern Ireland, UK
\and European Southern Observatory, 
Karl-Schwarzschild-Strasse 2, D-85748 Garching bei Muenchen, Germany
\and Ludwig Maximilian Universitaet, Professor-Huber-Platz 2, 80539 M\"unchen, Germany
\and INAF--Osservatorio Astronomico di Brera, 
via Brera 28, I-20128 Milano, Italy
\and Dipartimento di Fisica e Astronomia `G. Galilei', Universit\`a di Padova, 
vicolo dell'Osservatorio 3, I-35122 Padova, Italy
\and INAF--Osservatorio Astronomico di Padova, 
vicolo dell'Osservatorio 5, I-35122 Padova, Italy
\and Department of Astrophysics, University of Vienna,
Tuerkenschanzstrasse 17, A-1180 Vienna, Austria
\and Sterrewacht Leiden, Leiden University, 
Postbus 9513, 2300 RA Leiden, The Netherlands
\and Max-Planck-Institut fuer extraterrestrische Physik, 
Giessenbachstrasse, 85741 Garching bei Muenchen, Germany
\and Instituto de Astrof\'{\i}sica de Canarias, 
Calle V\'{\i}a L\'actea s/n, E-38200 La Laguna, Spain
\and Departamento de Astrof\'{\i}sica, Universidad de La Laguna, 
Calle Astrof\'{\i}sico Francisco S\'anchez s/n, E-38205 La Laguna, Spain
\and University of California Observatories, 
1156 High Street, Santa Cruz, CA 95064, USA
\and Max-Planck-Institut fuer Astronomie, 
Koenigstuhl 17, D-69117 Heidelberg, Germany
\and Department of Physics and Astronomy, Macquarie University,
Sydney, NSW 2109, Australia
\and Australian Astronomical Observatory, 
PO Box 915, Sydney, NSW 1670, Australia
\and Centre for Astrophysics Research, University of Hertfordshire, College
Lane, Hatfield AL10 9AB, UK
\and Sterrenkundig Observatorium, Universiteit Gent, 
Krijgslaan 281, B-9000, Gent, Belgium
}

\date{Received ....; accepted ......}

 
  \abstract
   {   The 31 brightest galaxies ($m_B \leq 15$ mag) inside the virial radius of the Fornax cluster were observed from the centres to the outskirts with the Multi Unit Spectroscopic Explorer on the Very Large Telescope. These observations provide detailed high-resolution maps of the line-of-sight kinematics and line strengths of the stars and ionised gas reaching 2--3 $R_{\rm e}$ for 21 early-type galaxies and 1--2 $R_{\rm e}$ for 10 late-type galaxies. The majority of the galaxies are regular rotators, with eight hosting a kinematically distinct core. Only two galaxies are slow rotators. The mean age, total metallicity, and [Mg/Fe] abundance ratio in the bright central region inside $0.5 R_{\rm e}$ and in the galaxy outskirts are presented. Extended emission-line gas is detected in 13 galaxies, most of them are late-type objects with wide-spread star formation. 
   The measured structural properties are analysed in relation to the galaxies' position in the projected phase space of the cluster. This shows that the Fornax cluster appears to consist of three main 
   groups of galaxies inside the virial radius: the old core; a clump of galaxies, which is aligned with the local large-scale structure and was accreted soon after the formation of the core; and a group of galaxies that fell in more recently.} 
 

\maketitle
%

\section{Introduction}
 {One of the most ambitious goals of astronomy is to study the formation history of the structures in the universe, 
from the large scale environments such as clusters and groups of galaxies down to the cluster/group members. 
From the observational side, this was done by analysis of the light distribution, the stellar and gaseous kinematics and the  stellar population 
properties. All these observables can be compared with  theoretical predictions on galaxy formation \citep[see, e.g.][]{Cappellari2016}.
Further advances in the understanding of galaxy formation and evolution, in particular regarding early-type galaxies (ETGs), 
resulted from integral-field spectroscopy, 
which allows accurate  mapping of the stellar and gas kinematics as well as of the stellar-population properties of thousands of nearby galaxies 
(see e.g. the SAURON, ATLAS3D, CALIFA, SAMI, and ManGA surveys described in \citealt{deZeeuw2002}, \citealt{Cappellari2011}, \citealt{Sanchez2012}, \citealt{Croom2012}, and \citealt{Bundy2015}, respectively). 
By combining two-dimensional spectroscopy and the large data-sets, these studies derived important constraints on 
the role of the environment in shaping the galaxy structure and stellar populations.}

 {In the deep potential well of clusters of galaxies, the gravitational interactions and merging 
between systems and/or with the intra-cluster medium, play a fundamental role in defining the structure of galaxies 
and affecting the star formation.
Galaxy harassment and ram-pressure stripping are thought to be partly responsible for the  morphology-density relation 
\citep{Dressler1997,vdWel2010,Fasano2015}, where ETGs dominate the central regions of the clusters, while late-type galaxies (LTGs), i.e. spirals and irregulars, populate the outskirts.
Galaxy harassment results from the repeated high-speed tidal encounters of the galaxies inside the cluster, 
which can strip dark matter, stars, and gas from galaxies and generates faint streams detectable along 
the orbit of the galaxy through the cluster \citep{Moore1998, Mastropietro2005, Smith2015}.
Ram-pressure stripping is the interaction of  gas-rich galaxies with the hot interstellar medium, which
can sweep off their atomic gas and therefore halt the star formation \citep[e.g.][]{Chung2009, Davies2016, Merluzzi2016, Poggianti2017}. 
This is the mechanism that might be responsible for converting late-type galaxies with ongoing star formation into quiescent early-type systems \citep{Boselli2006}. The ram-pressure stripping can also remove the hot, ionised gas within the halo of a galaxy (the so-called 
"strangulation") so that the supply of cold gas is lost and, as a consequence, the star formation is 
stopped  \citep[e.g.][]{Dekel2006, Peng2015}.}

 {A useful tool to investigate the many processes acting in clusters is the projected phase-space (PPS) diagram, which combines  
the cluster-centric velocities and radii in 
one plot \citep[see][and references therein]{Smith2015, Jaffe2015, Rhee2017, Owers2019}. 
According to \citet{Rhee2017}, the galaxies that are infalling for the first time into the cluster are found at projected distances beyond the virial radius 
with lower velocities than galaxies that are passing close to the pericentre, which are close to the cluster centre. 
The galaxies that fell into the cluster several gigayears ago are located in the regions of lower relative velocities.
The PPS diagram turned out to be an efficient tool to map the galaxy evolution inside 
a cluster and, in particular, to study the star formation history \citep{Mahajan2011, Muzzin2014} and the effect 
of the ram-pressure stripping \citep[e.g.][]{Hernandez2014, Jaffe2015, Yoon2017, Jaffe2018}.}

 {In addition to the role of the environment, it has been clearly demonstrated that the mass of the galaxy 
is the other key parameter that regulates the evolution and the star formation history  \citep[see e.g.][and references therein]{Peng2010, Cappellari2013, Greene2019}.  
The present day massive  (M$_{*} \simeq 3 \times 10^{11}$~M$_{\odot}$) and passive ETGs built their stellar mass by the  
gradual accretion of satellites forming the extended stellar halo around a compact spheroid 
\citep[e.g.][]{Kormendy2009, vanDokkum2010, Huang2018, Spavone2017}.
Therefore, the stellar content consists of the "in-situ" component, formed by collapse of the initial pro-galactic cloud, and the "ex-situ" 
population coming from the mass assembly in the stellar halo.
The two components have different properties in the light distribution, kinematics and stellar populations.
From the observational side, recent deep imaging surveys of nearby clusters have enabled extensive analyses of the light and colour distribution of galaxies in dense environments, out to the regions of the stellar halos where the imprints of the mass assembly reside \citep[e.g.][]{Duc2015, Capaccioli2015, Trujillo2016, Mihos2017}. 
Investigations of mass assembly in the outer regions of galaxies have been conducted also by means of stellar kinematics and population properties \citep[e.g.][]{Coccato2010, Coccato2011, Ma2014, Barbosa2018, Veale2018, Greene2019} and kinematics of discrete tracers like globular clusters (GCs) and planetary nebulae (PNe) \citep[e.g.][]{Coccato2013, Longobardi2013, Spiniello2018, Hartke2018}.
In particular, studies of the stellar population gradients out to the region of the stellar halos reveal a different age and chemical composition 
as function of the galacto-centric radius. They indicate a star formation history in the central in-situ component which differs from that in the galaxy outskirts
\citep[e.g.][]{Greene2015, McDermid2015, Barone2018, Ferreras2019}. 
This is consistent with the cosmological simulations of the massive ETGs that predict 
different metallicity profiles as function of the mass assembly history, with shallower profiles in the outskirts when repeated  mergers occur \citep{Cook2016}.}


 {Detailed analysis of integral-field measurements of the stellar kinematics showed that the massive ETGs, 
above the critical mass $M_{\rm crit} \simeq 2 \times
10^{11}$~M$_{\odot}$, are slow rotators (SRs) and are found in the dense environment, as the cluster core \citep{Cappellari2013b}.
They seem to result from an evolutionary track that differs from that of the fast rotators (FRs) ETGs. They are formed via intense star formation at high 
redshift at the centre of a massive dark matter halo, whereas the FRs originate from the transformation of star-forming disc galaxies, 
where a bulge grows and star formation then quenches. }

 {Therefore, clusters of galaxies are excellent sites to study many processes of galaxy 
transformation, even considering that the environmental effects can start to act already in the less dense environments, as groups, which, according to the hierarchical assembly framework, will merge into the cluster potential \citep[e.g.][]{DeLucia2012, Vija2013, Cybulski2014, Hirschmann2015}.  
In this context, the Fornax3D (F3D) project \citep[hereafter S18]{Sarzi2018} provides a unique and complete integral-field spectroscopic dataset for the 
33 galaxies (23 ETGs and 10 LTGs) brighter than $m_B \leq15$~mag inside the virial radius of the Fornax cluster (Table~\ref{tab:sample}). 
Observations were taken with the Multi Unit Spectroscopic Explorer \citep[MUSE,][]{Bacon2010} on the ESO Very Large Telescope. 
The science objectives of the F3D project, the observational strategy, and the data reduction are described in  \citetalias{Sarzi2018}. }

 {Since the data set span a quite large range of stellar masses ($10^9 < M_\ast < 10^{12}$ M$_\odot$) and morphological types, and cover the galaxies 
from their bright central regions to the faint outskirts, where the surface brightness is $\mu_B \geq 25$ mag~arcsec$^{-2}$,
F3D allows to trace the galaxy evolution and mass assembly inside the cluster. 
By analysing the stellar and ionised-gas kinematics for all the F3D galaxies and the line-strength maps for the ETGs,
the main goals of the present paper is to provide a global view of the assembly history of the cluster. 
In addition, taking advantage of the longer integration times in the MUSE pointings on the galaxy outskirts, F3D data allow to derive
a preliminary stellar population content of the stellar halo in the ETGs cluster members. }

 {First results on the galaxy structure and evolution inside the cluster using F3D data were published by \citet[][hereafter P19a and P19b]{Pinna2019a, Pinna2019b}. From the detailed analysis of the stellar kinematics and stellar population 
of the  edge-on lenticular galaxies FCC~153, 170, and 177, these studies addressed the thick-disc origin in the Fornax cluster.
 The first stellar population map and analysis for one of the brightest galaxies in the sample, FCC~167, is provided by \citet{MartinNavarro2019}, who
 studied the spatial variation of the initial mass function in this galaxy. In the same object,
by combining the ionised-gas maps from F3D and ALMA data, \citet{Viaene2019} analysed the properties of the dust in the centre and discussed
the origin of the interstellar medium in ETGs in the general framework of galaxy formation.}

 {This paper is organised as follows. Sect.~\ref{sec:fornax} gives a brief review of the main recent studies on the Fornax cluster 
from deep surveys and previous integral-field data. 
Sect.~\ref{sec:data} provides a brief summary of the observations and data reduction.
Sect.~\ref{sec:analysis} describes the derivation of the stellar kinematics, the properties of the ionised gas, and the measurements of the line-strengths and stellar population parameters. 
Sect.~\ref{sec:results} describes the results and includes the derivation of averaged kinematic and line-strength properties for the F3D ETGs. Sect.~\ref{sec:PPS} and Sect.~\ref{sec:outskirts} discuss the results in the context of the cluster environment. 
Sect.~\ref{sec:concl} contains a summary of the main conclusions.   
The resulting maps together with a description of the properties of the individual galaxies are presented in Appendix~\ref{sec:kin_map} and \ref{sec:description}, respectively.}


%

\section{The Fornax cluster}\label{sec:fornax} 


The Fornax cluster is one of the best sites to study the properties of galaxies in an environment dominated by the gravitational potential of the cluster. 
It is the second most massive galaxy concentration within 20 Mpc \citep{Blakeslee2009}, after the Virgo cluster, with a virial mass of $M_{\rm vir}= 7 \times 10^{13}$ 
M$_\odot$ \citep{Drinkwater2001}. Extensive multi-wavelength observations are available for Fornax, including the Herschel survey \citep{Davies2013}, Chandra 
and XMM-Newton imaging of the cluster core \citep{Scharf2005, Su2017}, and Galaxy Evolution Explorer UV-imaging \citep{Martin2005}. 
In the optical wavelength range, the Hubble Space Telescope data \citep{Jordan2007}, the Dark Energy Camera data from 
the Next Generation Fornax Cluster Survey \citep{Munoz2015}, and the Fornax Deep Survey (FDS) with the VLT Survey Telescope \citep[VST,][]{Venhola2017, Iodice2018} are the deepest and widest datasets mapping the Fornax cluster out to the virial radius \citep[$R_{{\rm vir}} \sim 0.7$~Mpc,][]{Drinkwater2001}. 
Upcoming data from the Atacama Large Millimeter/submillimeter Array \citep[ALMA,][]{Zabel2019} and neutral hydrogen data from the MeerKat survey \citep{Serra2016} will provide a complete census of the cool interstellar medium in Fornax. 
In addition, long-slit, multi-object, and integral-field spectroscopy with several instruments mapped the stellar kinematics 
and populations of the cluster members \citep{Bedregal2006, Scott2014}.

The wealth of data available for the Fornax cluster allows to trace the cluster assembly history in great detail. The cluster hosts a vast population of dwarf and ultra compact galaxies \citep{Munoz2015, Hilker2015a, Schulz2016, Venhola2017, Venhola2018, Eigenthaler2018}, an intra-cluster population of GCs \citep{Schuberth2010, Dabrusco2016, Cantiello2018, Pota2018} and PNe \citep{Napolitano2003, McNeil2012, Spiniello2018}.
 {It has a complex structure, indicative of continuing mass assembly} \citep{Drinkwater2001, Scharf2005, Iodice2018}.
The core is in an evolved state \citep{Jordan2007}, since most of the bright ($m_B<15$~mag) cluster members have transformed into ETGs, 
more so than in the Virgo cluster \citep{Ferguson1989}. 
The FDS data suggest that the bulk of the gravitational interactions between galaxies takes place in the west-northwest core region, where most of the bright 
ETGs are located and where the intra-cluster baryons (i.e., diffuse light, GCs, and PNe) are found \citep{Dabrusco2016, Iodice2018, Pota2018, Spiniello2018}. 
The west-northwest clump of galaxies may be a group falling into the cluster, which modified the structure of the galaxy outskirts (making asymmetric stellar 
halos), and produced the intra-cluster baryons, concentrated in this region of the cluster.

The ROSAT and XMM-Newton data have shown that the Fornax X-ray halo contains  several components whose centroids are offset with respect to the optical 
galaxies.  {This is likely due to the sloshing movement of NGC~1399 and other bright galaxies within the central dark matter halo 
traced by the X-ray gas, combined with the infall of the NGC~1316 group into the cluster. In this process, the collisional component (i.e., the hot gas) 
is lagging behind the non-collisional ones (i.e., stars, GCs, and galaxies) due to ram pressure effects \citep{Paolillo2002, Su2017, Sheardown2018}.}
A comparison of the optical and X-ray data reveals that the reddest and most massive galaxies in the high-density region of the cluster, where the X-ray emission is still detected, have been depleted of their gas content by processes such as harassment or suffocation induced by ram pressure stripping, so the star formation stopped earlier than for the galaxies in the low-density region of the cluster \citep{Iodice2018}.

 {
Stellar kinematics and stellar population gradients from long-slit data were provided by \citet{Bedregal2006, Bedregal2011} for 9 S0 galaxies. The measurements extend well beyond the bulges of these systems and allowed to map the 
kinematics and stellar population gradients in some detail for the first time.  
This revealed the existence of substructures, 
indicating possible interactions in the past. From the stellar population analysis, the resulting age and metallicity gradients are correlated, 
suggesting differences in the star formation history between the central and outer parts of the galaxies: the star formation ceased first 
in the outskirts, transforming the system in a bulge-dominated S0  \citep{Bedregal2011}.
\citet{Scott2014} published integral field measurements for 10 members of the Fornax cluster (some of them overlapping with the galaxies of \citealt{Bedregal2006}) and mainly focused on the distribution of SRs and FRs inside the virial radius of the cluster.
Compared to the more massive clusters, like Virgo and Coma, Fornax shows a decreasing fraction of SRs towards smaller projected environmental density 
at large cluster-centric radi. One of the only two SRs found by \citet{Scott2014} is the bright cluster galaxy NGC~1399 in the core of the cluster.}

 {In addition to the previous works on the Fornax cluster, the F3D data provide high-resolution spatially-resolved measurements of the 
internal structure, kinematics of stars and ionised gas, and of the stellar populations and the ionisation state of the gas for all the 
bright galaxies inside the virial radius, therefore for a complete sample in morphology, stellar masses and radial distribution. 
Moreover, compared to the recent long-slit and integral-field data available for Fornax cited above, the MUSE mosaics allow to map  
the stellar kinematics and stellar populations for the ETGs at larger radii (Fig. \ref{fig:Re_coverage}, right panel), 
i.e. two times more extended than the integral-field data of \citet{Scott2014} 
and extending more than 1--2 $R_{\rm e}$ covered by the long-slit data of \citet{Bedregal2006}.}


\section{Observations and data reduction}\label{sec:data}





The MUSE observations were carried out in service mode between July 2016 and December 2017. 
The wide-field mode ensured a field of view of $1 \times 1$~arcmin$^2$ with
a spatial sampling of $0.2 \times 0.2$ arcsec$^2$.
The wavelength range from 4650 to 9300 $\AA$ was covered with a spectral sampling of 1.25 \AA~pixel$^{-1}$ and a nominal spectral resolution of $\rm FWHM_{\rm inst} = 2.5$ \AA\ at 7000 \AA . 
 {The measured spectral resolution was on average $\rm FWHM_{\rm inst} = 2.8$ \AA\ with little variation ($<0.2$ \AA ) with wavelength and position over the field of view. 
This is slightly larger than the nominal value due to the combination of different offset exposures taken at different position angles.
As described in \citet{Sarzi2018}, the measured spectral resolution was derived, 
for each galaxy, by using the twilight datacubes. 
They were combined using the same observational pattern (including 
rotations and offsets) as the galaxy exposures. 
The resulting line-spread function (LSF) was derived from the combined twilight cubes with a solar-spectrum 
template. The process was divided in a number of wavelength intervals and for each spaxel in order to measure the variation of the 
fitting parameters with wavelength and position over the field of view.
For the analysis presented in this paper, the parametric LSF derived for the MUSE 
Ultra Deep Field 10 by \citet{Bacon2017} was adopted.}

\begin{figure}[t!]
\includegraphics[width=\hsize]{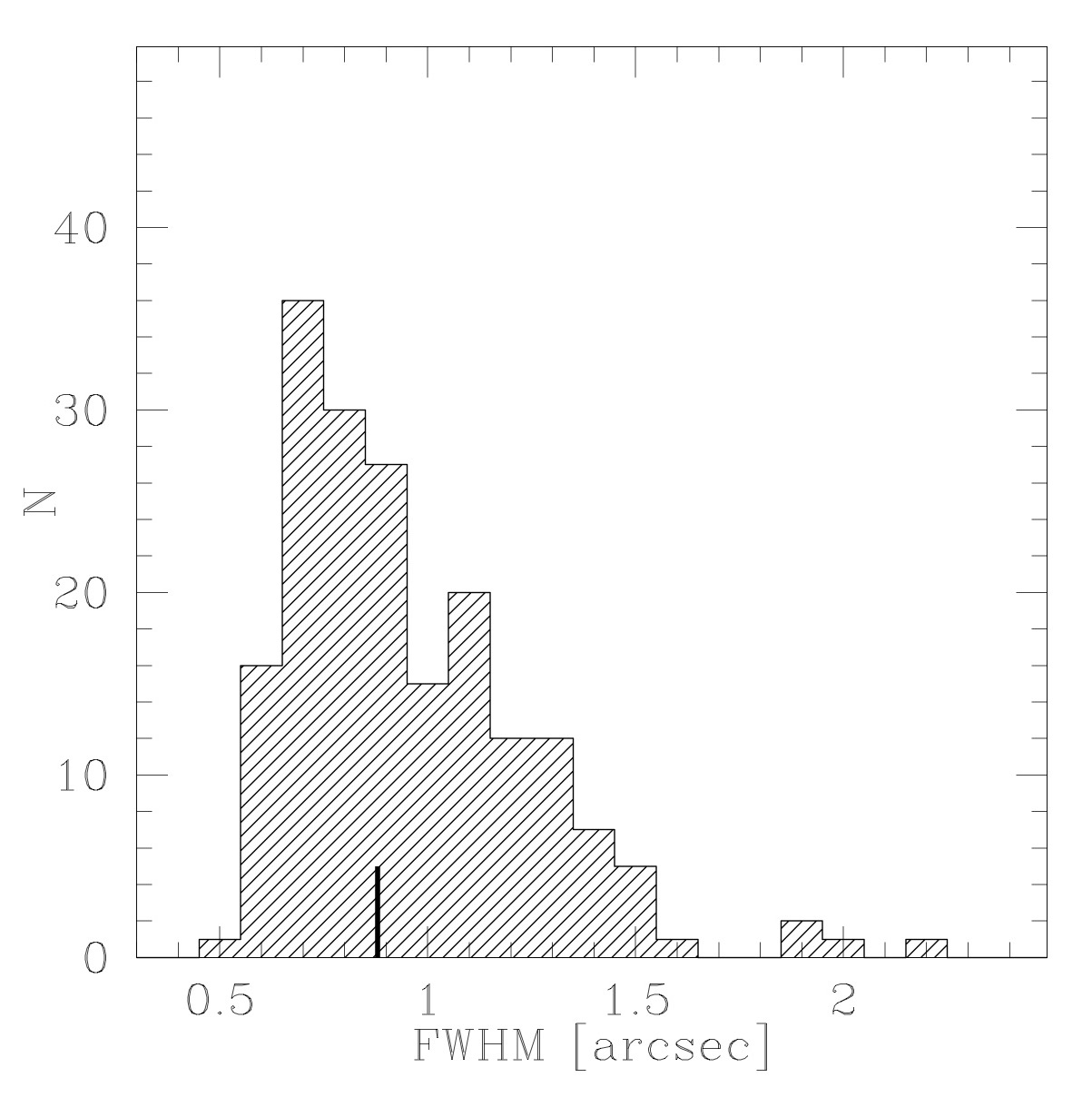}
\caption{Distribution of the seeing FWHM measured on the F3D MUSE pointings. The vertical thick segment marks the median value.}
\label{fig:PSF}
\end{figure}

The total integration times for the central and middle pointings are about 1 hour, 
while an integration of 1.5 hour was necessary to reach the limiting surface brightness of $\mu_B \simeq 25$ mag~arcsec$^{-2}$ in the outer pointings. 
A dither of a few arcseconds and a rotation by $90^\circ$ were applied to the single exposures, in order to minimise the signature of the 24 MUSE slices 
on the field of view. Sky frames were acquired immediately before/after each science exposure in order to perform sky modelling and subtraction on 
the single spaxels.
The observations were done in good seeing conditions with a median $\rm FWHM = 0.88$~arcsec (Fig.~\ref{fig:PSF}).
When available, some field stars were used to measure the FWHM.
For central pointings, the FHWM of the nucleus was measured when there were no stars available.

\begin{figure*}[t!]
\includegraphics[width=9cm]{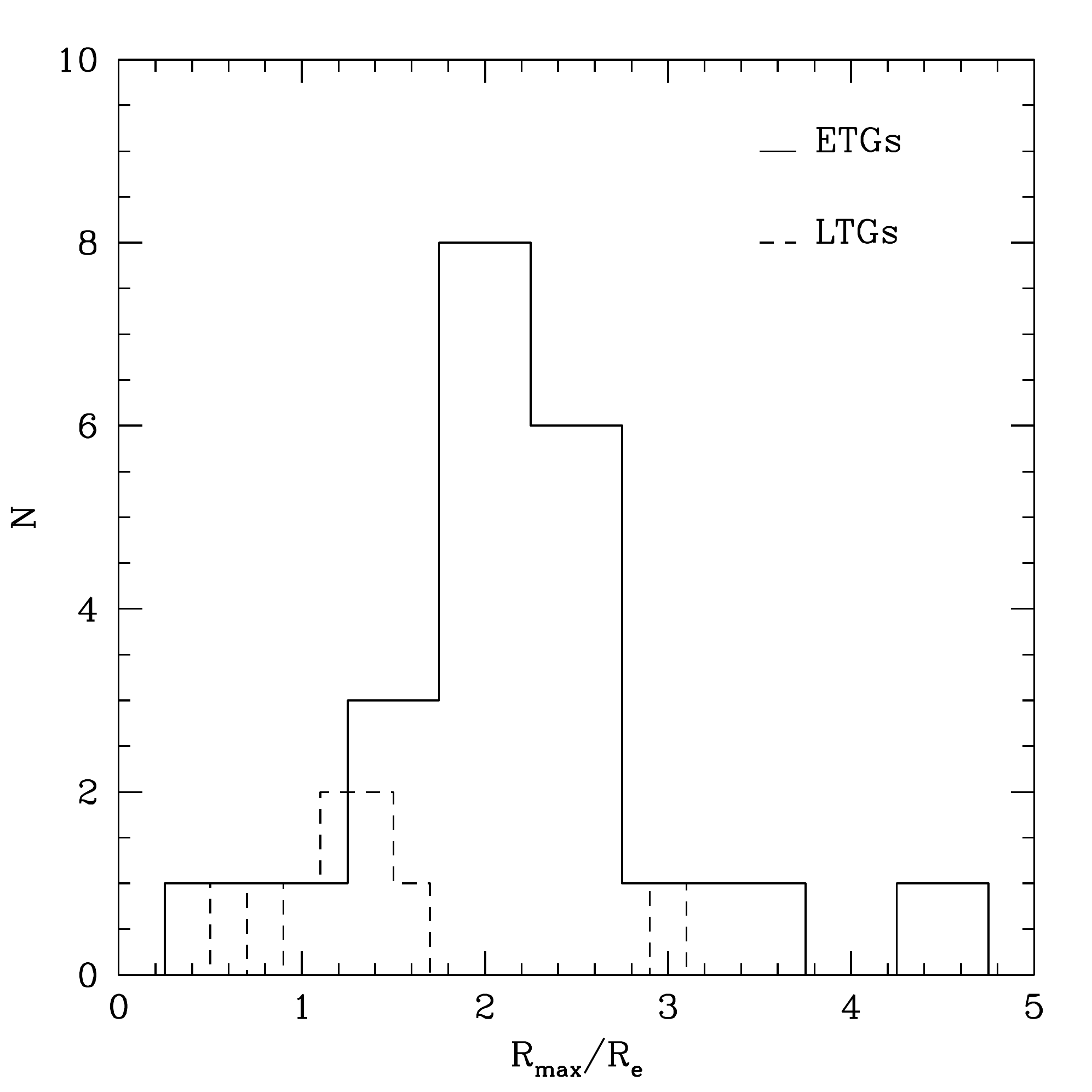}
\includegraphics[width=9cm]{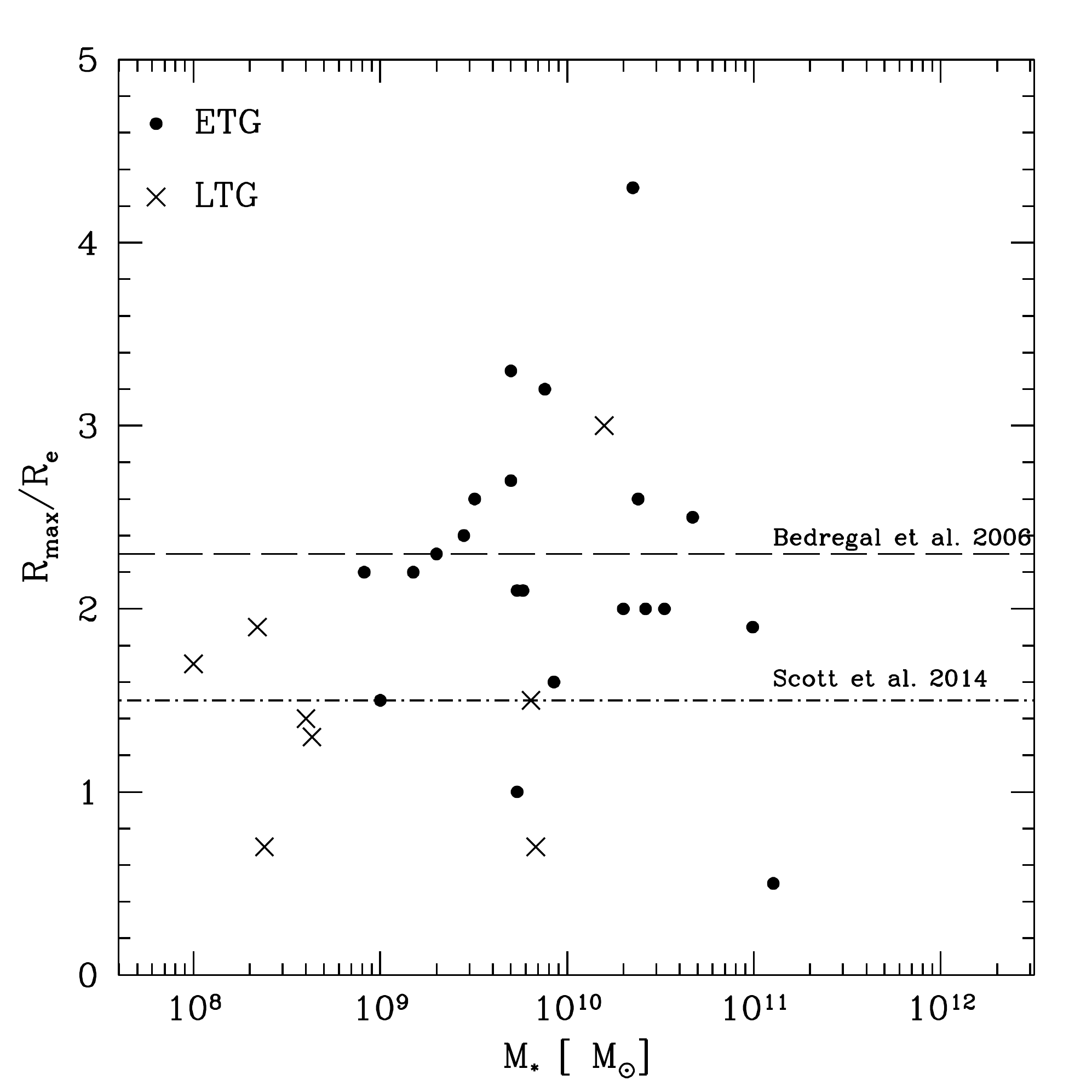}
\caption{{\it Left:\/} Distribution of the spatial coverage of the F3D MUSE data for ETGs (solid line) and LTGs (dashed line) in units of effective radius, as measured along the major axis. {\it Right:\/} Spatial coverage of the F3D ETGs (circles) and LTGs (crosses) as a function of the total stellar mass.
 {The long-dashed and dash-dotted lines correspond to the average spatial coverage from the long-slit data of \citet{Bedregal2006} and integral-field data of \citet{Scott2014}, respectively. The adopted values of $R_{\rm e}$ for the comparison with literature are given in Table~\ref{tab:sample}.}}
\label{fig:Re_coverage}
\end{figure*}

The data reduction was performed with the MUSE pipeline version 2.2 \citep{Weilbacher2012, Weilbacher2016} under the ESOREFLEX environment
\citep{Freudling2013}. The main steps included bias and overscan subtraction, flat fielding to correct the pixel-to-pixel response variation, wavelength calibration, determination of the line spread function, and illumination correction. By using the dedicated sky frames, the sky subtraction was done by fitting and subtracting a sky model spectrum on each spaxel of the field of view. 
The flux calibration and the first-order correction of the atmospheric telluric features were obtained using the spectro-photometric standard star observed at twilight.
For each galaxy of the sample, the single pointings were aligned using reference stars and then combined to produce the final MUSE mosaics. 
These mosaics map the galaxy structure and stellar population out to 2--3 $R_{\rm e}$ for the ETGs and out to 1--2 $R_{\rm e}$ for the LTGs (Fig.~\ref{fig:Re_coverage}, left panel). The MUSE data extend out to 2--4 $R_{\rm e}$  (Fig.~\ref{fig:Re_coverage}, right panel) for the more massive F3D galaxies (with a total stellar mass $M_\ast \geq 10^{9}$--$10^{10}$ M$_{\odot}$). Table~\ref{tab:sample} lists 
the radial extent of the MUSE data along the major axis of each galaxy. 

The quality of the F3D MUSE data was extensively assessed for FCC~167 in \citetalias{Sarzi2018}. They demonstrated that the residuals of the fitted spectra do not show any systematic feature over the whole adopted wavelength range. 
From the original sample listed in \citetalias{Sarzi2018}, FCC~267 was not observed and we do not have the data for this galaxy.  
The data analysis for FCC~213 (NGC~1399) will be the subject of a forthcoming paper including the F3D middle and halo pointings and the MUSE archival data covering the central regions.

      
      
\placetabone

\section{Data analysis}\label{sec:analysis}
   

   
The methods described in \citetalias{Sarzi2018}, \citetalias{Pinna2019a}, and \citetalias{Pinna2019b} 
were used to derive the kinematic and line-strength maps. 
These were obtained with the modular pipeline for the analysis of integral-field spectroscopic data developed by \citet{Bittner2019}. 
Specific details on the main steps of the data analysis are provided in the following subsections.  

\subsection{Stellar kinematics}
 
The derivation of the stellar kinematics was done with the Penalised Pixel-Fitting code (pPXF; \citealt{Cappellari2004}, \citealt{Cappellari2017}). 
 {It is based on a penalised maximum likelihood approach} and it extracts a line-of-sight velocity distribution (LOSVD) parametrised by the mean velocity $v$, velocity dispersion $\sigma$, and Gauss-Hermite moments $h_3$ and $h_4$ \citep{Gerhard1993,vanDerMarel1993}. 
The MILES single stellar population (SSP) models \citep{Vazdekis2012, Vazdekis2015} with a spectral resolution 
of ${\rm FWHM} = 2.51$ $\AA$ \citep{FalconBarroso2011} were used as spectral templates. 
 {This model library covers a large range in age (from 30 Myr to 14 Gyr), total metallicity ($-2.27 \leq$ [$M$/H] $\leq 0.4$ dex), and $\alpha$-element overabundance ($[\alpha/{\rm Fe}]=0$ and 0.4 dex),} thus minimising effects of template mismatch. 
 {All the MILES templates were broadened to the MUSE spectral resolution before performing the fit.}
The analysis was conducted on the full MUSE rest-frame wavelength range between 4800 and 9000 \AA, in order to take into account 
also the NaD and CaT lines.

The emission lines in this wavelength range were masked during the analysis of the stellar 
kinematics. In addition, a fourth-order multiplicative Legendre polynomial was included in 
the fit to account for small differences in the continuum shape between the F3D data and 
spectral templates. 

A reliable extraction of stellar kinematics, in particular of the higher-order Gauss-Hermite moments, requires a relatively high signal-to-noise ratio $S/N$ \citep[e.g.][]{Gadotti2005}. Thus, the Voronoi binning scheme of \citet{Cappellari2003} was applied to spatially bin the data to an approximately constant $S/N=40$. Spaxels which surpass this $S/N$ threshold remain unbinned.  Previous studies on F3D data (\citetalias{Sarzi2018,Pinna2019a,Pinna2019b}) have shown that $S/N=40$ provides sufficient signal for the analysis whilst preserving a high spatial resolution.
In addition, a minimum $S/N$ threshold is applied by removing all spaxels below the isophote level which has an $\langle S/N \rangle \sim 3$. 
This avoids systematic effects in the low surface-brightness regime of the observations. 
The heliocentric systemic velocity from the stellar velocity field and the average velocity dispersion inside the r-band effective radius are given in Table~\ref{tab:sample}.

Uncertainties in the stellar kinematics were estimated by performing Monte Carlo simulations \citep[e.g.][]{Cappellari2004, Wegner2012}. This procedure is discussed in detail in \citetalias{Pinna2019b} for FCC~170. They found errors of 6~km~s$^{-1}$ for $v$, 9~km~s$^{-1}$ for $\sigma$, and $0.03$ for both $h_3$ and $h_4$. Similar error estimates were also obtained by \citetalias{Sarzi2018} for FCC~167 and \citetalias{Pinna2019a} for FCC~153 and FCC 177. 
For two galaxies of the sample, FCC~113 and FCC~285, which are LTGs with strong emission lines, the true velocity dispersion is 
so much below the instrumental resolution that pPxf has real problems getting a realistic value. For these two objects, 
the pPxF fit was constrained to determine only the first two moments of the LOSVD, i.e., velocity and $\sigma$. 
 The error estimates on the $\sigma$ for these two objects, derived from the Monte Carlo simulations are quite large ($\geq 80$\%), therefore, the average values $\sigma_e$ are not taken into account in the present study.
 

The radial profile of the kinematic position angle PA$_{\rm kin}$ was measured by means of the Kinemetry code \citep{Krajnovic2006}, and  compared with the radial profile of the photometric position angle PA$_{\rm phot}$ from \citet{Iodice2018} for ETGs and \citet{Raj2019} for LTGs. T
he mean value of PA$_{\rm kin}$ and the difference $\Delta$PA with respect to the mean value of PA$_{\rm phot}$ are given in Table~\ref{tab:sample}. The kinematic PA is poorly estimated for galaxies with an irregular rotation pattern and/or very low values of rotational velocity like FCC~90 (Fig.~\ref{fig:FCC090map}). For galaxies characterised by the presence of strong dust lanes in their inner regions, the mean value of PA$_{\rm kin}$ was calculated for radii larger than 5 arcsec. 

The stellar kinematic maps of $v$, $\sigma$, $h_3$, and $h_4$ derived from $S/N=40$ Voronoi-binned MUSE data and the radial profiles of PA$_{\rm kin}$ and PA$_{\rm phot}$ for the F3D galaxies are shown in Appendix~\ref{sec:kin_map}. 

\subsection{Ionised-gas distribution and kinematics}
  
The derivation of the ionised-gas distribution and kinematics was done with the analysis approach discussed in \citetalias{Sarzi2018}. A simultaneous spaxel-by-spaxel fit was done for both the stellar and ionised-gas contribution to the MUSE spectra with the Gas 
and Absorption Line Fitting code \citep[GandALF,][]{Sarzi2006, FalconBarroso2006} in the wavelength range between 4800 and 6800 \AA\ and using a two-component reddening correction to adjust for the stellar continuum shape and observed Balmer decrement. 
 {In an effort to decrease the impact of template-mismatch on the emission-line measurements, all the models of the MILES library were used in order to achieve the best possible fit to the stellar continuum.}
As illustrated in \citetalias{Sarzi2018} and \citet{Viaene2019} this provides a rich set of ionised-gas parameter maps. For the purpose of this work, only the maps of the observed flux and velocity of the H$\alpha$ emission and of the classification of the ionised-gas emission according to the standard  {Baldwin, Phillips \& Terlevich (BPT)} diagnostic diagram \citep{Baldwin1981} were analysed. These are shown in 
Appendix~\ref{sec:kin_map} for the 13 F3D objects with a significant amount of diffuse ionised-gas emission, 
which generally occurs in the central regions. 
The remaining galaxies display only a number of unresolved sources of  [\ion{O}{iii}]$\lambda$5007 emission. These are PNe, which will be discussed in a forthcoming paper. 

The star-formation rate (SFR) was calculated through the conversion SFR~( M$_{\odot}$~y$^{-1}$)$\,=\,5.5 \times 10^{-42}$ L$_{\rm H\alpha}$ (erg s$^{-1}$) provided
by \citet{Calzetti2012} and 
from the total de-reddened H$\alpha$ fluxes assuming a distance of 20 Mpc \citep{Blakeslee2009}, in the absence of secure
distance measurements for the majority of the line-emitting objects in the F3D sample.
Table~\ref{tab:emission} provides two estimates of the spatially integrated SFR: SFR$_{\rm up}$, which combines  all spaxels classified as exhibiting 
'star-forming' or 'transition' emission-line
ratios in the BPT diagnostics (this is considered as an upper limit on the SFR, assuming that all transition spaxels are powered by O-type stars); and a second estimate, SFR, which includes only the spaxels classified as 'star-forming', and is considered as a lower limit. 

Table~\ref{tab:emission} also provides measurements for the equivalent
width of the de-reddened\footnote{\bf In the context of the Gandalf 
fits, reddening does have an impact on the equivalent width of emission lines 
since nebular emission and the stellar continuum can suffer from different amount of extinction.} 
H$\alpha$ emission when observed against the
integrated MUSE spectrum of the galaxy. This gives an indicative
measure for the specific star-formation rate in the regions with
ionised-gas emission, which generally lie within the central MUSE
pointing. As in the case of SFR estimates, the upper-limits combine 'star-forming' and 'transition' spaxels, and the lower limits are  based on 'star-forming' spaxels only. 



\subsection{Line-strength indices and stellar population properties}
 
The line-strength indices maps were derived from the MUSE spectra by adopting a threshold of $S/N=200$ for the  
Voronoi binning. In addition, to avoid regions where the sky subtraction could have left residuals,
the analysis was performed on a restricted wavelength range from 4800 to 5500 \AA. 
The emission-line analysis was repeated to obtain emission-subtracted spectra. 
The line-strength indices of H$\beta$, Fe5015, Mg$b$, Fe5270, and Fe5335 were computed in the LIS system \citep{Vazdekis2010, Vazdekis2015} by 
adopting the routines of \citet{Kuntschner2006}.

Subsequently, the measured line-strength indices were compared to those predicted by the MILES model library \citep{Vazdekis2012} 
to provide the stellar age, [$M$/H], and [Mg/Fe] abundance. 
The best-fitting SSP model was determined by means of a Markov-Chain-Monte-Carlo algorithm following the prescriptions by \citet{MartinNavarro2018}.  

 {Since the line-strength analysis is based on the emission-line subtraction that comes from a GandALF fit with the SSP models,
for galaxies having strong emission lines, i.e., almost all LTGs of the F3D sample, an imperfect flux calibration could introduce a bias in the estimate of the emission 
lines. Therefore, due to such an additional complication,  
the stellar population analysis of the LTGs will be presented in detail in a forthcoming paper (Mart\'in-Navarro et al., in prep.).}

The maps of the H$\beta$, Fe5015, and Mg$b$ line-strength indices for all the F3D ETGs are shown in Appendix~\ref{sec:kin_map}.


\section{Results}\label{sec:results}

This section presents an overview of the global properties of the F3D galaxies in terms of mean stellar kinematic properties, ionised-gas emission and star formation activity, and mean stellar population properties. A detailed description of the individual galaxies is provided in Appendix~\ref{sec:description}.

\subsection{Mean stellar kinematic properties}
\label{sec:kin}
   




The simplified scheme for kinematical classification presented by \citet{Schulze2018}, based on \citet{Krajnovic2011},
was adopted to describe the average stellar kinematics maps.
Visual inspection reveals that  
the velocity fields of 22 out 31 F3D galaxies ($71\%$) constitute a regular spider diagram and therefore 
these galaxies are classified as regular rotators. 
Most of the ETGs and all the LTGs, except for FCC~306, are regular rotators. 
FCC~306 is the faintest galaxy of the sample (Table~\ref{tab:sample}) and it resembles a prolate 
rotator because its axis of rotation coincides with the photometric major axis (Fig.~\ref{fig:FCC306map}). 
The kinematical classification for all the F3D galaxies is given in Table~\ref{tab:sample}. 
A kinematically distinct core is observed in five F3D galaxies ($16\%$). All of them are ETGs, whose velocity field shows a central rotating 
component but low or even no rotation at larger radii. In all but one of the galaxies with a distinct core, the inner structure is 
characterised by a higher velocity dispersion and a redder colour \citep[$g-i\sim1.2$~mag,][]{Iodice2018} with respect to the 
surrounding regions. FCC~301 hosts an embedded disc,
which was also identified by \citet{Bedregal2006} with long-slit spectroscopy. 
It has a lower velocity dispersion (Fig.~\ref{fig:FCC301map}) and bluer colour \citep[$g-i\sim0.5$~mag,][]{Iodice2018} than the outer regions. 
A kinematically decoupled core is observed in two ETGs  ($6\%$). They are FCC~184 and FCC~193, which are the two brightest barred galaxies of the 
sample in the core of the cluster. 
 {As discussed in Sec.~\ref{sec:PPS}, the different kinematic and photometric properties of the decoupled components correlate with the 
location of their host galaxies 
inside the cluster and suggest a different formation process.}



\begin{figure*}[t!]
\includegraphics[width=9cm]{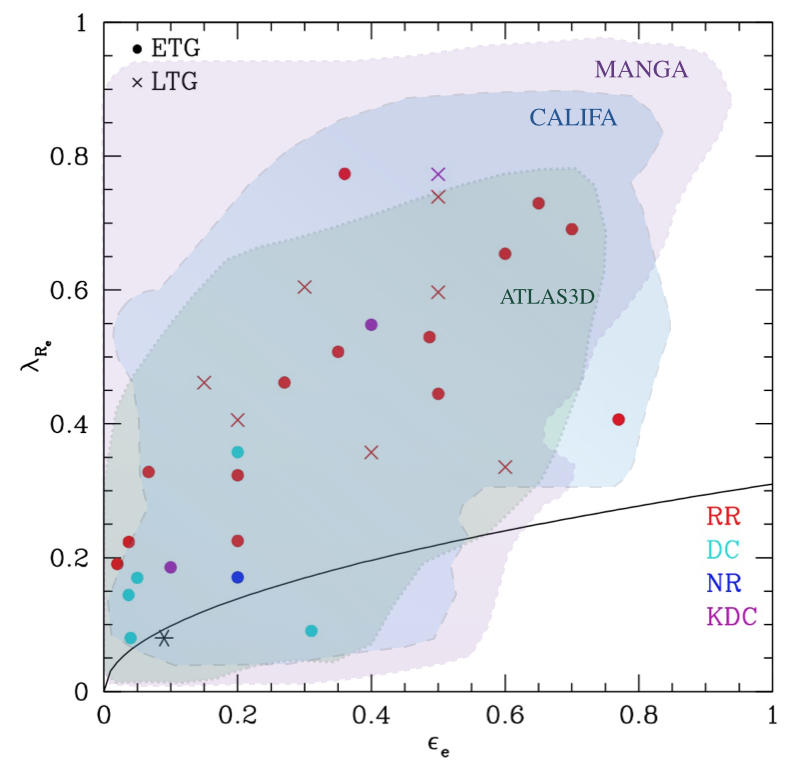}
\includegraphics[width=9cm]{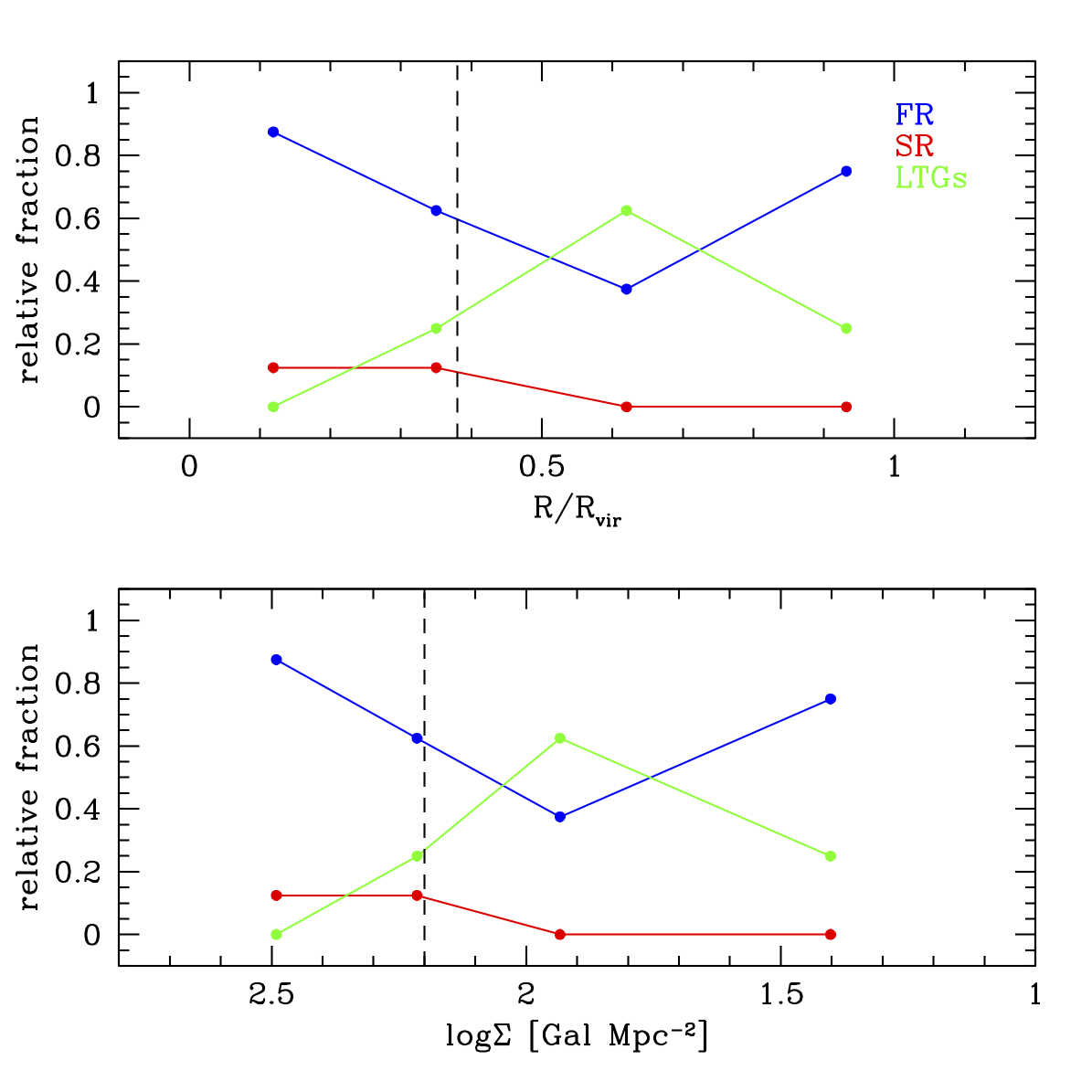}
\caption{{\it Left\/}: Specific stellar angular momentum of the F3D ETGs (circles) and LTGs (crosses) as a function of the ellipticity measured at the effective radius. Galaxies classified as regular rotators (RRs) are shown with red symbols. The light-blue symbols correspond to galaxies with a distinct core (DC). The blue circle corresponds to the only non-rotating ETG of the sample (NR). Galaxies with a kinematically decoupled core (KDC) are shown with magenta symbols. The black asterisk corresponds to FCC~213 (NGC~1399) and it is taken from \citet{Scott2014}. The black solid line divides FRs from SRs according to \citet{Emsellem2011}. 
 {The light green, light blue, and light magenta areas indicate the location of the galaxies from the ATLAS3D \citep{Emsellem2011}, CALIFA \citep{FalconBarroso2011},  and MANGA survey \citep{Graham2018}, respectively.}.
{\it Right:\/} Relative fraction of the F3D SRs (red circles), FRs (blue circles), and LTGs (green circles) as function of the projected radius from the centre of the Fornax cluster in units of virial radius ({\it top\/}) and as a function of the local projected density of the cluster ({\it bottom\/}). The vertical dashed line separates the high and low-density regions of the cluster.}
\label{fig:LambdaR}
\end{figure*}

Following \citet{Emsellem2011}, the mean of the specific stellar angular momentum $\lambda_{R_{\rm e}}$ 
of all the F3D galaxies was computed within $R_{\rm e}$.
It is plotted as a function of the ellipticity $\epsilon_{\rm e}$ measured at $R_{\rm e}$ in 
Fig.~\ref{fig:LambdaR} (left panel) to identify fast-rotating (FRs) and slow-rotating galaxies (SRs). The values of $\lambda_{R_{\rm e}}$ and $\epsilon_{\rm e}$ and the kinematic type of all the F3D galaxies are reported in Table~\ref{tab:sample}. 
In the $\lambda_{R_{\rm e}} - \epsilon_{\rm e}$ plane, the F3D galaxies are consistent with values derived for the galaxies in the
ATLAS3D and CALIFA surveys \citep{Emsellem2011,FalconBarroso2011}.
The Fornax cluster is dominated by FRs. Most of them are ETGs and the remaining ones are LTGs. 
Only two F3D galaxies are SRs: 
 {the brightest central galaxy FCC~213 (NGC~1399) and 
FCC~276. Data for FCC~213 comes from \citet{Scott2014}. 
The results are consistent with those found by \citet{Scott2014} on a smaller sample of cluster members.} 
All the ETGs with a distinct core, except for FCC~301, are close to the SR regime. This further indicates that the distinct core of FCC~301 has a different kinematic structure than the others, being similar to a fast-rotating inner disc.

The fractions of the slow and fast-rotating ETGs as well as of the LTGs are shown as function of the projected distance from the cluster centre in Fig.~\ref{fig:LambdaR} (top right panel) and as function of the local projected galaxy density of the cluster in Fig.~\ref{fig:LambdaR} (bottom right panel). 
In the region of transition from higher to 
lower galaxy density at $0.4 R_{\rm vir} \leq R_{\rm proj} \leq 0.8 R_{\rm vir}$ where $\Sigma \leq 40$~Gal~Mpc$^{-2}$ \citep{Iodice2018}, 
FRs and SRs decrease whereas LTGs increase. In the low-density region of the cluster ($R_{\rm proj} > 0.8 R_{\rm vir}$), 
there are no SRs. The fraction of FRs increases, and that of the LTGs decreases. These results 
improve upon the findings of \citet{Scott2014} for Fornax  since all the bright 
galaxies inside the virial radius are now taken into account. In particular, even considering the larger number of galaxies, the 
results confirm that the degree of segregation of the SRs towards the centre in Fornax is less pronounced than in the more massive Virgo and Coma clusters
\citep{Cappellari2011b,Houghton2013}.
The implications for the formation and segregation of the different kinematic types are addressed in 
Sec.~\ref{sec:concl}.

\begin{figure}
\includegraphics[width=9cm]{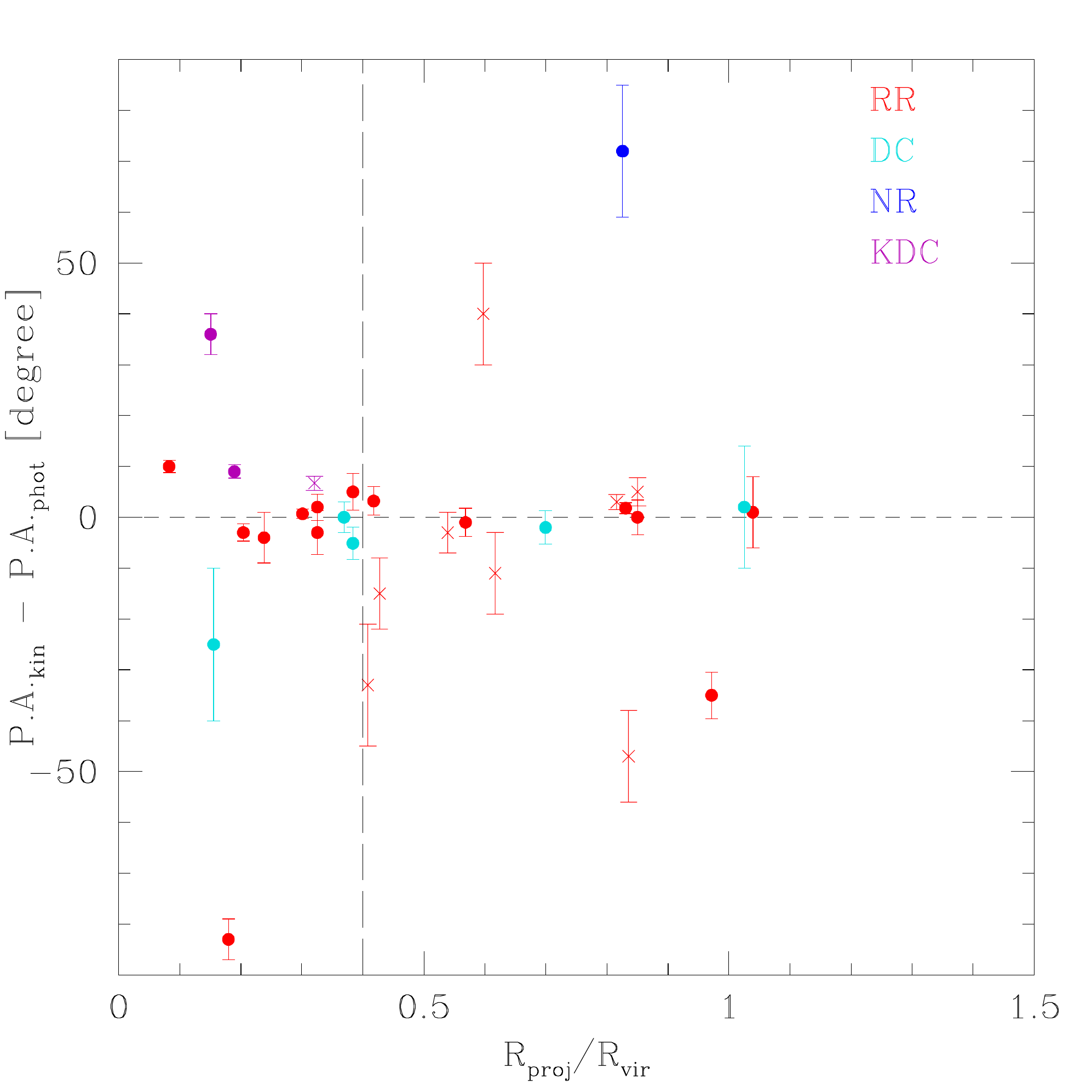}
\caption{Difference between the mean kinematic and photometric position angles for the F3D galaxies as function of the projected radius from the centre of the Fornax cluster in units of virial radius. Symbols are as in the left panel of Fig.~\ref{fig:LambdaR}. The vertical dashed line separates the high and low-density regions of the cluster.}
\label{fig:PA}
\end{figure}

The difference $\rm \Delta PA$ between the mean kinematic and photometric position angle of the F3D 
galaxies is shown in Fig.~\ref{fig:PA} as function of the projected distance from the cluster centre. On average, larger position angle differences ($\rm \Delta PA>2^\circ$) are found for the LTGs with disturbed morphology, as FCC~263, FCC~285, and FCC~306 (Figs.~\ref{fig:FCC263map}, \ref{fig:FCC285map}, and \ref{fig:FCC306map}). In particular, the kinematics for FCC~306 
suggests a prolate rotation for this object.
For the ETGs, larger position angle differences ($\rm \Delta PA>2^\circ$) are found for the 
galaxies in the high-density region of the cluster ($R_{\rm proj} \leq 0.4 R_{\rm vir}$). At larger cluster-centric 
distances, the measured values of $\rm \Delta PA$ are consistent with zero, except for the barred galaxy 
FCC~310 (Fig.~\ref{fig:FCC310map}). 
However, it should be noticed that the four barred galaxies of the sample are characterised by a larger $\rm \Delta PA$ in the region where the bar is prominent (FCC~184, Fig.~\ref{fig:FCC184map}; FCC~190, Fig.~\ref{fig:FCC190map}; FCC~193, Fig.~\ref{fig:FCC193map}; FCC~301, Fig.~\ref{fig:FCC310map}).

\placetabtwo

\subsection{Ionised-gas emission and star formation activity}
\label{sec:emission}


Extended ionised-gas emission was detected in 13 galaxies corresponding to $42\%$ of the cluster members inside the virial radius (Table~\ref{tab:emission}). Five galaxies are ETGs (FCC~90, FCC~119, FCC~167, FCC~184, and FCC~219), three of them close to the cluster core, whereas the remaining eight objects display pervasive star-formation activity. Seven of the eight are LTGs. The exception is FCC~90 which is classified as a peculiar S0. Three galaxies display star-forming regions localised in the nucleus (FCC~119, Fig.~\ref{fig:FCC119map}) or around it (FCC~179, Fig.~\ref{fig:FCC179map}; FCC~184, Fig.~\ref{fig:FCC184map}). The two remaining galaxies  are dominated by the typical diffuse LINER-like emission observed in ETGs (FCC~167, Fig.~\ref{fig:FCC167map}; FCC~219, Fig.~\ref{fig:FCC219map}). However, for FCC~167 \citet{Viaene2019} discussed the evidence for some circumnuclear star formation, as indicated by the presence of molecular gas and 
composite SF/AGN emission.
In FCC~90 the ionised-gas emission is entirely powered by star formation and it shows a concentrated distribution towards the centre, where some modest rotation is detected. A plume of material to the west of the galaxy appears to be unsettled (Fig.~\ref{fig:FCC090map}).
In FCC~312 the nebular emission is dominated by \ion{H}{II} regions, although some highly-excited material protrudes out of the disc near, but not exactly along, the direction of the galaxy minor axis (Fig.~\ref{fig:FCC312map}). 

For most of the F3D galaxies, the observed ionised-gas distribution and kinematics are globally consistent with a settled system in coherent rotation around the galaxy centre. Yet, a regular disc-like kinematics is observed only in approximately half of the galaxies (e.g. FCC~179, Fig.~\ref{fig:FCC179map}; FCC~290, Fig.~\ref{fig:FCC290map}; FCC~312, Fig.~\ref{fig:FCC312map}), whereas in many other objects there is evidence of unsettled material (e.g. FCC~90, Fig.~\ref{fig:FCC090map}; FCC~184, Fig.~\ref{fig:FCC184map}; FCC~219, Fig.~\ref{fig:FCC219map}).


\placetabthree

\begin{figure*}[t!]
\includegraphics[width=9cm]{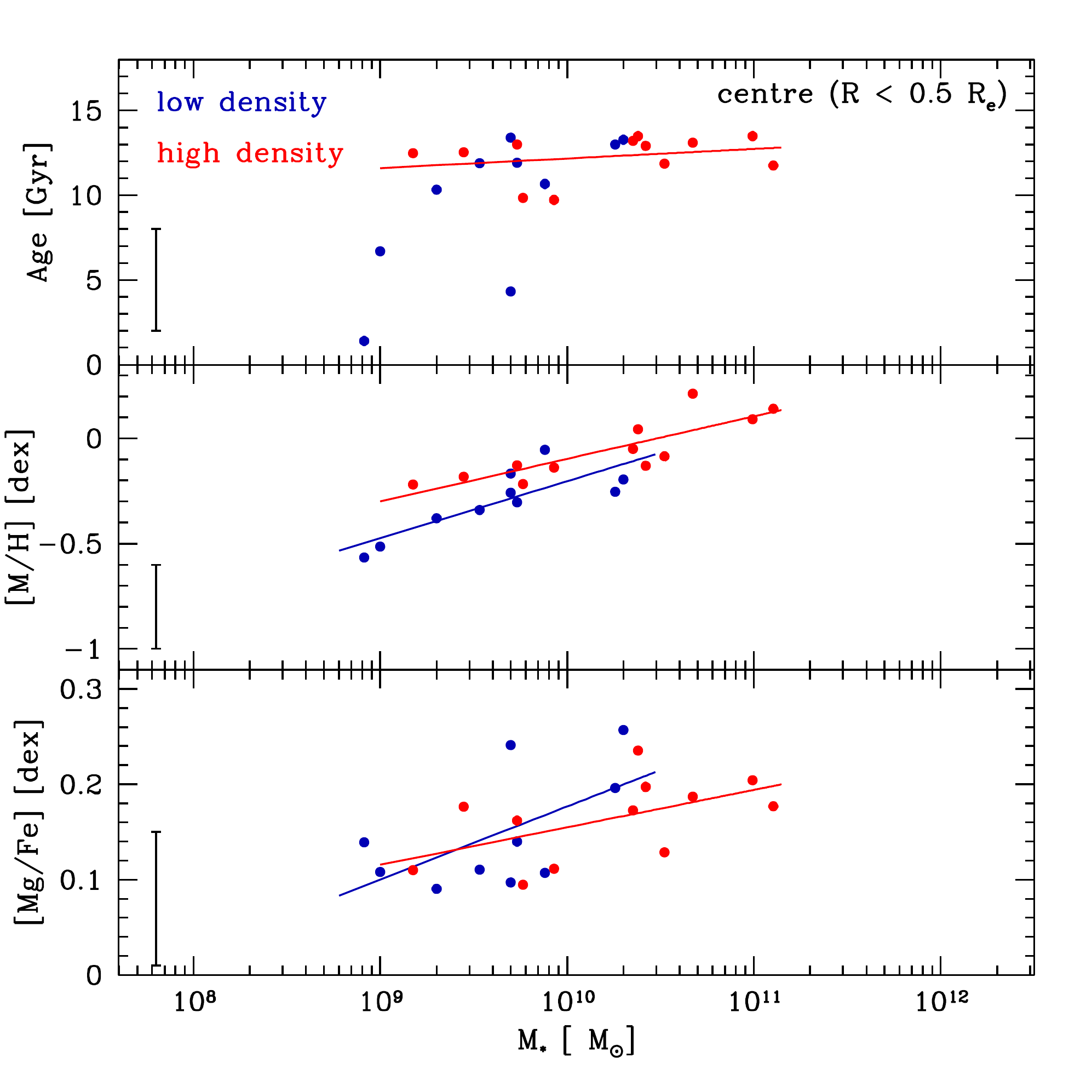}
\includegraphics[width=9cm]{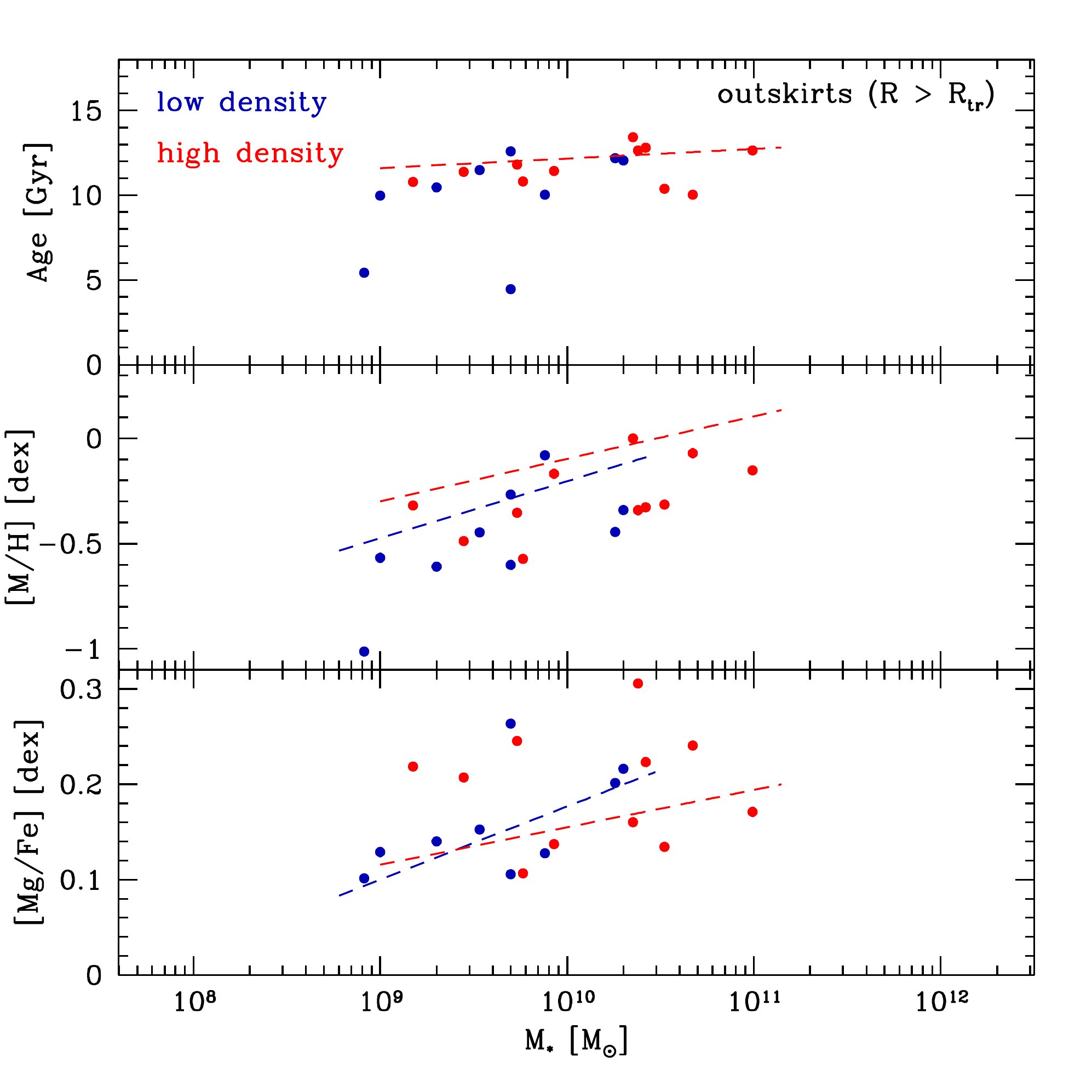}
\caption{{\it Left:\/} Mean age ({\it top\/}), total metallicity [$M$/H] ({\it middle\/}), and [Mg/Fe] abundance ratio ({\it bottom\/}) in the central parts of the F3D ETGs located in the high (red circles) and low-density region (blue circles) of the cluster as a function of the total stellar mass. 
The red and blue lines in the middle and bottom panels are the least-square fits of the values for the central parts of the ETGs in the high and low-density regions of the cluster, respectively. 
The error bars in the low-left corner indicate the average uncertainty derived for each quantity.
{\it Right:\/} Same as in the left panels but for the outskirts of the F3D ETGs. 
The least-square fitting lines of the left panels are shown as dashed lines for comparison.}
\label{fig:SP_all}
\end{figure*}

\subsection{Mean stellar population properties of ETGs}
\label{sec:SSP}
   

  

For all the F3D ETGs, the luminosity-weighted mean of the age, [$M$/H], and [Mg/Fe] abundance was derived from the line-strength indices in two galaxy regions: the bright central part inside $0.5 R_{\rm e}$ and the outskirts, outside the transition radius $R_{\rm tr}$ defined by \citet{Spavone2019} and listed in Table~\ref{tab:SP_analysis}. It varies between $0.6 R_{\rm e}$ and $3.6 R_{\rm e}$. \citet{Spavone2019} presented a multi-component photometric decomposition of the azimuthally-averaged surface-brightness profiles of the ETGs in FDS. They determined the transition radius between the bright central component (which consists mostly of stars formed in situ), a second component corresponding to the outskirts (where there is an increasing contribution from material accreted during the mass assembly process), and a third component corresponding to the envelope.

The central and outer galaxy regions are defined by taking into account the ellipticity and PA from FDS. 
This choice allows a direct comparison of 
the stellar population properties with the colour properties derived from FDS for the same galaxy regions. 
The deep FDS imaging shows that the ETGs in the high-density region of the Fornax 
cluster have redder $g-i$ colours than galaxies at larger cluster-centric distances. 
This behaviour persists in the colour distribution derived in the inner $0.5 R_{\rm e}$. 
 {For the F3D ETGs, the mean age, [$M$/H], and [Mg/Fe] are derived inside the same regions, i.e. in the inner $0.5 R_{\rm e}$. Values are listed in Table~\ref{tab:SP_analysis} and plotted as function of total stellar mass in Fig.~\ref{fig:SP_all} (left panels).}
The ETGs in the high-density region of the cluster ($R_{\rm proj} \leq 0.4 R_{\rm vir}$) have central mean ages in the range between 10 and 14 Gyr with an average value close to 13 Gyr (Fig.~\ref{fig:SP_all}, top left panel). 
The scatter in the mean age of the central parts increases for galaxies in the low-density region ($\sim1-11$~Gyr). 
The three younger galaxies are FCC~90 (at $R_{\rm proj} = 0.83 R_{\rm vir}$ with a central mean age of $\sim 1$ Gyr), FCC~255 (at $R_{\rm proj} = 0.85 
R_{\rm vir}$ with a central mean age of $\sim 4$ Gyr) and FCC~119 (at $R_{\rm proj} = 1.39 R_{\rm vir}$ with a central mean age of $\sim 7$ Gyr). 
FCC~90 is a recent infaller (Fig.~\ref{fig:PPS}) and the FDS data show that this galaxy is very blue close to the centre. 
The younger age and detection of ionised gas in the galaxy centre suggest the presence of ongoing star formation. 
The average of the mean age of all other ETGs in the low-density region of the cluster is $\sim 11$ Gyr, which is similar to the age 
of the galaxies in the high-density region. The more massive galaxies are always the older systems both in the low and high-density 
regions of the cluster.

The relation between the total stellar mass and central mean [$M$/H] is shown in Fig.~\ref{fig:SP_all} (middle left panel) and, as 
expected, it suggests that the more massive galaxies are more metal rich. 
In addition, there is an evident segregation in the central mean [$M$/H] as function of the cluster environment. 
In the same mass range ($1.5\times10^9 < M_\ast <3\times10^{10}$~M$_{\odot}$),  
the ETGs in the high-density region are indeed more metal rich ($[M/{\rm H}] = -0.1 \pm 0.1$ dex) with respect to the other galaxies at larger cluster-centric distances ($[M/{\rm H}] = -0.3 \pm 0.2$ dex). 
This suggests that the redder colours of the bright central regions of the ETGs in the high-density region of the cluster 
with respect to those at larger distances are due to a difference in [$M$/H]. 
A similar trend, but with a larger scatter, is also observed for [Mg/Fe] in Fig.~\ref{fig:SP_all} (lower left panel). 
The massive ETGs in the high-density region have higher central mean [Mg/Fe] ($[{\rm Mg/Fe}] = 0.16 \pm 0.04$ dex) 
than the cluster members at larger cluster-centric distances ($[{\rm Mg/Fe}] = 0.12 \pm 0.03$ dex).


The mean age, [$M$/H], and [Mg/Fe] in the outskirts of the F3D ETGs are listed in Table~\ref{tab:SP_analysis} and 
plotted as function of total stellar mass in Fig.~\ref{fig:SP_all} (right panels). 
It has to be noticed that the derived values should be considered as ``lower'' limits 
of the stellar population parameters of the total accreted component, since at the mapped 
galactocentric radii the presence of the underlying in-situ component is indistinguishably 
mixed with the accreted stellar component. 

On average, the outskirts of the ETGs in the high-density region of the cluster are only 
$\sim1$ Gyr younger than the inner parts of the galaxies and span a range between 10 and 13
Gyr with an average age of $\sim12$ Gyr (Fig.~\ref{fig:SP_all}, top right panel), 
except for FCC~90 which has ongoing central star formation. Therefore, taking into account the average 
uncertainties on the age estimates, the central regions and the galaxy outskirts have stellar populations of comparable age. 
The relation between the total stellar mass and mean [$M$/H] is still in place for the 
outskirts (Fig.~\ref{fig:SP_all}, middle right panel), but with a larger scatter.
%
By comparing the least-squares fit computed for the values in the central parts and taking into account the average uncertainties,
there is an indication that the ETGs in the high-density region have more metal-poor outskirts.
This is consistent with previous results from \citet{Bedregal2011} for a small sub-sample of S0 galaxies in Fornax.
By measuring the stellar population gradients, they found that, on average, steeper metallicity gradients  
towards the outskirts.
For ETGs in the low-density region of the cluster, the difference in 
mean [$M$/H] is still appreciable but it is more pronounced in the galaxies with a 
peculiar central structure (e.g. FCC~90 and FCC~301). 
This analysis allows to constrain the mean [$M$/H] of the outskirts as function of the 
cluster environment and provides evidence that ETGs in the high-density region have more 
metal-poor outskirts than those in the low-density region of the cluster. 
Moreover, the 
difference in mean [$M$/H] between the central parts and the outskirts is larger for the
more massive galaxies in the high-density region, i.e., the outskirts have a lower [$M$/H] 
(with an average $[M/{\rm H}] = -0.3 \pm 0.1$ dex) than the central in-situ dominated component. 
This is in agreement with the properties of the accreted populations identified 
in the edge-on galaxies, especially in their thick discs
(\citetalias{Pinna2019a}, \citetalias{Pinna2019b}).

Even if the scatter is larger, on average, [Mg/Fe] is higher in the galaxy outskirts than in the central parts. 
The [Mg/Fe] values remain almost the same in the outskirts of the galaxies in the low-density regions 
(Fig.~\ref{fig:SP_all}, lower right panel).


\section{The assembly history of the Fornax cluster}
\label{sec:PPS}

 
 
 


   
\begin{figure*}[t!]
\includegraphics[width=\hsize]{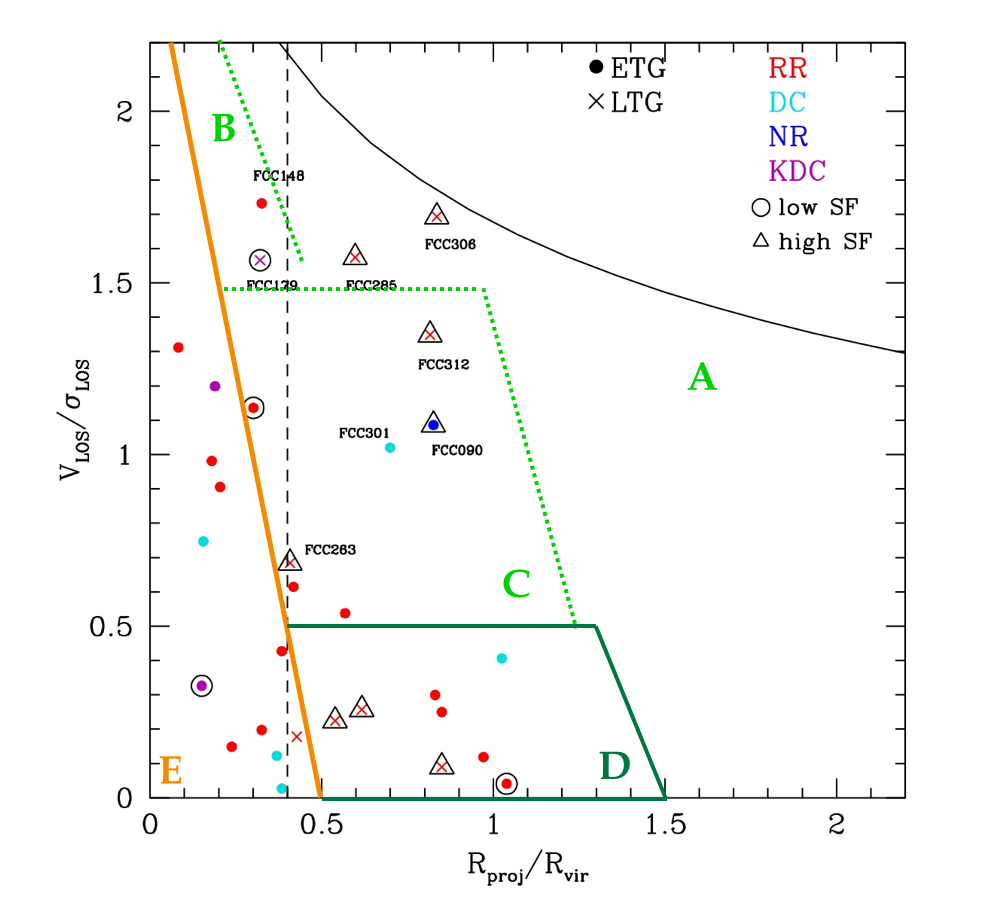}
\caption{Projected phase-space diagram of the F3D ETGs (filled circles) and LTGs (crosses). Symbols are as in the left panel of Fig.~\ref{fig:LambdaR}. Galaxies with low (black open triangles) and high star formation (black open circles) are also shown. 
The black solid line corresponds to the cluster escape velocity. The vertical dashed line separates the high and low-density regions of the cluster. According to \citet{Rhee2017}, region A is populated by the first infallers, regions B and C (together delimited by the dashed light-green, solid orange, and solid green lines) contain 80\% of the recent infallers, regions D (delimited by the solid green and solid orange lines) and E (delimited by the solid orange line) contain 40\% of the intermediate infallers, and the ancient infallers  are found in region E. The name of the galaxies classified as recent infallers are also indicated on the plot. }
\label{fig:PPS}
\end{figure*}

 
The projected phase-space (PPS) diagram of the F3D galaxies (Fig.~\ref{fig:PPS}) was 
investigated in order to trace the structure of the Fornax cluster and address its 
formation history. To this aim the line-of-sight radial velocity $V_{\rm los}$ of each 
F3D galaxy was derived relative to the cluster recession velocity, which was assumed to 
be that of FCC~213 (NGC~1399, $V_{\rm los} =1425$ km~s$^{-1}$), the brightest member in the core of 
the Fornax cluster. The ratio between $V_{\rm los}$ and the cluster velocity dispersion 
($\sigma_{\rm los} \simeq 300$ km~s$^{-1}$, \citealt{Drinkwater2001}) is plotted in the PPS
diagram as a function of the projected cluster-centric distance of the galaxy in units of 
virial radius $R_{\rm proj}/R_{\rm vir}$. This distance is a lower limit for  
the three-dimensional distance of the galaxy from the cluster centre. The PPS unveils the 
assembly process of the cluster and it traces the location of galaxies that have reached 
the `virialised' region. This is constrained by the escape velocity of the cluster, which
is shown with a solid line in Fig.~\ref{fig:PPS}. 
It was derived by assuming a mass density profile for the dark matter halo as 
described in \citet{Navarro1997} and following the prescriptions of \citet{Rhee2017}. 

By selection, all the F3D galaxies are located inside the virial radius of the Fornax cluster and have 
consistently smaller values of $V_{\rm los}/\sigma_{\rm los}$ with respect to the cluster 
escape velocity. Most of them are characterised by small values of both $V_{\rm 
los}/\sigma_{\rm los}$ and $R_{\rm proj}/R_{\rm vir}$, i.e., these are galaxies residing 
close to the cluster core. 
 {According to the simulations by \citet{Rhee2017}, this is the 
PPS regime of the ancient infallers, which entered the cluster potential in an early stage of the cluster formation, at an indicative time $\geq 8$ Gyr 
ago (region E in Fig.~\ref{fig:PPS}). 
At a larger $R_{\rm proj}/R_{\rm vir}$, galaxies in the PPS reside in the regime of
intermediate infallers (region D in Fig.~\ref{fig:PPS}), which settled into the
cluster later, some time between 4 and 8 Gyr ago.} 
The PPS region of the ancient infallers is populated by ETGs and it hosts most of the 
galaxies with a kinematically distinct core or a kinematically decoupled core. 
The only LTG in this region is FCC~179, for which there is no detection of the ionised-gas emission (Table~\ref{tab:emission}).
Here there are few galaxies with smaller values of the H$\alpha$ equivalent width and with ionised-gas emission 
close to the galaxy centre (Table~\ref{tab:emission}). 
Most of the intermediate infallers (region D in Fig.~\ref{fig:PPS}) are still ETGs, but $\sim30\%$ of the LTGs are found as well in this
region of the PPS diagram that show high ionised-gas emission (Table~\ref{tab:emission}).
Taking into account the ranges given by the simulations, it is reliable to consider the three galaxies (FCC~153, FCC~263, FCC~277) 
located very close to the regions E and D as intermediate infallers.
For simplicity, we grouped as recent infallers all galaxies that
 are expected to have large values of both $V_{\rm los}/\sigma_{\rm 
los}$ and $R_{\rm proj}/R_{\rm vir}$ in the PPS diagram. This corresponds to regions A, B and C in  
Fig.~\ref{fig:PPS}. According to \citet{Rhee2017}, here there are galaxies that entered the cluster from 1 to 4 Gyr ago.
Four of the seven galaxies in this regime are LTGs, while the remaining ones are ETGs. 
Most of these galaxies are actually located in the low-density region of the cluster, except for the peculiar S0 galaxy FCC~148 
and the spiral galaxy FCC~179, which are closer in projection to the cluster core. 
Among the recent infallers, FCC~148 is the only ETG showing a blue nucleus and ionised-gas emission 
and FCC~301 hosts a decoupled core with different kinematic and photometric properties 
with respect to the other F3D decoupled cores. 
These peculiar properties are a further indication for a different formation mechanism of the nuclear structures in these galaxies. 
Half of the galaxies in this PPS regime show large values of the H$\alpha$ equivalent width (Table~\ref{tab:emission}). 
 
\begin{figure*}[t!]
\includegraphics[width=\hsize]{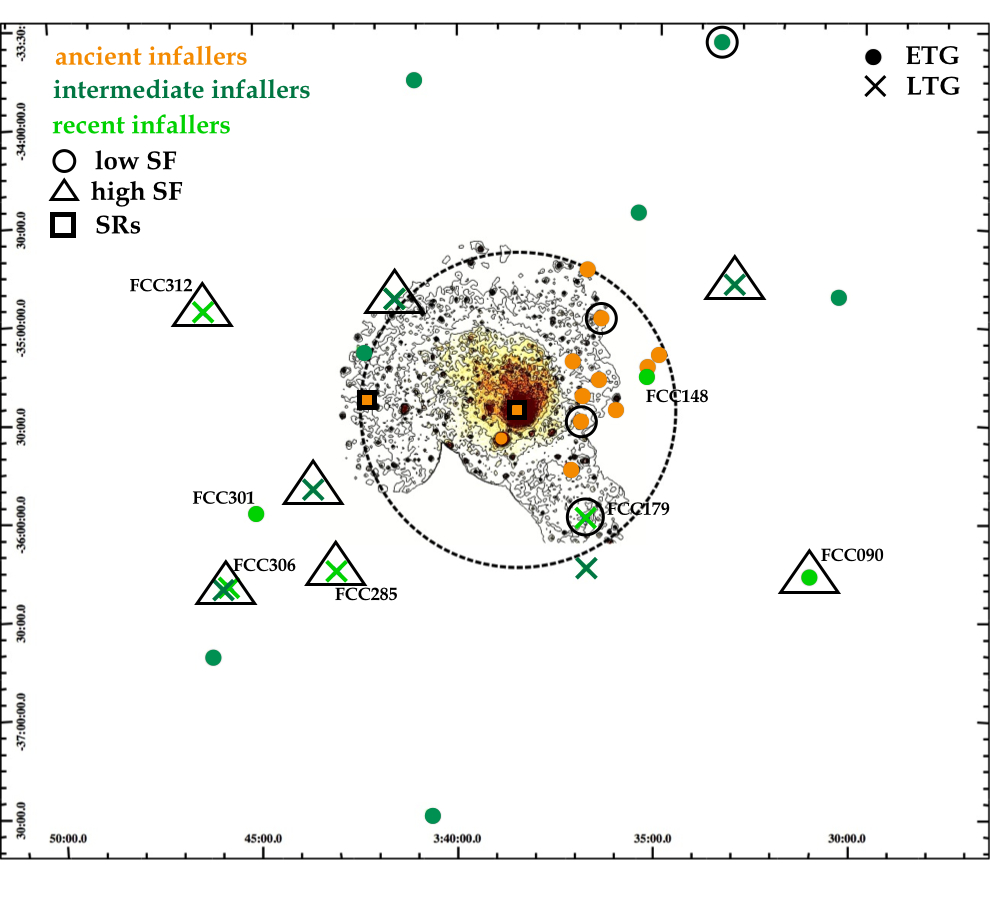}
\caption{Distribution of the F3D ETGs (circles) and LTGs (crosses) onto the sky plane. The right ascension and declination (J2000.0) are given in degrees on the horizontal and vertical axes of the field of view, respectively. The background image and contours map the X-ray emission in the energy range 0.4--1.3 KeV as measured by XMM-Newton \citep{Frank2013}. The dashed circle indicates the transition from the high-to-low density region of the cluster at $0.4 R_{\rm vir}$.
Orange, green, and light-green symbols represent the ancient, intermediate, and recent infallers, respectively, according to the analysis of the PPS diagram shown in Fig.~\ref{fig:PPS}. Galaxies with low (black open triangles) and high star formation (black open circles) are also shown. The slow rotators (SRs) are marked with open black squares. As in Fig.~\ref{fig:PPS}, all the recent infallers are labelled. }
\label{fig:Xray}
\end{figure*}

Figure~\ref{fig:Xray} shows the distribution of the F3D galaxies on the sky: 
the ancient infallers are mostly distributed along the north-south 
filament on the west side of the cluster, and all of them are located inside the high-density region of the cluster.
The intermediate and recent infallers are found at larger cluster-centric distances ($R_{\rm proj}\geq 0.3$~Mpc) 
in the low-density region of the cluster. Most of these galaxies are characterised by ionised-gas emission and evidence for ongoing star formation. 
The comparison with the X-ray emission detected in the Fornax cluster \citep[e.g.][]{Frank2013} 
shows that all the ancient infallers are located 
inside the X-ray halo whereas the galaxies with active star formation are found at larger cluster-centric radii, far away from the X-ray emission.
 
Combining the two-dimensional distribution of the F3D galaxies shown in Fig.~\ref{fig:Xray} with the PPS information 
and the MUSE results on stellar kinematics, star formation, and stellar populations, 
the Fornax cluster appears to consist of three main groups, where galaxies have common properties:
the {\it core}, the {\it north-south clump}, and the {\it infalling galaxies}.
 
The core is dominated by the potential of the brightest and massive cluster member FCC~213 (NGC~1399), which is one of the only two SRs 
inside the virial radius \citep[Fig.~\ref{fig:LambdaR} and see also][]{Scott2014}. 
This coincides with the peak of the X-ray emission (Fig.~\ref{fig:Xray}).

The clump on the north-northwest side of the cluster contains the redder \citep{Iodice2018} and more metal-rich galaxies of the F3D sample 
(Table~\ref{tab:SP_analysis}) and all of them are FRs (Table~\ref{tab:sample}). The majority of the kinematically 
distinct cores are found in the galaxies of this clump, 
and two out of a total of three show ionised-gas emission in the centre. 
On average, the larger differences between the kinematic and photometric position angles 
are found in the ETGs populating this group (Fig.~\ref{fig:PA}). 
Finally, the outskirts of these galaxies have lower metallicity than their bright central regions (Fig.~\ref{fig:SP_all}).

The third group of galaxies in the Fornax cluster consists of the recent and intermediate infallers. 
In projection, they are distributed nearly symmetrically around the 
core in the low-density region of the cluster ($R_{\rm proj} \geq 0.4$~Mpc). 
The majority of them are LTGs with ongoing star formation (Table~\ref{tab:emission}). 
In this region, galaxies have on average lower [$M$/H] and [Mg/Fe] with respect to clump galaxies (Table~\ref{tab:SP_analysis} 
and Fig.~\ref{fig:SP_all}). 
Most of the LTGs show signs of interaction with environment and/or minor merging events in the form of tidal tails and 
disturbed molecular gas as found in FCC~312 \citep{Zabel2019, Raj2019}, or havedisturbed ionised-gas as FCC~263 
(see Appendix~\ref{sec:description}). 
This suggests that the structure of these galaxies has been modified upon entering the cluster potential \citep{Boselli2006}. 

\begin{figure}[t!]
\includegraphics[width=\hsize]{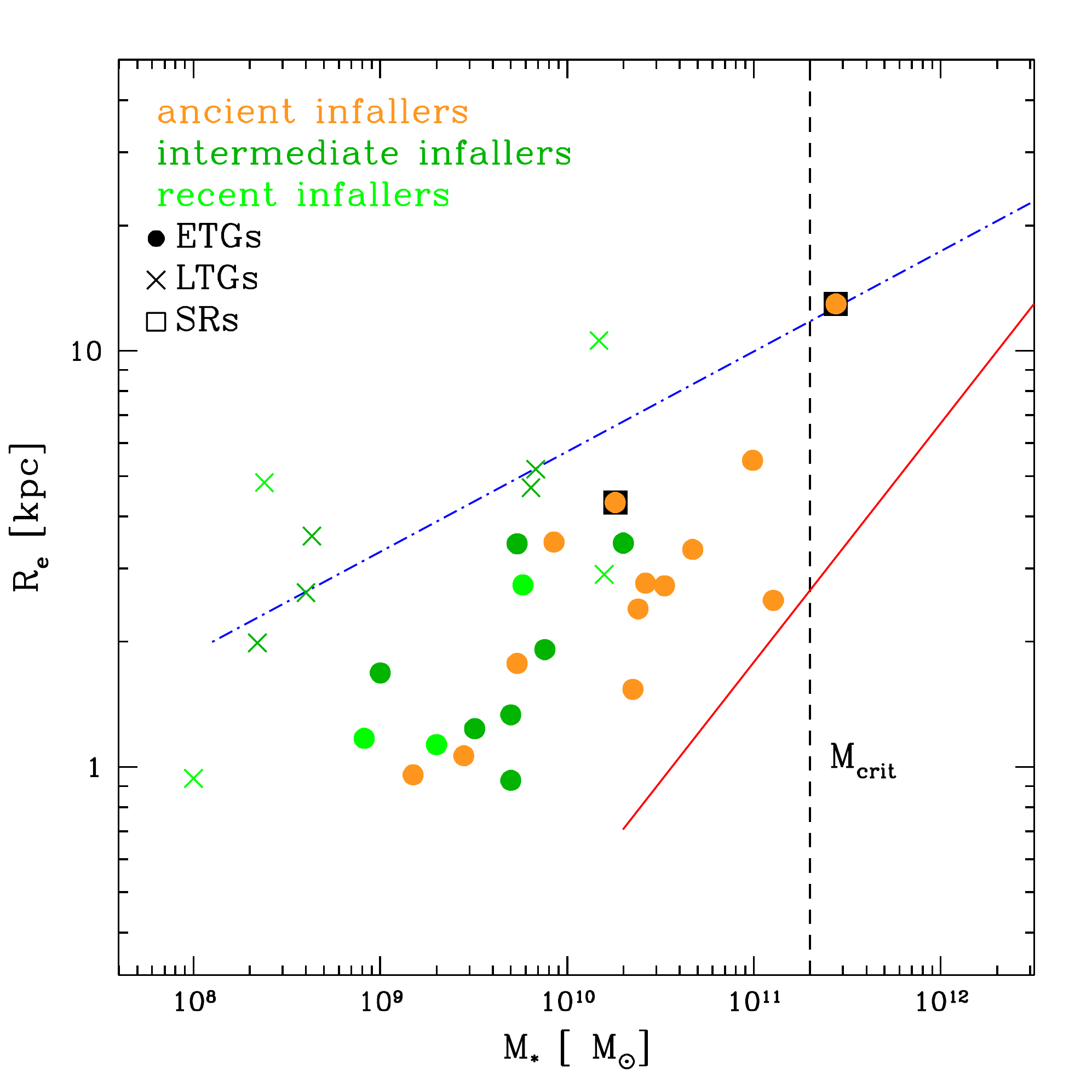}
\caption{Mass-size distribution of the F3D ETGs (circles) and LTGs (crosses).
Orange, green, and light-green symbols represent the ancient, intermediate, and recent infallers, respectively, according to the analysis of the PPS diagram shown in Fig.~\ref{fig:PPS}. Galaxies with low (black open triangles) and high star formation (black open circles) are also shown. The slow rotators (SRs) are marked with open black squares. The critical mass, defined by \citet{Cappellari2013b}, is marked by the black dashed line. For reference, the blue dash-dotted line and the solid red line indicate the upper limit of the spiral galaxies and the lower limit of the ETGs from ATLAS3D sample \citep{Cappellari2013}.}
\label{fig:mass_size}
\end{figure}

Clusters have a significant fraction (25--40\%) of their galaxies accreted through galaxy groups \citep{McGee2009, DeLucia2012}.
The clump may result from the accretion of a group of galaxies during the gradual build-up of the cluster, 
which induced the observed asymmetry in the spatial distribution of the bright galaxies \citep{Iodice2018}. 
All galaxies in the clump have similar colours, age, [$M$/H] and are fast rotator ETGs, with
stellar masses in the range $0.3-9.8 \times 10^{10}$~M$_{\odot}$ (see Tab.~\ref{tab:sample} and Fig.~\ref{fig:mass_size}). 
Galaxy interactions are more frequent in groups than in clusters due to their lower velocity dispersion \citep{Tremaine1981}, 
therefore the cluster members accreted as part of groups can undergo pre-processing that modifies their structure and internal dynamics. 
This would explain the high fraction of kinematically decoupled components in the galaxies belonging to the clump, and also the thick discs 
observed in the three edge-on S0 galaxies studied by \citetalias{Pinna2019a} and \citetalias{Pinna2019b}. 
The deep FDS imaging revealed that the region of the cluster where the clump is located hosts the bulk of the gravitational interactions between galaxies. It is the only region inside the 
virial radius where intra-cluster baryons were found, including faint stellar filaments between some of the galaxies \citep{Iodice2016, Iodice2017b}.

The low star-formation activity detected in the centre of only two clump galaxies suggests that harassment or 
suffocation induced by ram pressure stripping depleted the gas content. 
This is further supported by the presence of X-ray emission in this region of 
the cluster (Fig.~\ref{fig:Xray}). 
Therefore, the star formation in clump galaxies stopped earlier than for those that entered later into the cluster potential. 
These are the galaxies in the third group, consisting of recent and intermediate infallers which show ongoing star formation 
and are located outside the X-ray emission. 
According to numerical simulations \citep{Hwang2018}, galaxies entering the cluster can lose the cold gas but still retain star-formation 
activity during the collision phase due to the interaction with massive cluster members and/or with the hot gas. This could be the case for 
FCC~179 (Fig.~\ref{fig:FCC179map}), a late-type galaxy with spiral arms and central star formation, which recently arrived in the high-density region of the cluster (Fig.~\ref{fig:PPS}) where the X-ray emission is still present (Fig.~\ref{fig:Xray}).

On a larger scale, the sub-clustering structure of the Fornax cluster was already suggested by \citet{Nasonova2011}, 
who studied the distribution of galaxies in the vicinity of the cluster. 
This evidence supports the hypothesis that the assembly of the Fornax cluster is still 
ongoing \citep{Dunn2006}. The galaxies residing in the clump close to the cluster centre 
are distributed north-south and follow the cosmic web filament connecting the 
Fornax-Eridanus large-scale structure \citep{Nasonova2011}. These findings also support
the idea that the clump in the Fornax cluster, which consists of galaxies with similar properties
in terms of morphology, kinematics, and stellar populations, is one of the smaller galaxy 
groups in the surrounding large-scale structure that, according to the PPS analysis, 
merged into the potential well of the massive cluster member NGC~1399, more than 8 Gyr ago. 
The stellar masses of the ETGs populating the clump, as well as of all others cluster members,
are an order of magnitude lower than the critical mass  M$_{crit} \simeq 2 \times 10^{11}$~M$_{\odot}$ (see Fig.~\ref{fig:mass_size})
that defines the transition region in the mass-size relation from FRs ETGs to SRs in the high-density environments  \citep{Cappellari2013b}.
Therefore, according to \citet{Cappellari2013b}, the galaxies residing in the 
clump were not massive enough to form or maintain their own SR before becoming part of the Fornax cluster, whereas 
NGC~1399 has a stellar mass larger than M$_{crit}$ and, consistently, it is the SR in the core (Fig.~\ref{fig:Xray}).
Compared to the other nearby massive Virgo and Coma clusters, Fornax appears to be in an intermediate evolutionary phase. 
As observed in Coma \citep{Cappellari2013b}, the slow rotator in the core of the Fornax cluster is segregated in mass, 
i.e., with M$_{*} \geq$~M$_{crit}$, from the fast rotators (Fig.~\ref{fig:mass_size}). 
In contrast, the Virgo cluster is a dynamically young and unrelaxed cluster \citep{Mihos2017}, with 
 a higher fraction of spiral galaxies than Coma and Fornax \citep{Cappellari2011}, and with a larger numbers of SRs associated
 with substructures identified in the cluster, consistent with being in an early phase in the cluster assembly.
 As observed in Coma and Virgo, the fast rotators ETGs have smaller R$_e$ than the spirals, which are still on a parallel sequence
 to the ETGs, as in Virgo but less pronounced in Coma \citep[see][and Fig.~\ref{fig:mass_size}]{Cappellari2013b}.
 The other SR found in Fornax, FCC~276 located on the east side of the cluster in the transition region from high to lower density,
 has comparable properties to the SRs in less dense environments.

In conclusion, the structure of the cluster that emerges from the above analysis is based on combining 
the two-dimensional distribution of the F3D galaxies with the PPS information and with the 
structural properties of the galaxies (morphology, colors, kinematics, and stellar population). 
There is a clear segregation between galaxies classified as ancient infallers, not only in terms of projected location 
inside the cluster but mainly in the average observed properties, and the galaxies belonging to the other two classes 
(intermediate and recent infallers).
This observational evidence reflects the assembly history of the cluster and it could be considered as a baseline 
for future simulations designed to trace the formation of the Fornax cluster.


\section{The mass assembly in the outskirts of ETGs}\label{sec:outskirts}

 
 {The stellar population analysis in ETGs provides stringent constraints on the 
properties of the galaxy outskirts and are extremely important for a direct comparison 
with theoretical predictions to constrain the mass assembly of the stellar halo.
The study of the [$M$/H] and [Mg/Fe] maps of the F3D galaxies is beyond the scope of this paper and will be the subject of a forthcoming 
work (Mart\'in-Navarro et al., in prep.). 
Nevertheless, it is possible to draw some conclusions on the mass assembly of the galaxies in the Fornax cluster.} 
According to 
\citet{Spavone2019}, the galaxies in the low-density region of the cluster have a quite small accreted mass fraction ($<40\%$), while 
galaxies in the high-density region have a remarkably large accreted mass fraction ($>80\%$). According to simulations 
\citep{Pillepich2018}, these quantities correlate with total stellar mass, i.e., more massive galaxies have a larger 
stellar-halo mass fraction. Given that, the small difference in stellar population 
properties between the central parts and outskirts 
in the less massive galaxies located in the low-density region suggests that inner and outer measurements are tracing the same stellar 
component, i.e., there is no significant contribution from the accreted material. 
This is also consistent with the assembly history of 
the cluster, as inferred from the PPS diagram (Fig.~\ref{fig:PPS}), which suggests that the galaxies in the low-density region are 
intermediate and recent infallers, which did not undergo the strong tidal interactions responsible for gas stripping and accretion 
from the outside.

In contrast, the galaxies in the high-density region of the cluster, i.e., the massive members residing in north-south clump, 
have a higher fraction of accreted material that dominates most the galaxy structure \citep{Spavone2019}. 
The FDS data show that the mass assembly in this region of the cluster is an ongoing process. 
In these galaxies, the outskirts are more metal poor ($-0.4 \leq [M/{\rm H}] \leq 0 $ dex) than the central in-situ dominated parts. 
According to the Auriga simulations of Milky Way-mass galaxies \citep{Monachesi2019} for galaxies in the range of masses 
$(2$--$10)\times10^{10}$~M$_{\odot}$, which is the same mass range of the massive galaxies of the F3D sample, the observed [$M$/H] values are consistent with an 
accreted stellar mass of $\sim 2 \times 10^{10}$~M$_{\odot}$. This is in agreement with the high fraction of accreted mass estimated from the analysis 
of the light distribution \citep{Spavone2019}.

\section{Concluding remarks}
\label{sec:concl}

The superb capabilities of MUSE at the Very Large Telescope provide high-quality, integral-field spectroscopic data to map the structure of all the bright ($m_B\leq15$) galaxies within the virial radius of the Fornax cluster. The stellar and ionised-gas kinematics for all galaxies in the F3D sample were derived and the resulting maps are presented in this paper (Figs.~\ref{fig:FCC083map}--\ref{fig:FCC312map}). Furthermore, for the ETGs the average stellar population properties from measurements of line-strength indices have been estimated.

The 31 sample galaxies were mapped in high resolution from the brightest central regions to the outskirts, 
where the surface brightness $\mu_B \geq 25$~mag arcsec$^{-2}$, and out to 2--3 effective radii for the 21 ETGs and 
1--2 effective radii for the 10 LTGs in the sample.  The analysis shows that 

\begin{itemize}
    
    \item while the Fornax cluster is dominated by ETGs inside the virial radius, the majority of the galaxies in this region are fast rotators, except for FCC~213 (NGC~1399), the brightest cluster member in the core and FCC~276, a bright spheroidal galaxy on the east side of the cluster in the low-density region (Table~\ref{tab:sample} and Fig.~\ref{fig:LambdaR});

    \item $42\%$ of the cluster members inside the virial radius have ionised-gas emission, where five galaxies are ETGs (Table~\ref{tab:emission}), with three of them at projected distances close to the cluster core;
    
    \item the analysis of the average stellar population properties of the ETGs inside the virial radius of the cluster shows that the inner and brightest parts of the ETGs in the high-density region are more metal-rich than the galaxy outskirts. Galaxies in the low-density region are as old as those at smaller cluster-centric distance, but have lower [$M$/H] and, on average, show a smaller difference in the mean stellar population properties between the inner parts and the outskirts (Table~\ref{tab:SP_analysis} and Fig.~\ref{fig:SP_all}).

\end{itemize}


 
\noindent 
The structure of the Fornax cluster has been traced by combining 
the two-dimensional distribution of the sample galaxies with the PPS information (Fig.~\ref{fig:PPS}) and with the 
structural properties of the galaxies (morphology, colors, kinematics, and stellar population, see Fig.~\ref{fig:Xray}).
The cluster shows three well-defined groups of galaxies: the core, the north-south clump and the infalling galaxies, 
which are added to the southwest merging group centred on NGC~1316 \citep{Drinkwater2001}. 
Galaxies in each group have different properties in the light and colour distributions, in the kinematics, and in the stellar populations.


The clump galaxies are the redder and more metal-rich galaxies of the sample  
(Table~\ref{tab:SP_analysis}). The phase-space diagram of numerically simulated clusters 
\citep{Rhee2017} suggests that the clump galaxies entered the cluster potential more than 
8 Gyr ago. All are fast-rotators, with many of them showing distinct nuclear components and
kinematically decoupled cores, and two out of a total of three show ionised-gas emission in
the centre. On average, galaxies populating this group show the larger differences 
between kinematic and photometric position angles (Fig.~\ref{fig:PA}). The outskirts of the
galaxies in this clump have lower metallicity than the bright central regions 
(Fig.~\ref{fig:SP_all}), which is an indication that the mass assembly of metal-poor 
satellites continues in the outskirts. 

The third group in the cluster contains the intermediate and recent infallers, i.e., galaxies that, 
according to the phase-space diagram, entered onto the cluster potential less than 4 Gyr ago. 
They are  distributed nearly symmetrically around the core, in the low-density region of the cluster ($R_{\rm proj} \geq 0.3$~Mpc).
The majority are LTGs with ongoing star formation (Table~\ref{tab:emission}). 
In this region, galaxies have on average lower [$M$/H] and [Mg/Fe] with respect to galaxies in the clump (Table~\ref{tab:SP_analysis} and Fig.~\ref{fig:SP_all}). 
Taking into account the position of the three groups inside the cluster and the average observed properties for the galaxies, 
it is reasonable to conclude that they are tracing the assembly history of the cluster.

Future work on the F3D data will extend the analysis presented in this paper to map the stellar population distributions in each galaxy of the sample (Mart\'in-Navarro et al., in prep.). 
In addition, ongoing works aim at performing the dynamical modelling to 
investigate the galaxy structure and  to constrain the star-formation history. 
The F3D data represent a rich mine for the study of galaxy structure and to compare the observed properties with predictions 
from galaxy formation models in relation with environment.

\begin{acknowledgements}
Based on observations collected at the European Organisation for Astronomical Research in the Southern Hemisphere under ESO programme 296.B-5054(A).
{\bf The F3D team wish to thank the anonymous referee for a constructive report that  led to an improvement of the paper.}
The F3D team is grateful to E.\ Dalla Bont\`a, T.\ Davis, K.\ Fahrion, L.\ Morelli, A.~Poci, M.~A.\ Raj, T.\ Spriggs, and L.\ Zhu for their contributions and discussions.
E.I.\ and M.S.\ acknowledge financial support from the VST funds (P.I.\ P.\ Schipani).
E.I. wishes to thank ESO in Garching and the Department of Physics and Astronomy in Padua for the hospitality and support during the several working visits related to the Fornax3D project.
P.T.d.Z. is grateful to the Department of Physics and Astronomy in Padua for support of a working visit. 
F.P., J.F.B., and G.v.d.V. acknowledge support from grant AYA2016-77237-C3-1-P from the Spanish Ministry of Economy and Competitiveness (MINECO). 
FP acknowledges Fundación La Caixa for the
financial support received in the form of a post-doc contract.
E.M.C. acknowledge financial support from Padua University through grants DOR1715817/17, DOR1885254/18 and BIRD164402/16.
R.McD. is the recipient of an Australian Research Council Future Fellowship (project number FT150100333). 
I.M.N. acknowledges support by the Deutsche Forschungsgemeinschaft under the grant MI 2009/1-1. 
G.v.d.V. acknowledges funding from the European Research Council (ERC) under the European Union’s Horizon 2020 research and innovation programme under grant agreement No 724857 (Consolidator Grant ArcheoDyn).
%
\end{acknowledgements}

%

   \bibliographystyle{aa} 
   \bibliography{f3d} 

%

\begin{appendix} 

\section{Maps for the stellar kinematics, line-strength indices, and ionised-gas properties for the F3D galaxies}\label{sec:kin_map}

\begin{figure*}[t!]
\begin{tabular}{ccc}
\includegraphics[width=6cm]{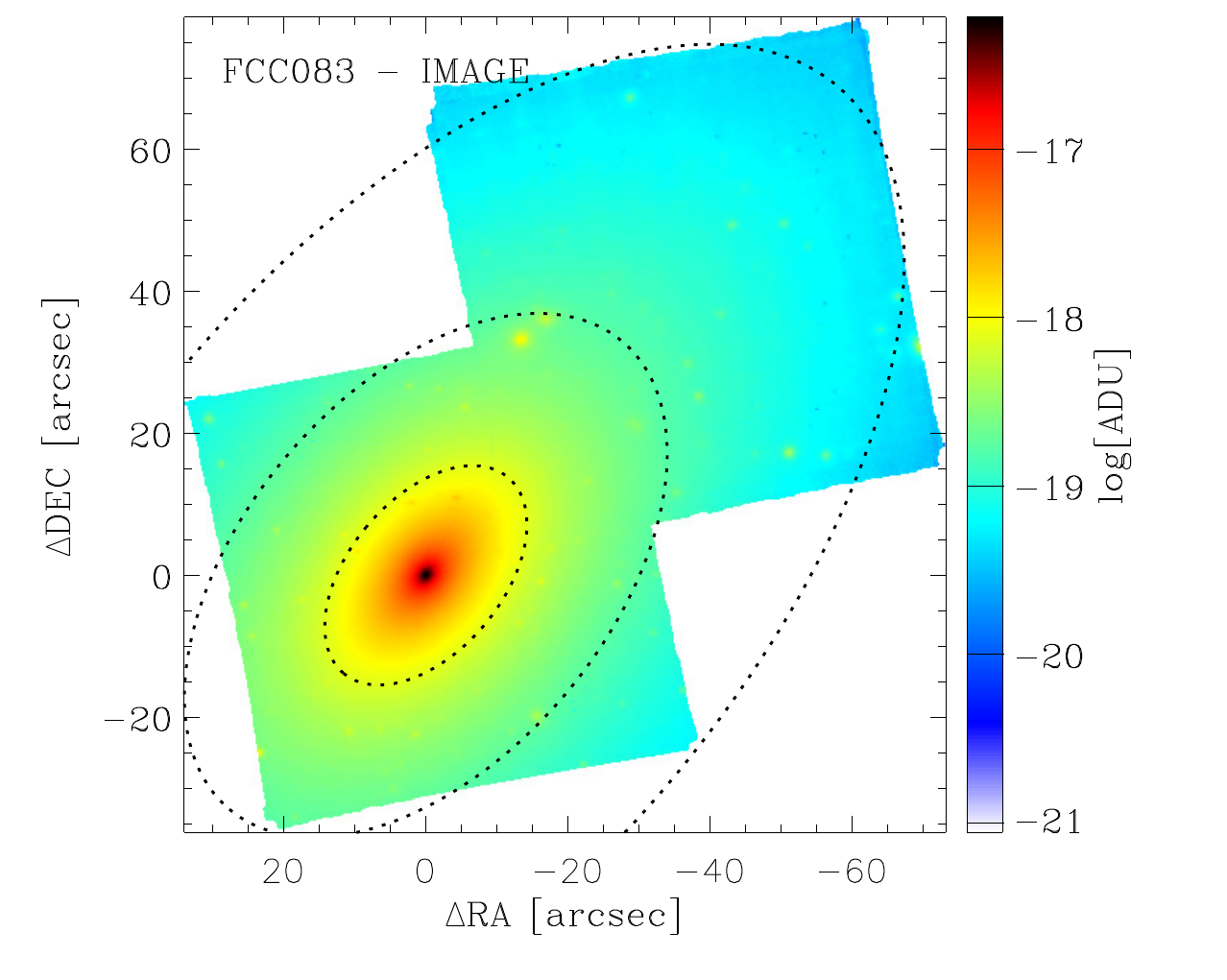} &
\includegraphics[width=6cm]{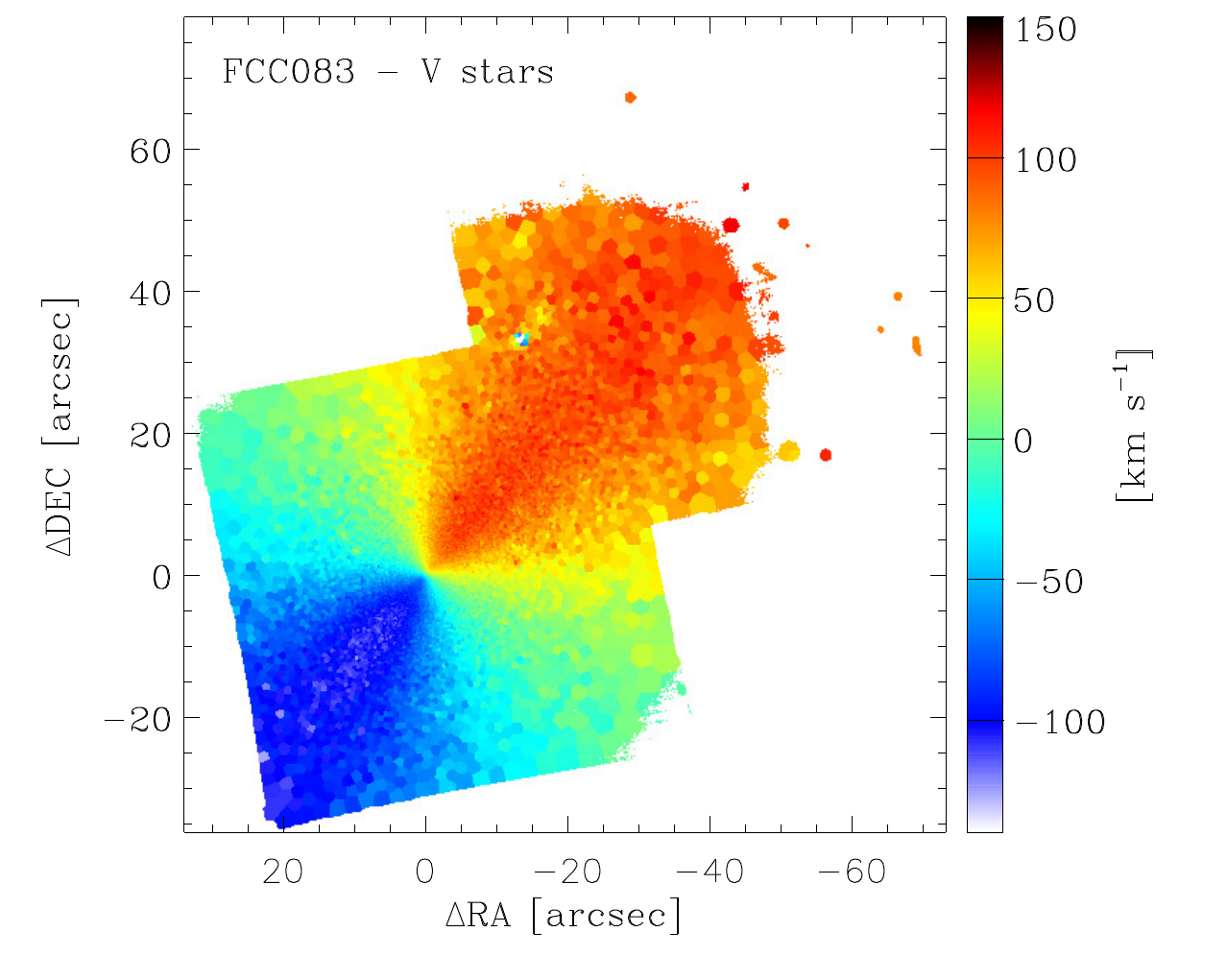} &
\includegraphics[width=6cm]{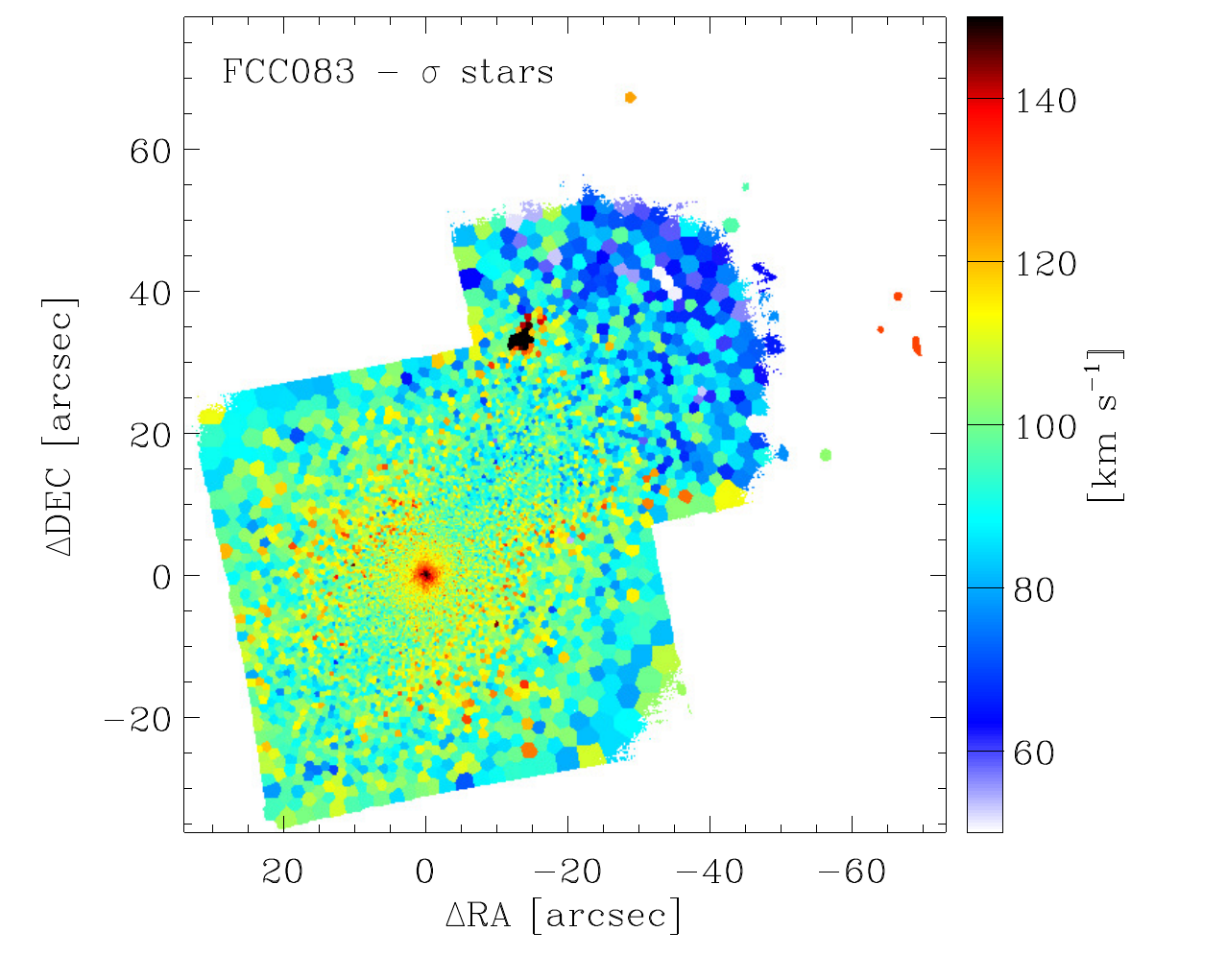} \\
\makebox[0pt][c]{\hspace{-0.5cm}\raisebox{0.3cm}{\includegraphics[width=4.7cm]{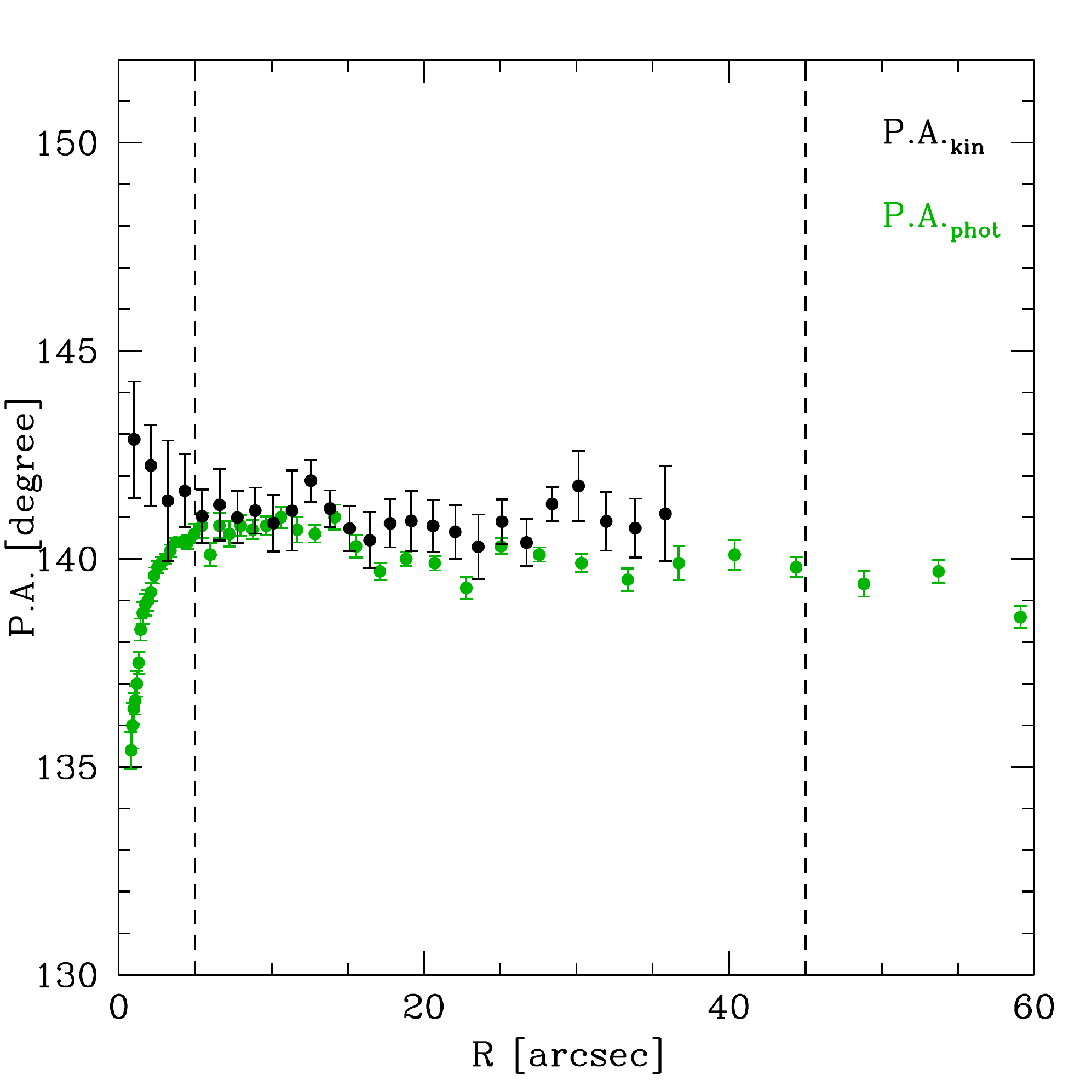}}} &
\makebox[0pt][c]{\hspace*{0.2cm}\includegraphics[width=6.25cm]{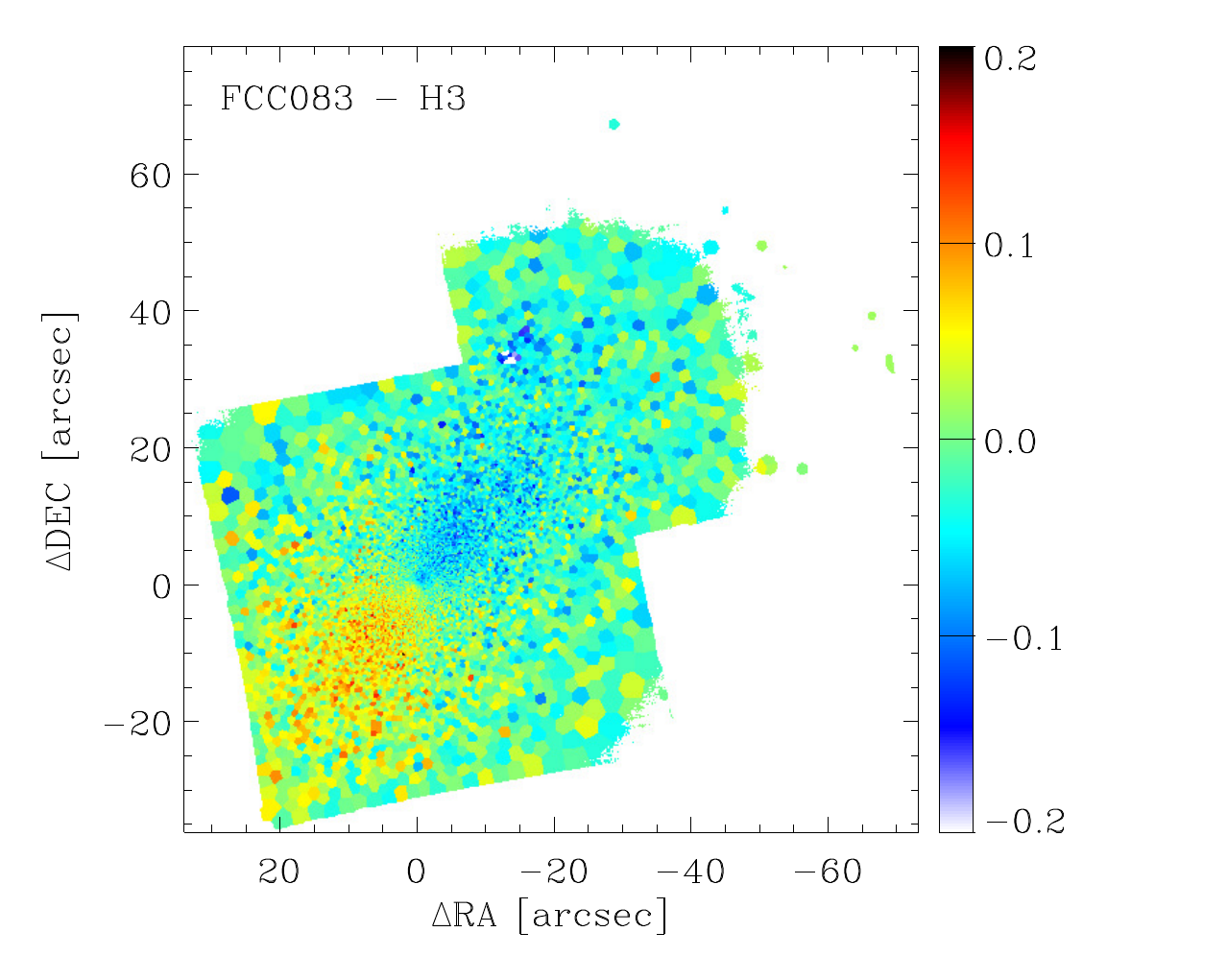}} &
\makebox[0pt][c]{\hspace*{0.2cm}\includegraphics[width=6.25cm]{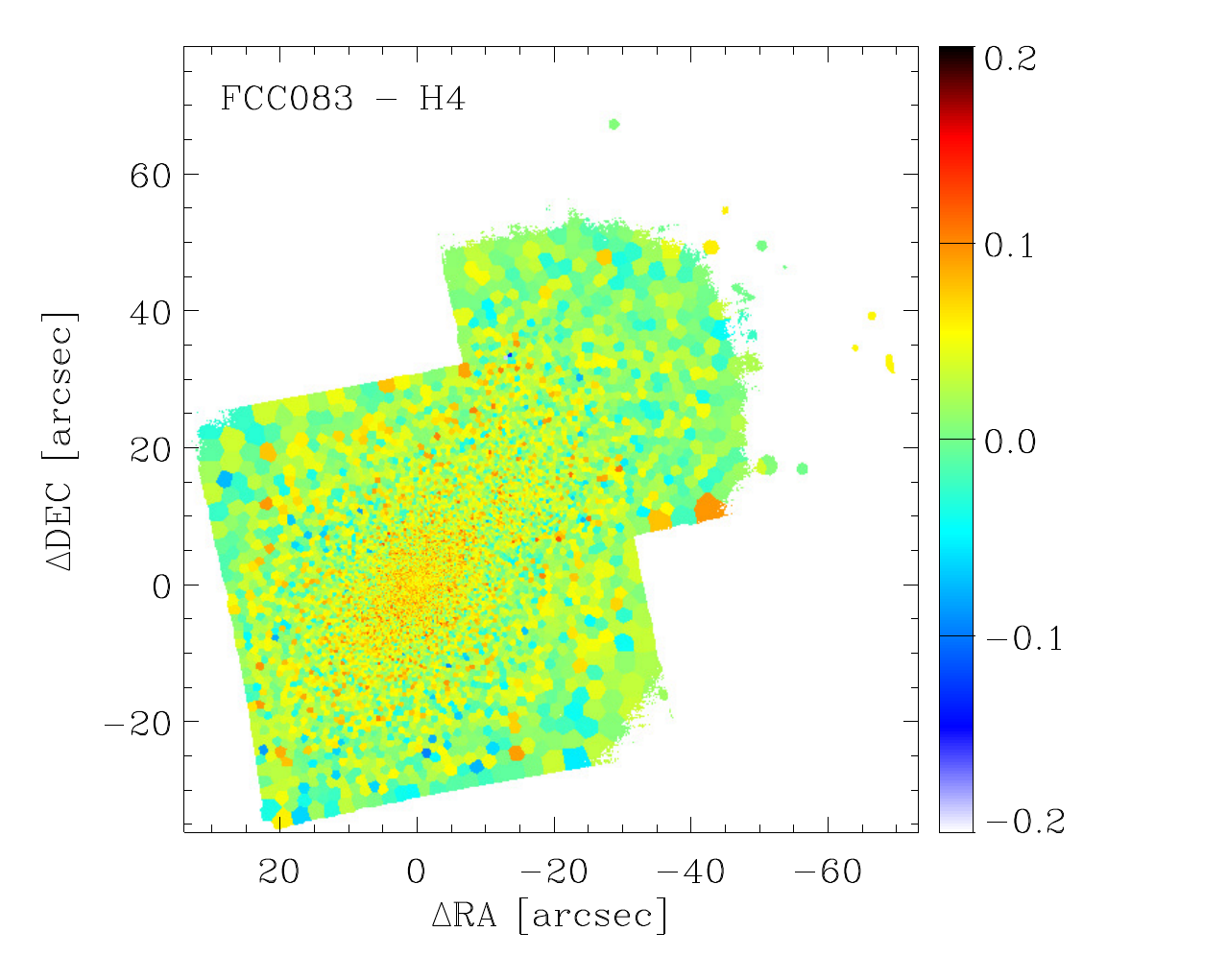}} \\
\includegraphics[width=6cm]{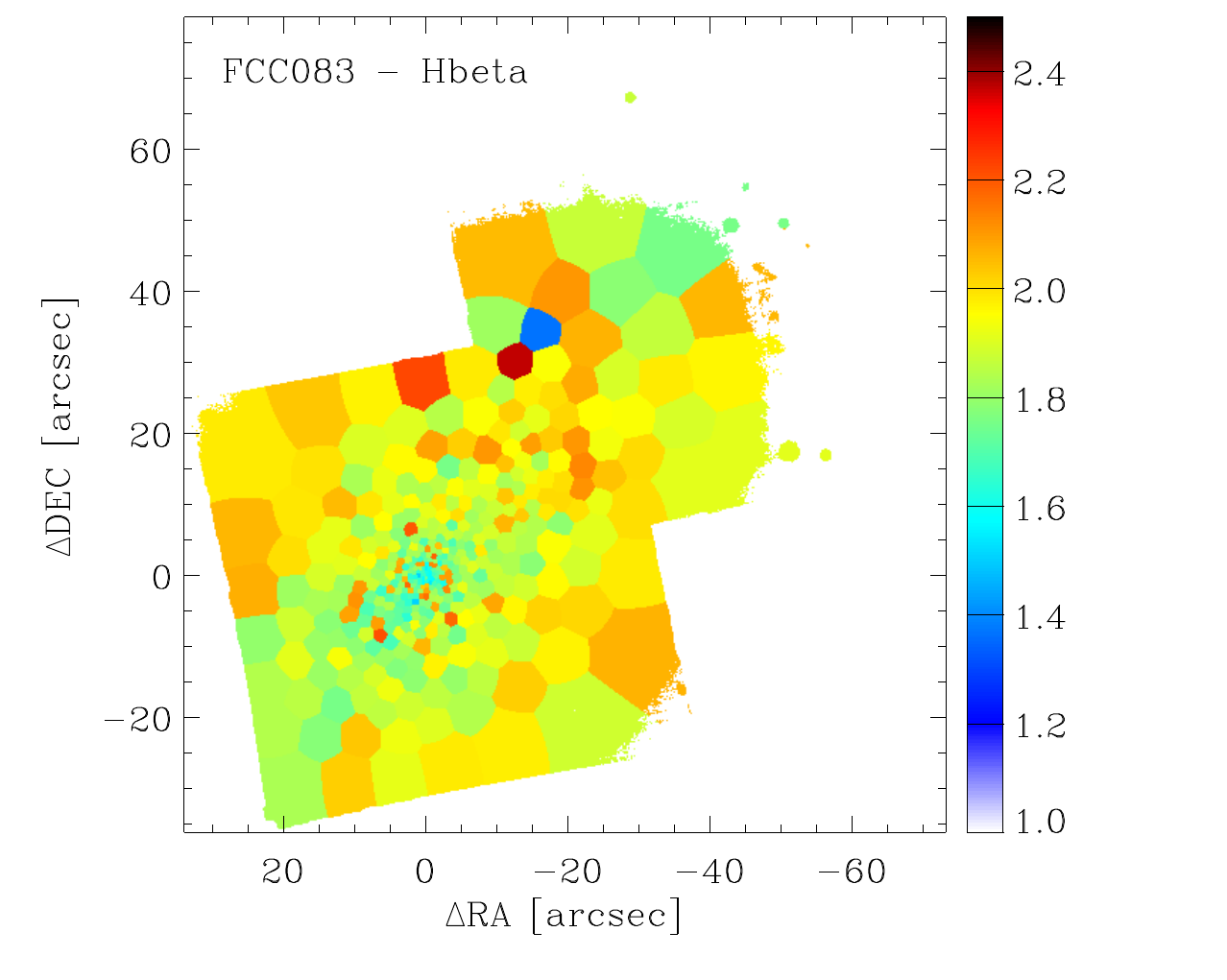} &
\includegraphics[width=6cm]{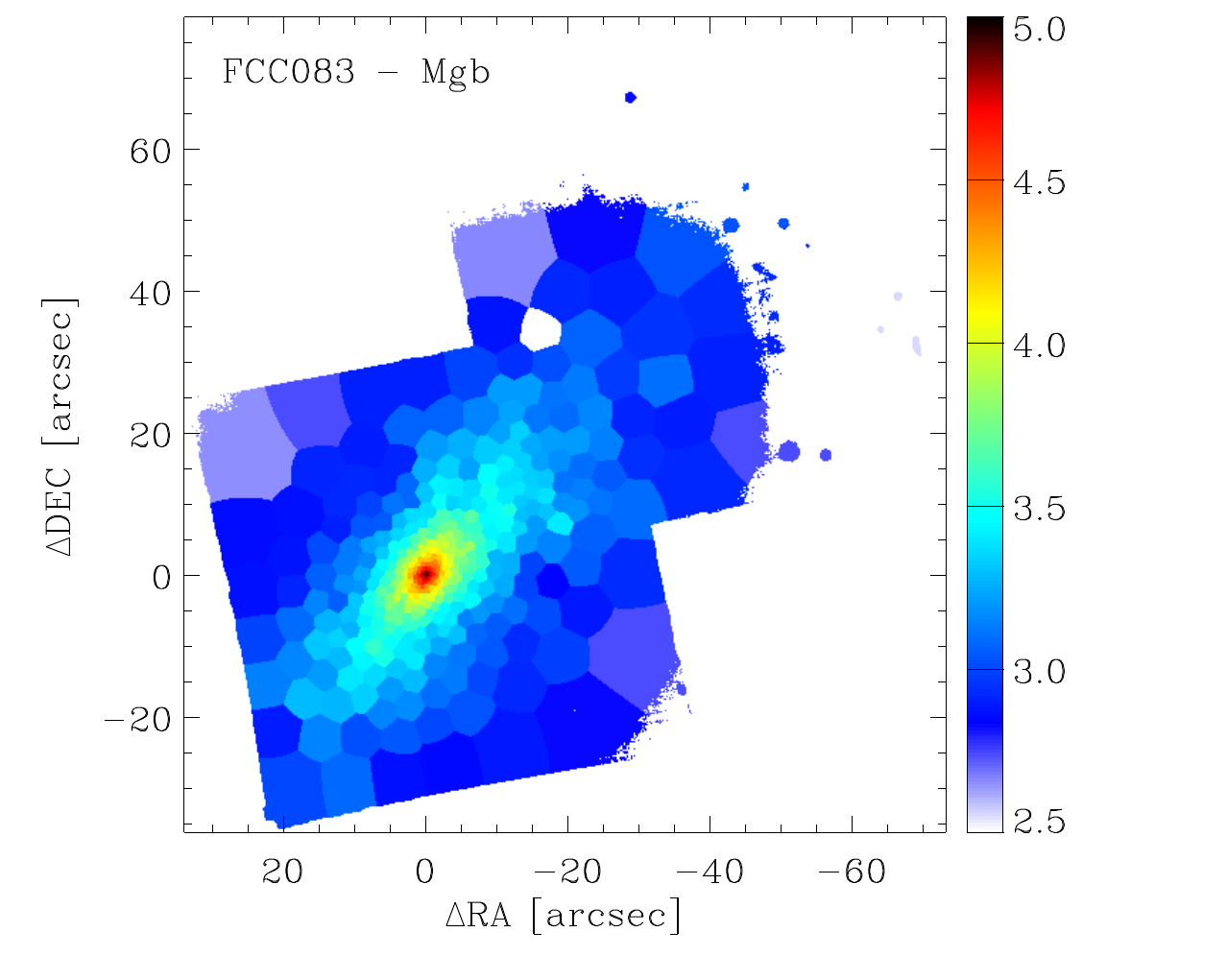} &
\includegraphics[width=6cm]{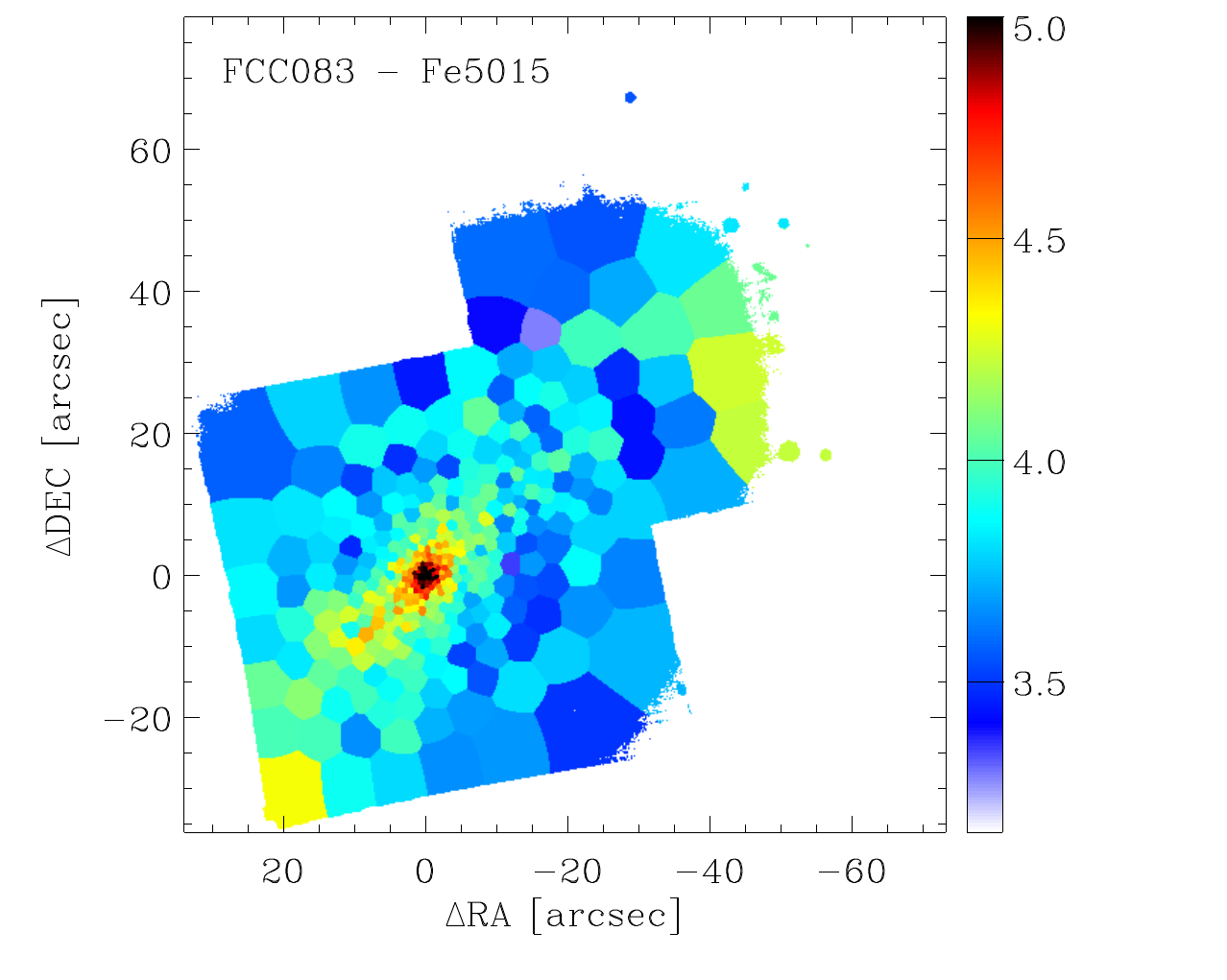} \\
\end{tabular}
\caption{Stellar analysis for FCC~83.
{\it First row panels:\/} MUSE reconstructed image ({\it left\/}). The dotted ellipses correspond to the isophotes at $0.5 R_e$, $R_{\rm tr}$, and $\mu_B=25$~mag arcsec$^{-2}$, respectively. 
Maps of the mean velocity ({\it middle\/}) and velocity dispersion ({\it right\/}) of the stellar LOSVD.
{\it Second row panels:\/} Radial profiles of the kinematic (black circles) and photometric (green circles) position angle ({\it left\/}). The vertical dashed lines mark the radial range where the average position angles are computed. Maps of the third ({\it middle\/}) and fourth Gauss-Hermite coefficient ({\it right\/}) of the stellar LOSVD.
{\it Third row panels:\/} Maps of the H$\beta$ ({\it left\/}), Mg$b$ ({\it middle\/}), and Fe5015 line-strength index ({\it right\/}).}
\label{fig:FCC083map}
\end{figure*}

\begin{figure*}[t!]
\begin{tabular}{ccc}
\includegraphics[width=6cm]{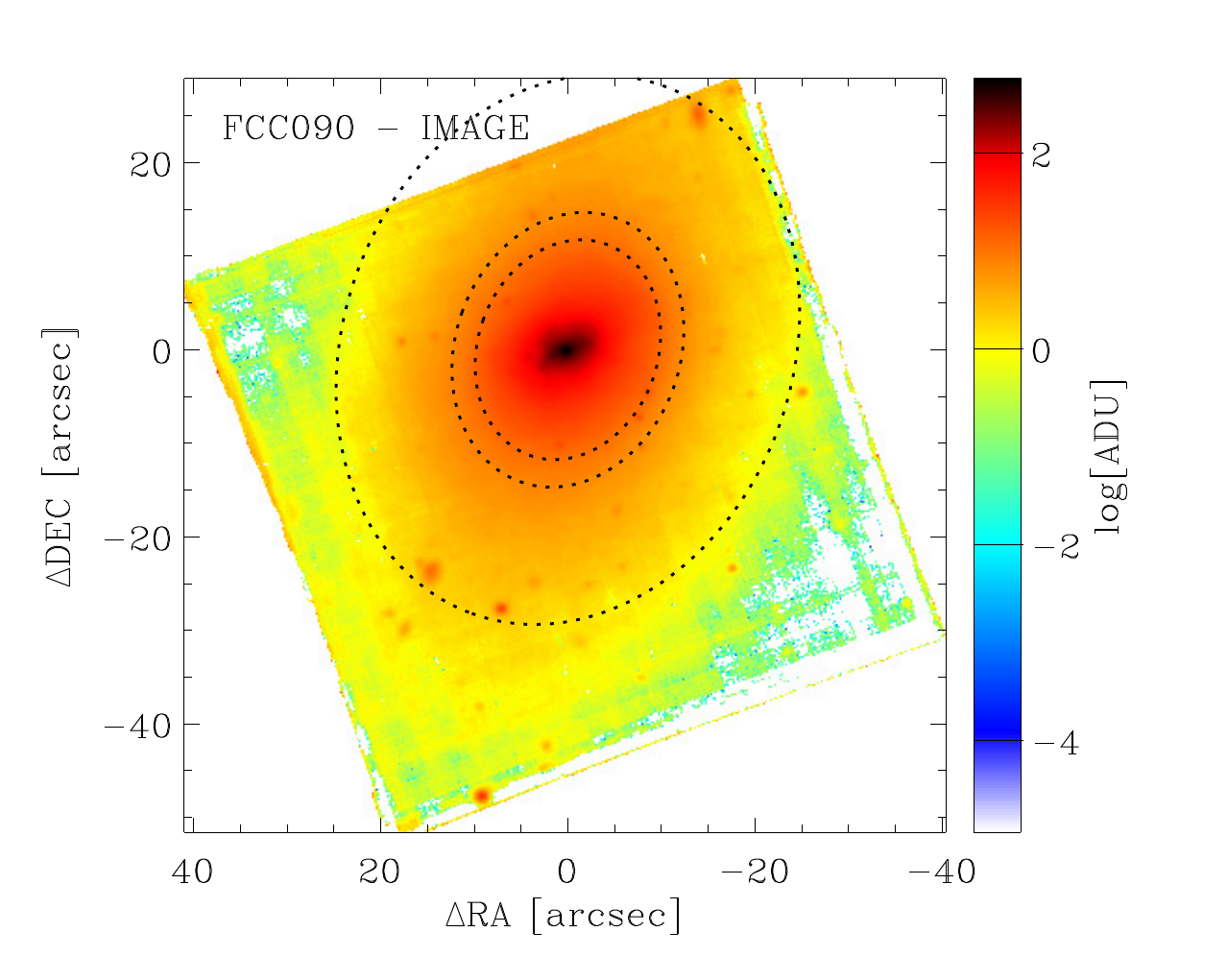} &
\includegraphics[width=6cm]{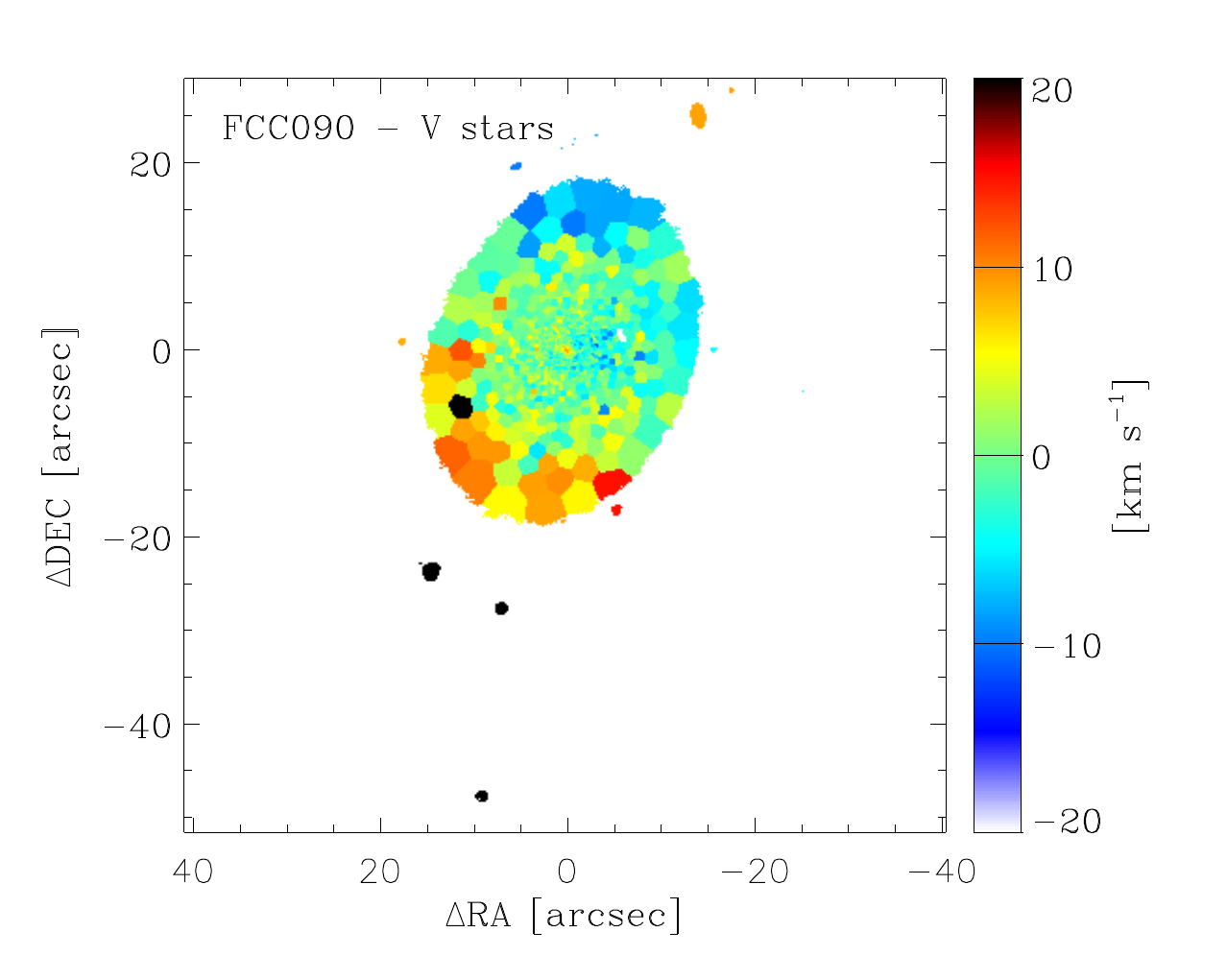} &
\includegraphics[width=6cm]{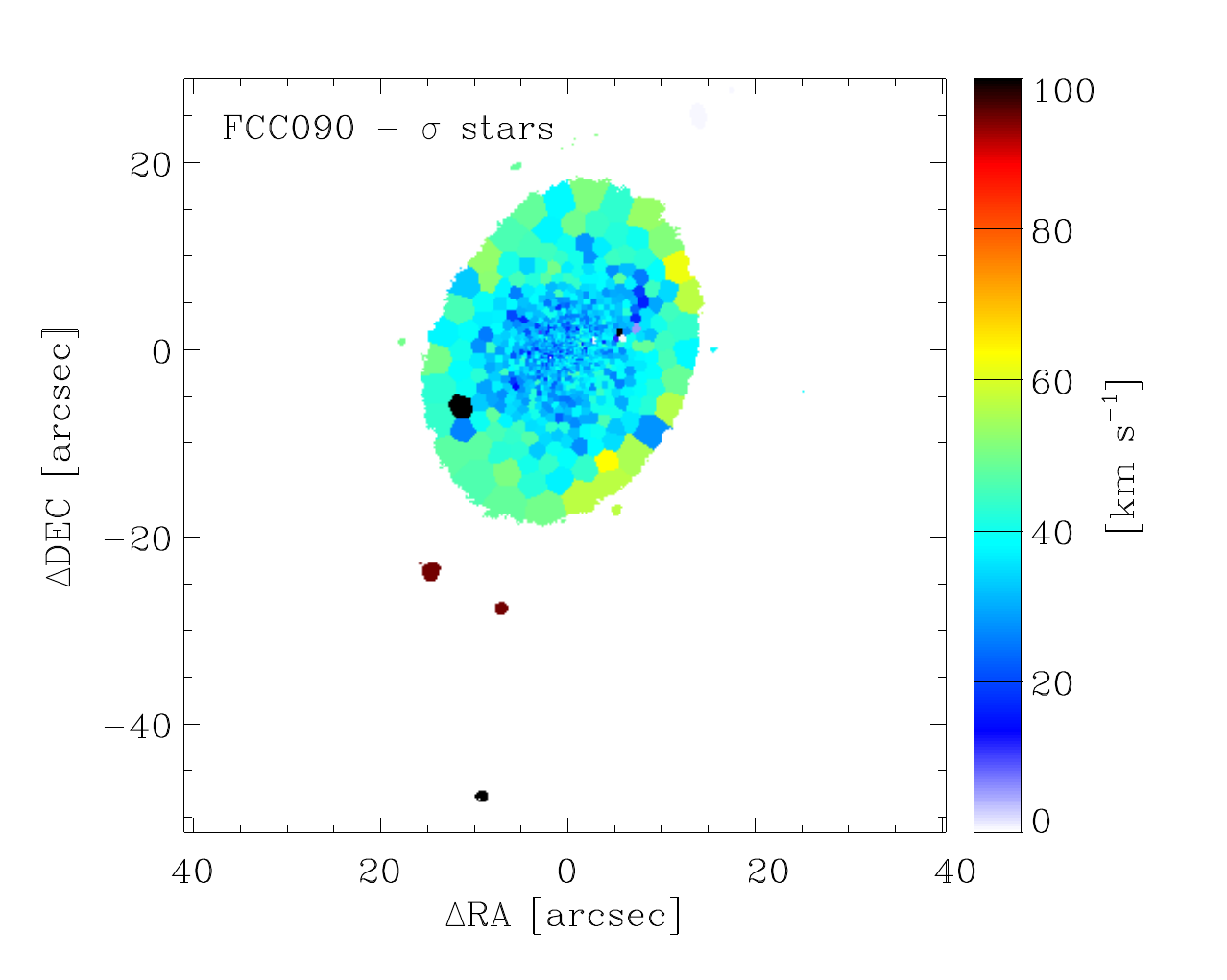} \\
\makebox[0pt][c]{\hspace{-0.65cm}\raisebox{0.2cm}{\includegraphics[width=4.5cm]{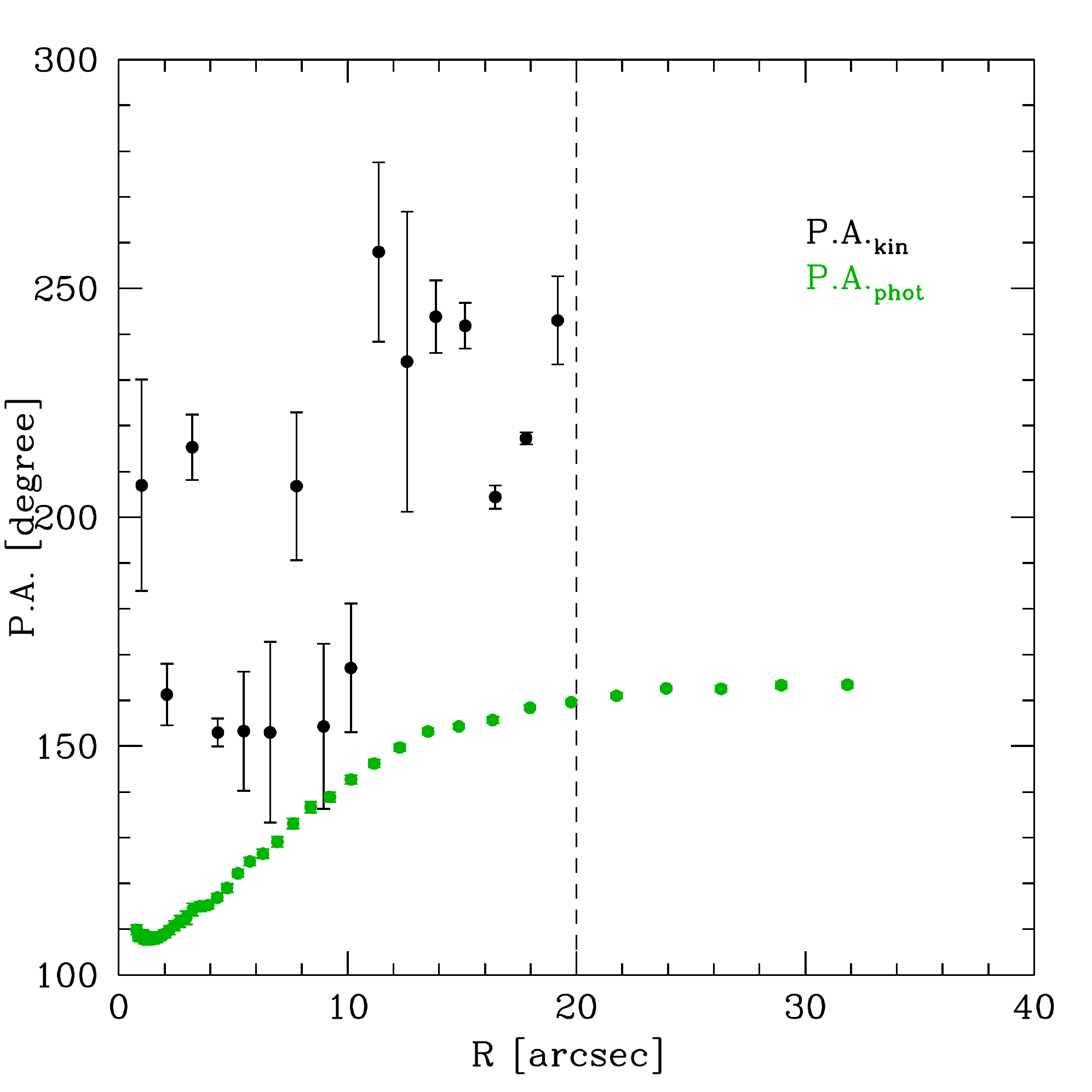}}} &
\includegraphics[width=6cm]{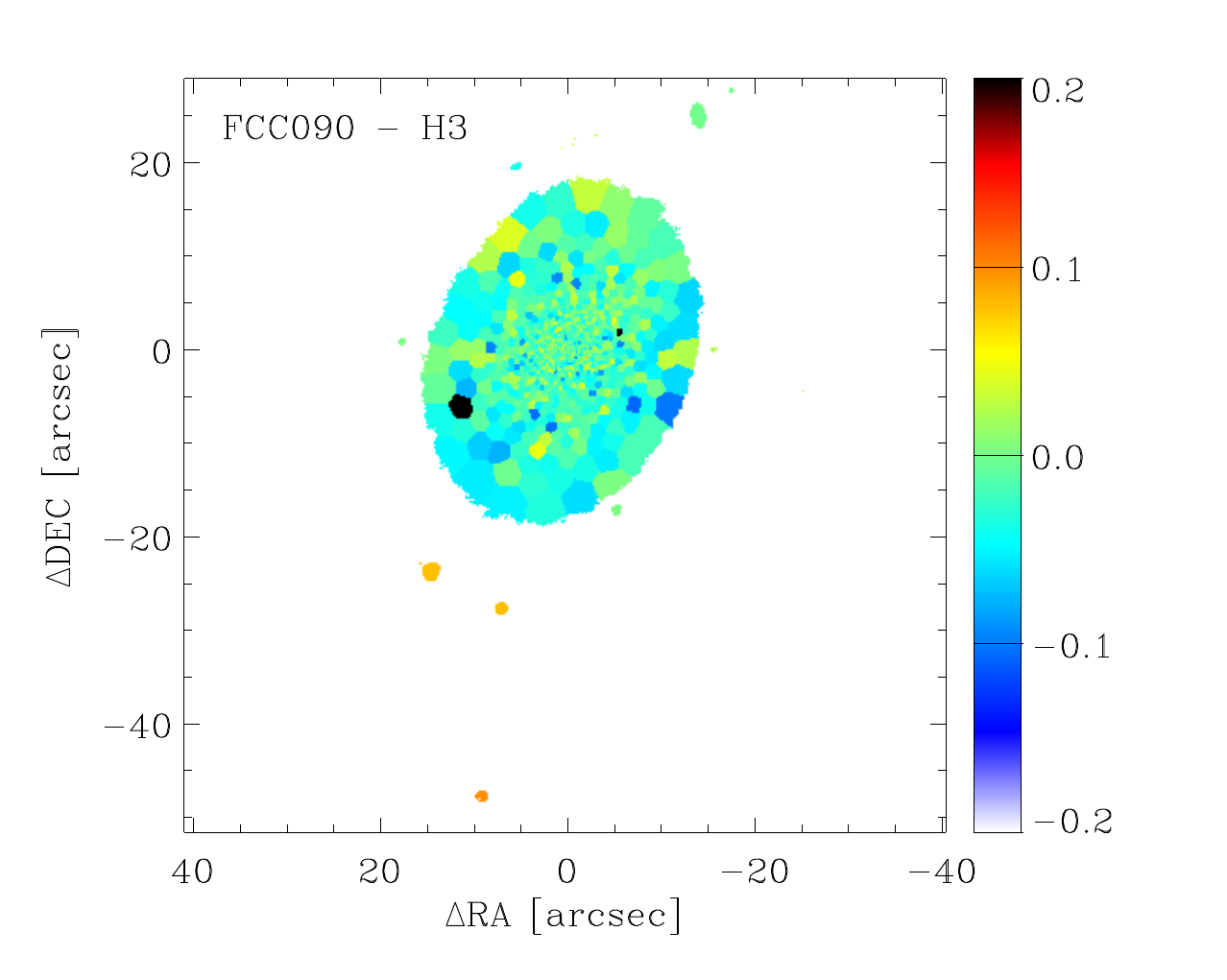} &
\includegraphics[width=6cm]{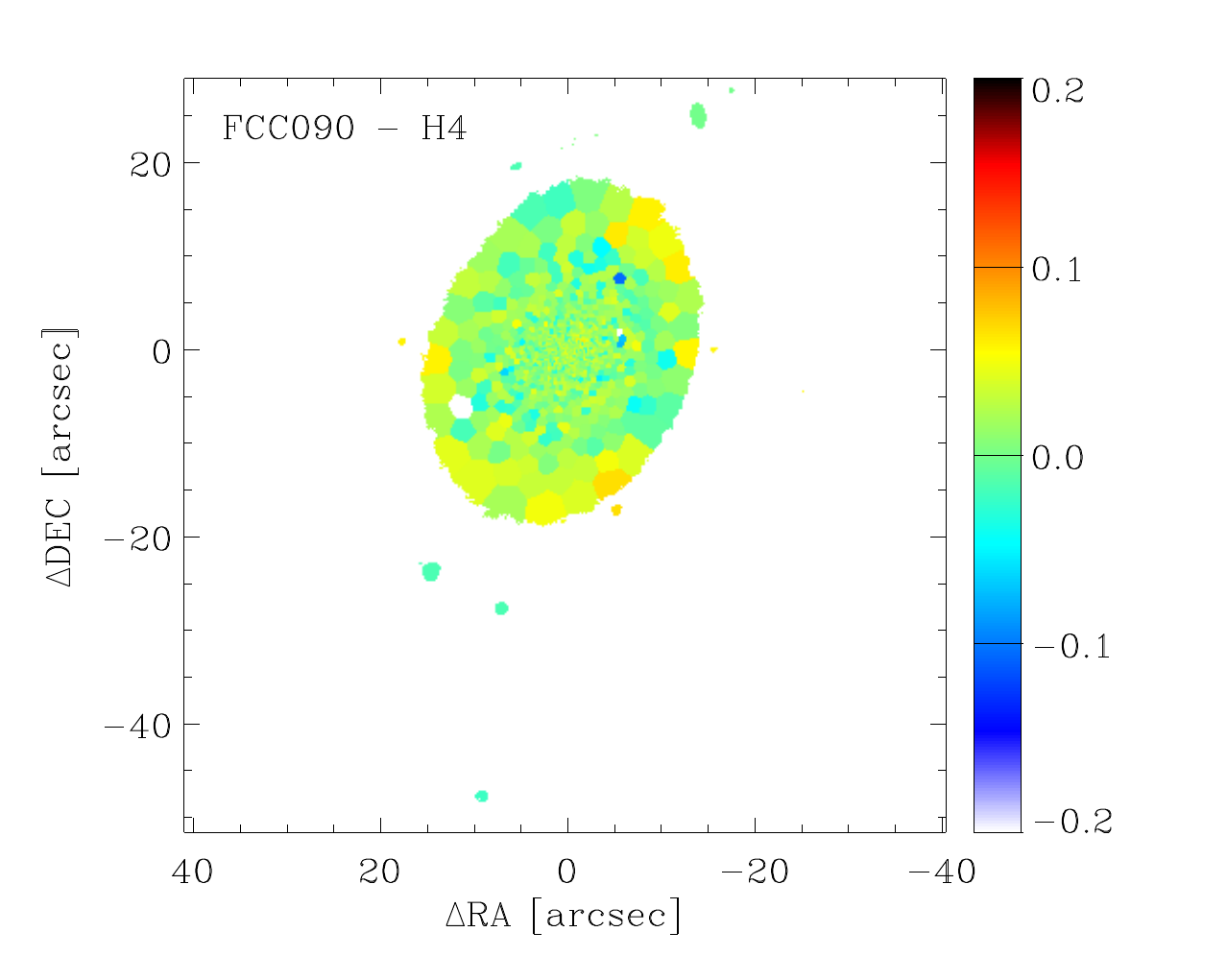} \\
\includegraphics[width=6cm]{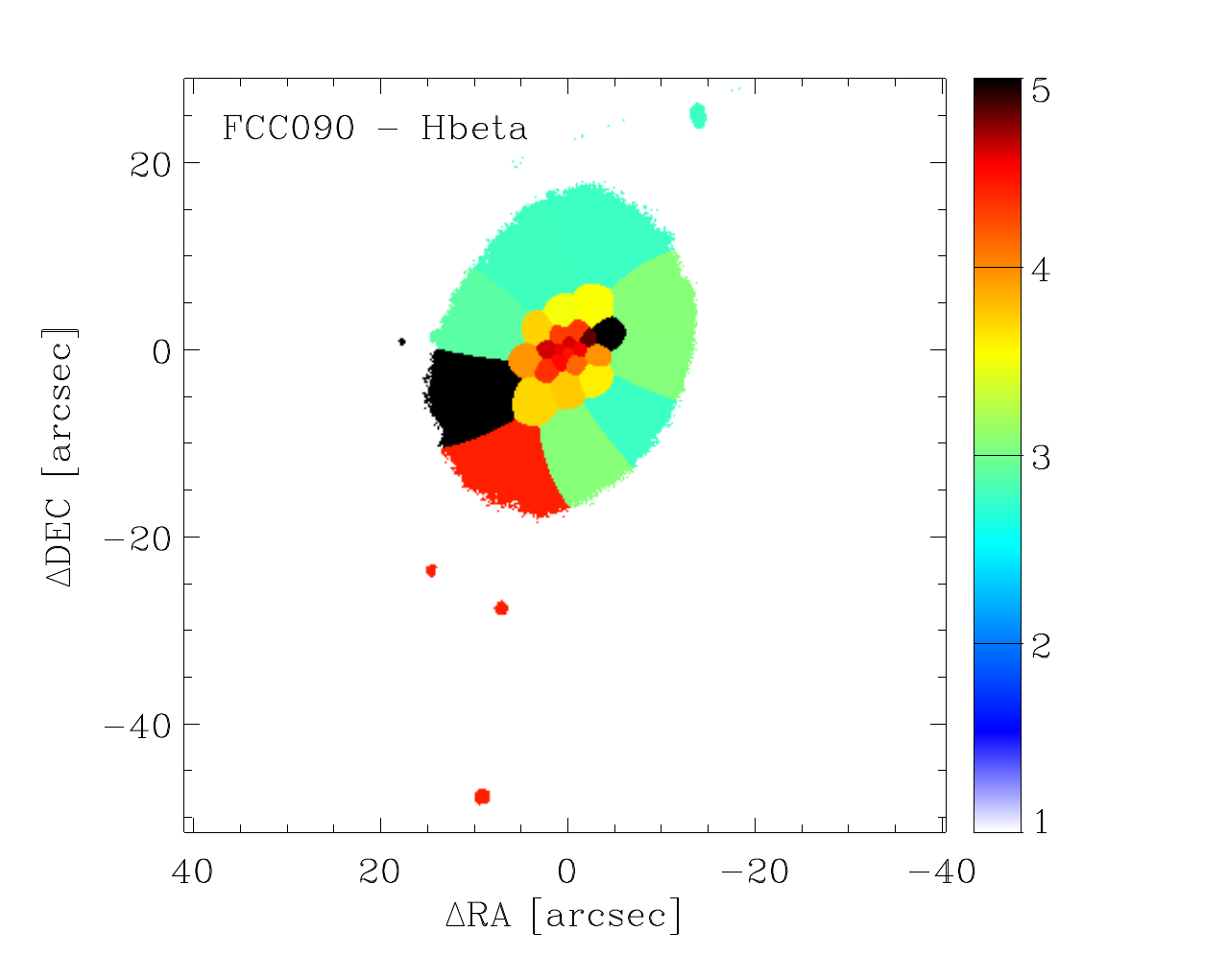} &
\includegraphics[width=6cm]{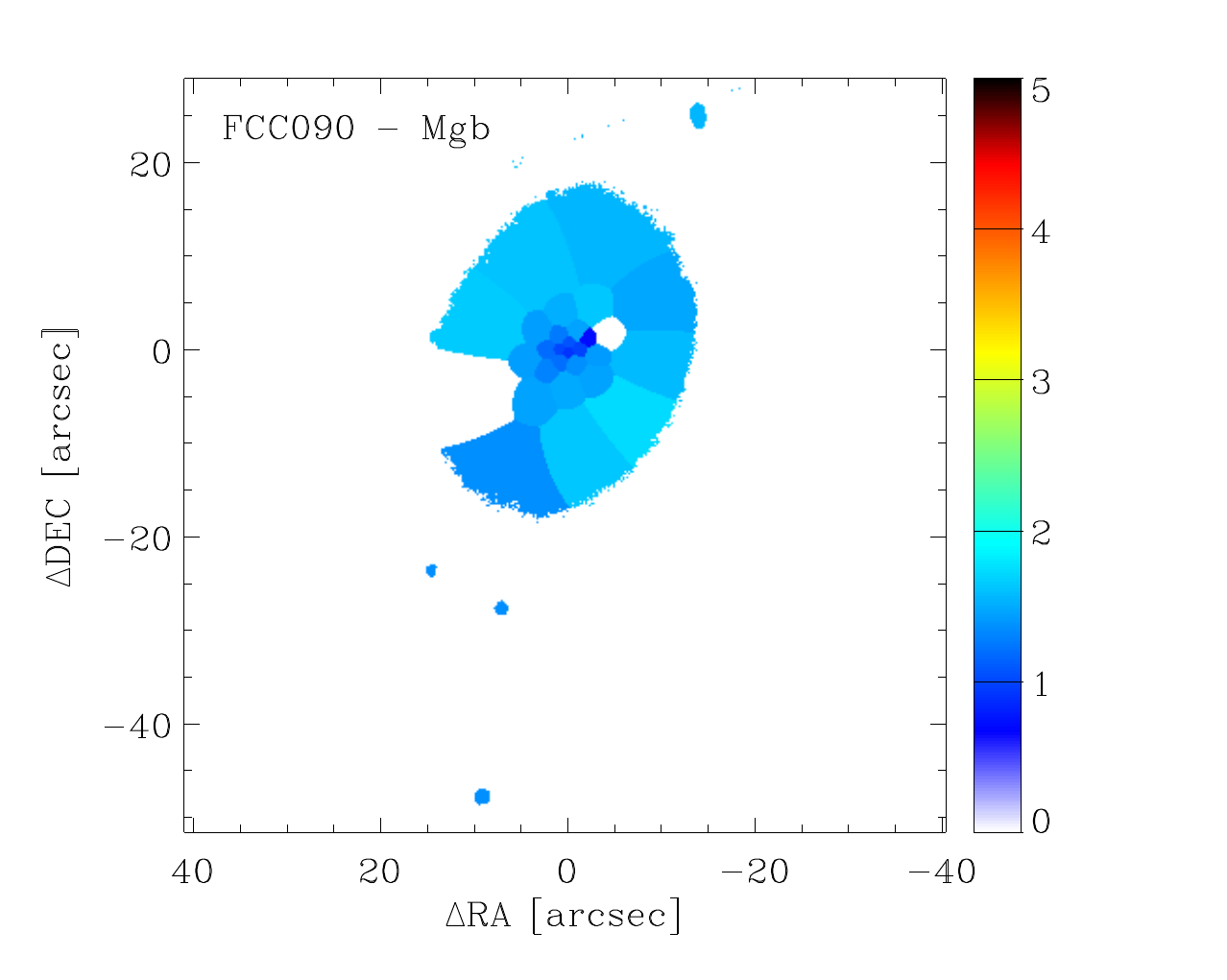} &
\includegraphics[width=6cm]{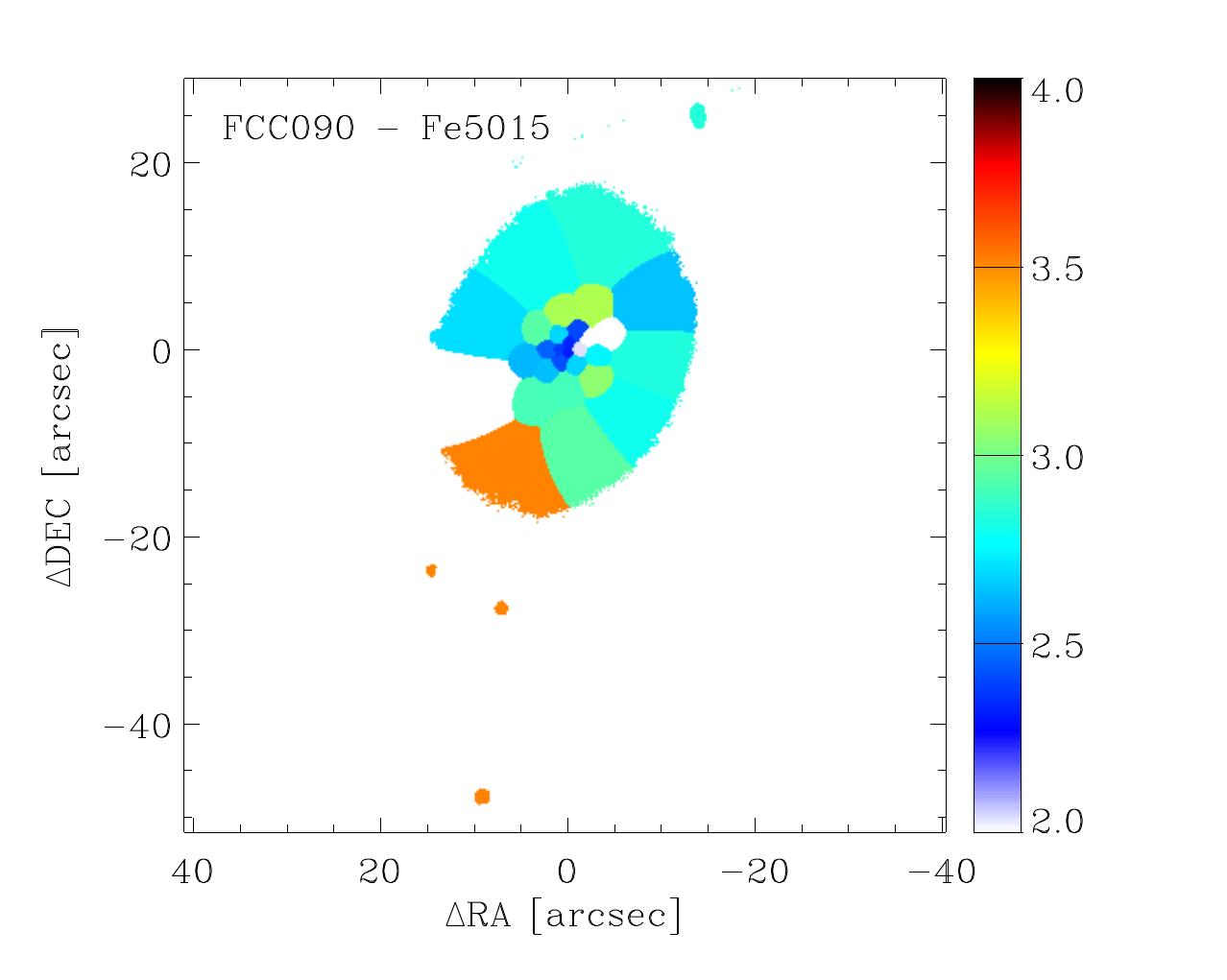} \\
\includegraphics[width=6cm]{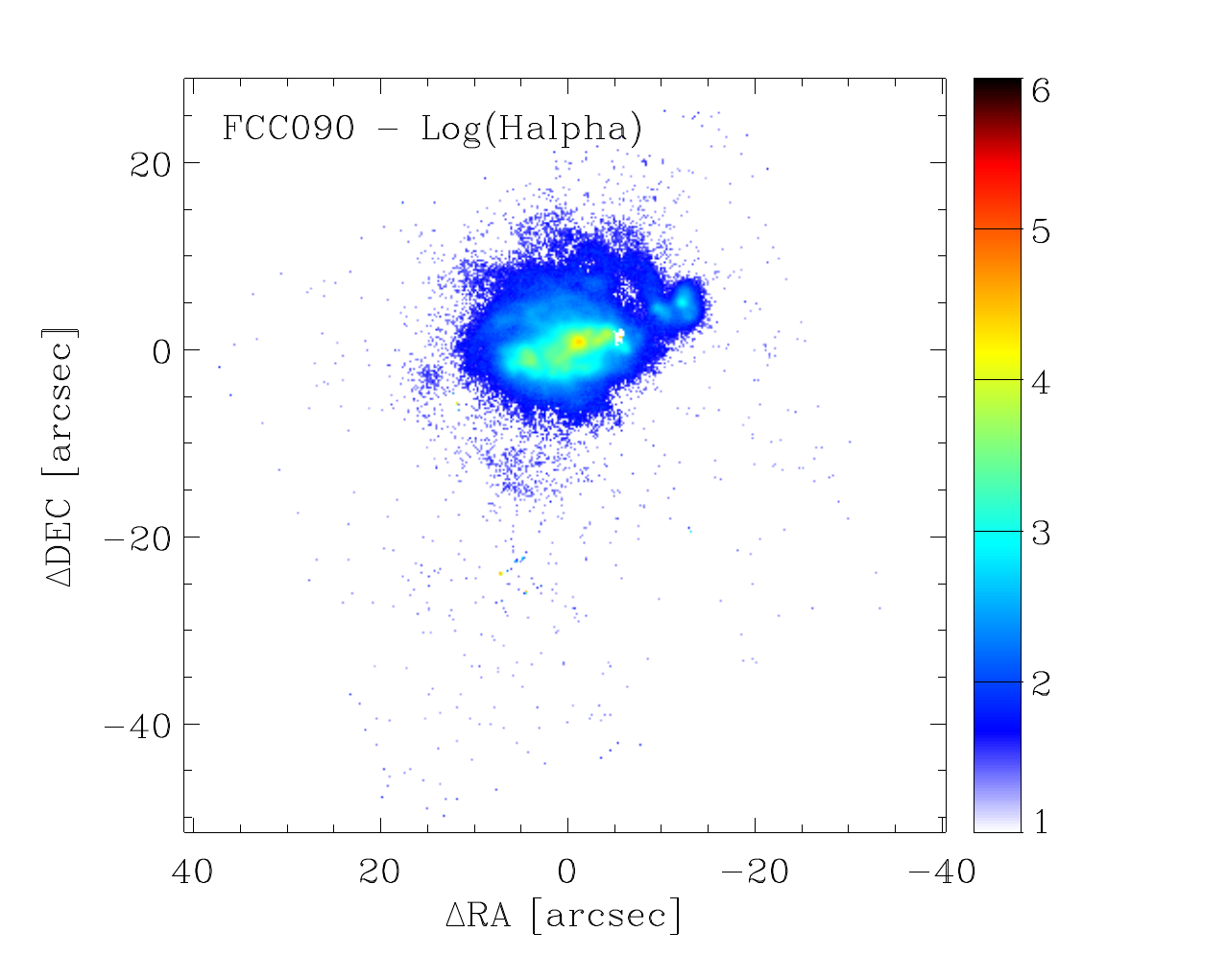} &
\includegraphics[width=6cm]{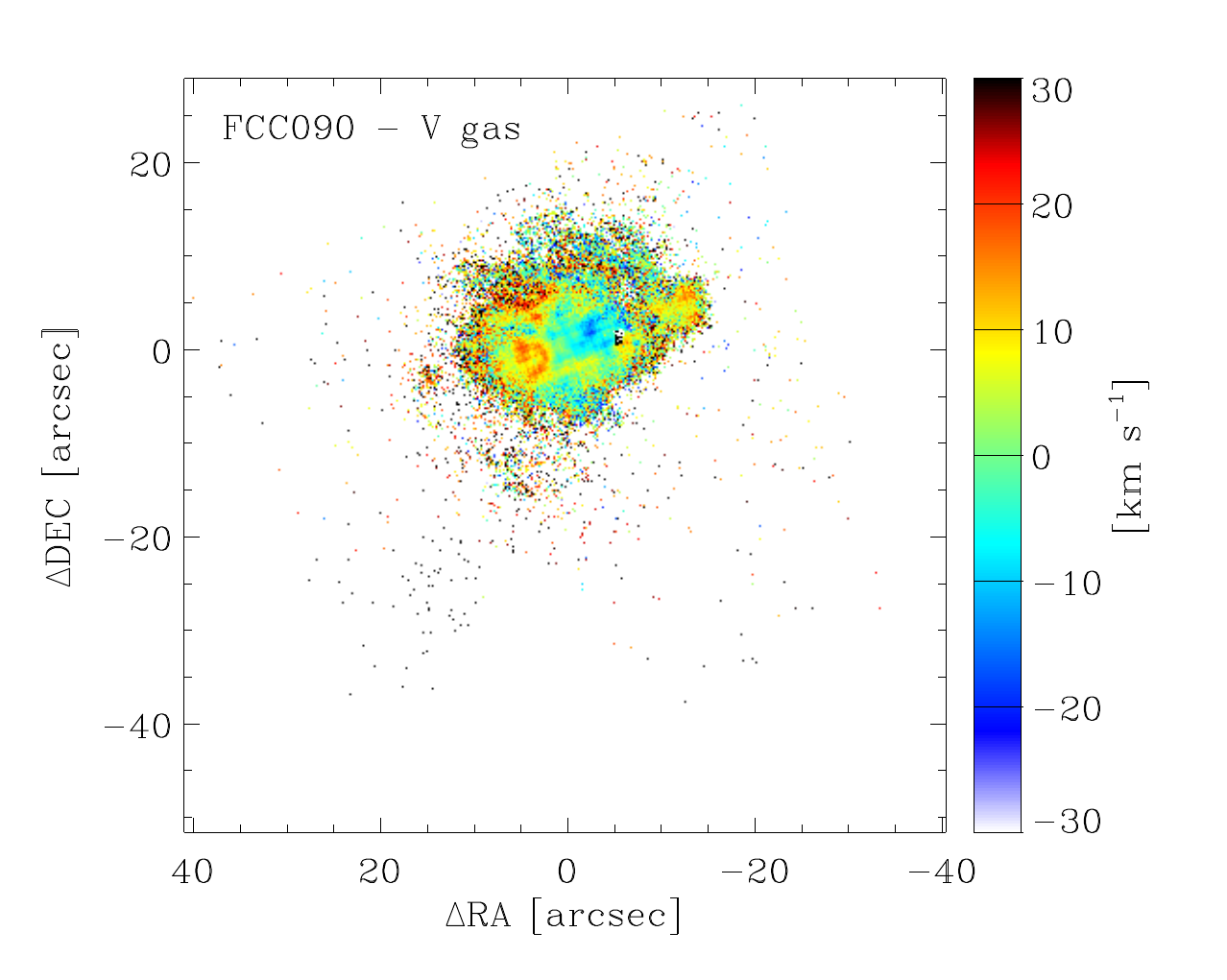} &
\includegraphics[width=6cm]{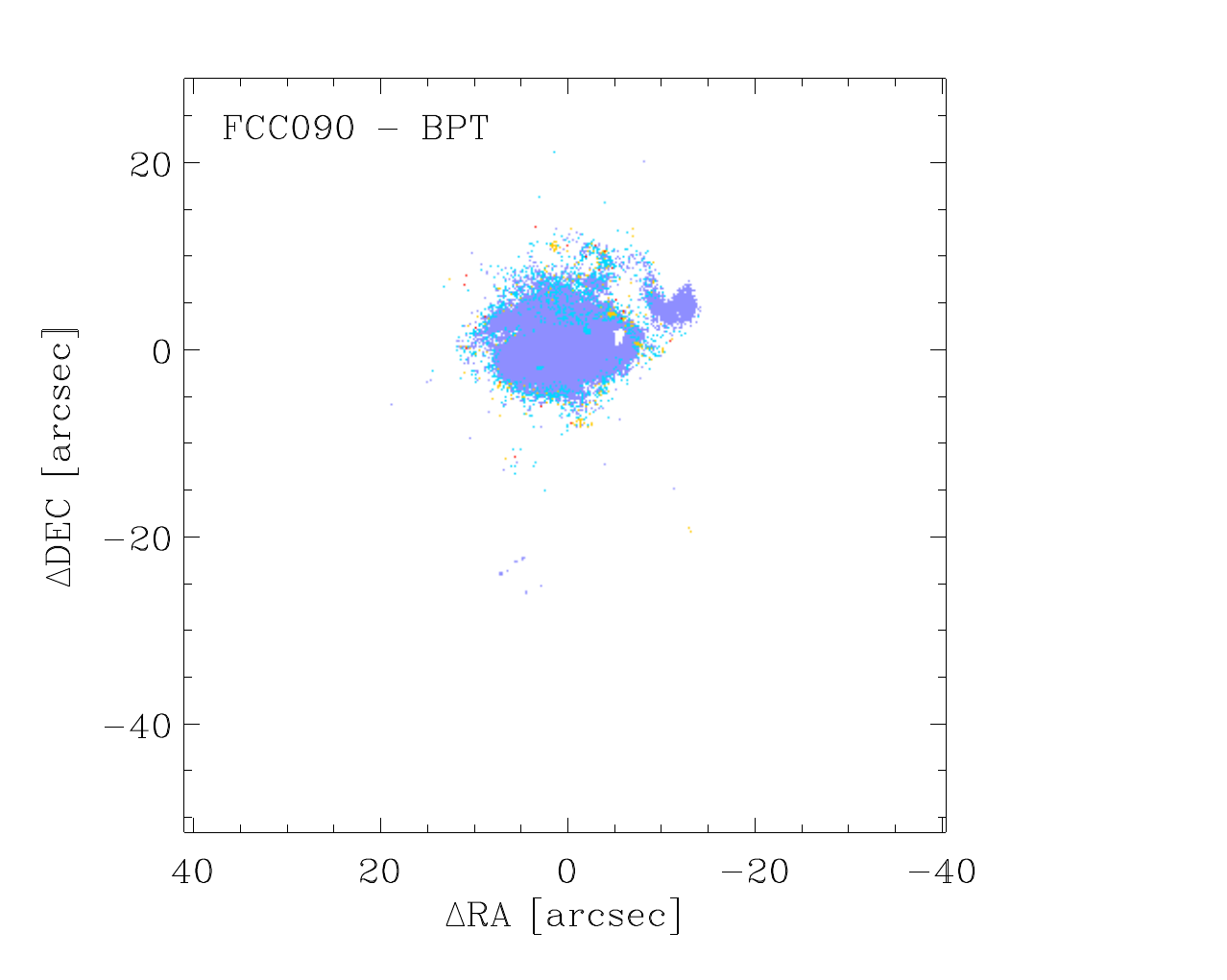}\\
\end{tabular}
\caption{Stellar and ionised-gas analysis for FCC~90.
{\it First row panels:\/} MUSE reconstructed image ({\it left\/}). 
The dotted ellipses correspond to the isophotes at $0.5 R_e$, $R_{\rm tr}$, and $\mu_B=25$~mag arcsec$^{-2}$, respectively. 
Maps of the mean velocity ({\it middle\/}) and velocity dispersion ({\it right\/}) of the stellar LOSVD.
{\it Second row panels:\/} Radial profiles of the kinematic (black circles) and photometric (green circles) position angle ({\it left\/}). The vertical dashed lines mark the radial range where the average position angles are computed. Maps of the third ({\it middle\/}) and fourth Gauss-Hermite coefficient ({\it right\/}) of the stellar LOSVD.
{\it Third row panels:\/} Maps of the H$\beta$ ({\it left\/}), Mg$b$ ({\it middle\/}), and Fe5015 line-strength index ({\it right\/}).
{\it Fourth row panels:\/} Map of the H$\alpha$ flux ({\it left\/}). Data are shown in logarithmic scale and in units of $10^{-20}$ erg~cm$^{-2}$~s$^{-1}$ where the $S/N$ of the line is sufficiently elevated to exclude a false positive detection. 
Map of the H$\alpha$ velocity ({\it middle\/}) with respect to the systemic velocity, as derived from the median gas-velocity value in the central regions of the galaxy.
\bf{Map of the ionised-gas classification ({\it right\/}) into \ion{H}{ii} regions (blue), highly-ionised or AGN regions (yellow), LINER-like emission regions (red), and regions where the nebular emission is powered by some combination of massive-star radiation and other ionisation sources (green)}.  }
\label{fig:FCC090map}
\end{figure*}

\begin{figure*}[t!]
\begin{tabular}{ccc}
\includegraphics[width=6cm]{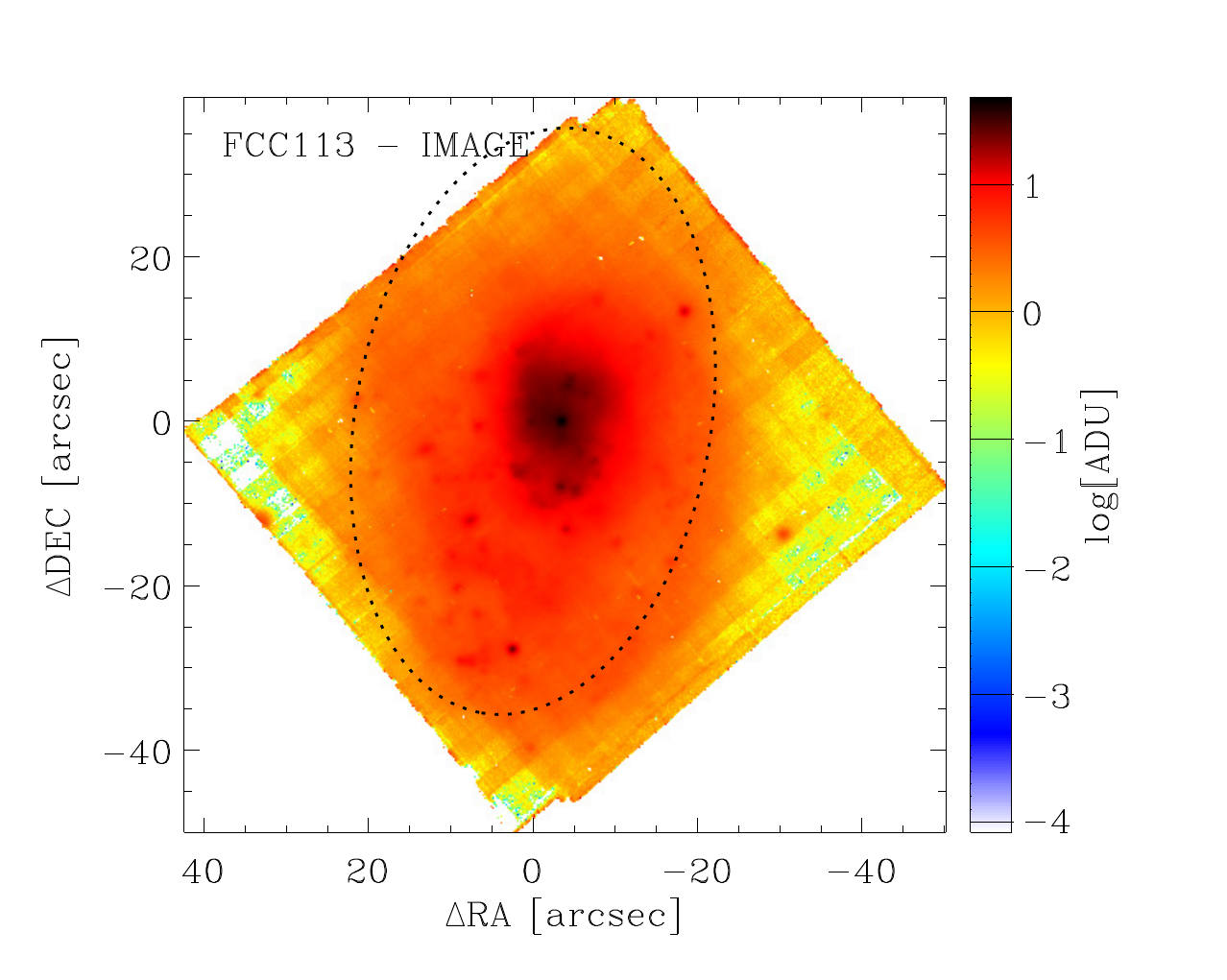} &
\includegraphics[width=6cm]{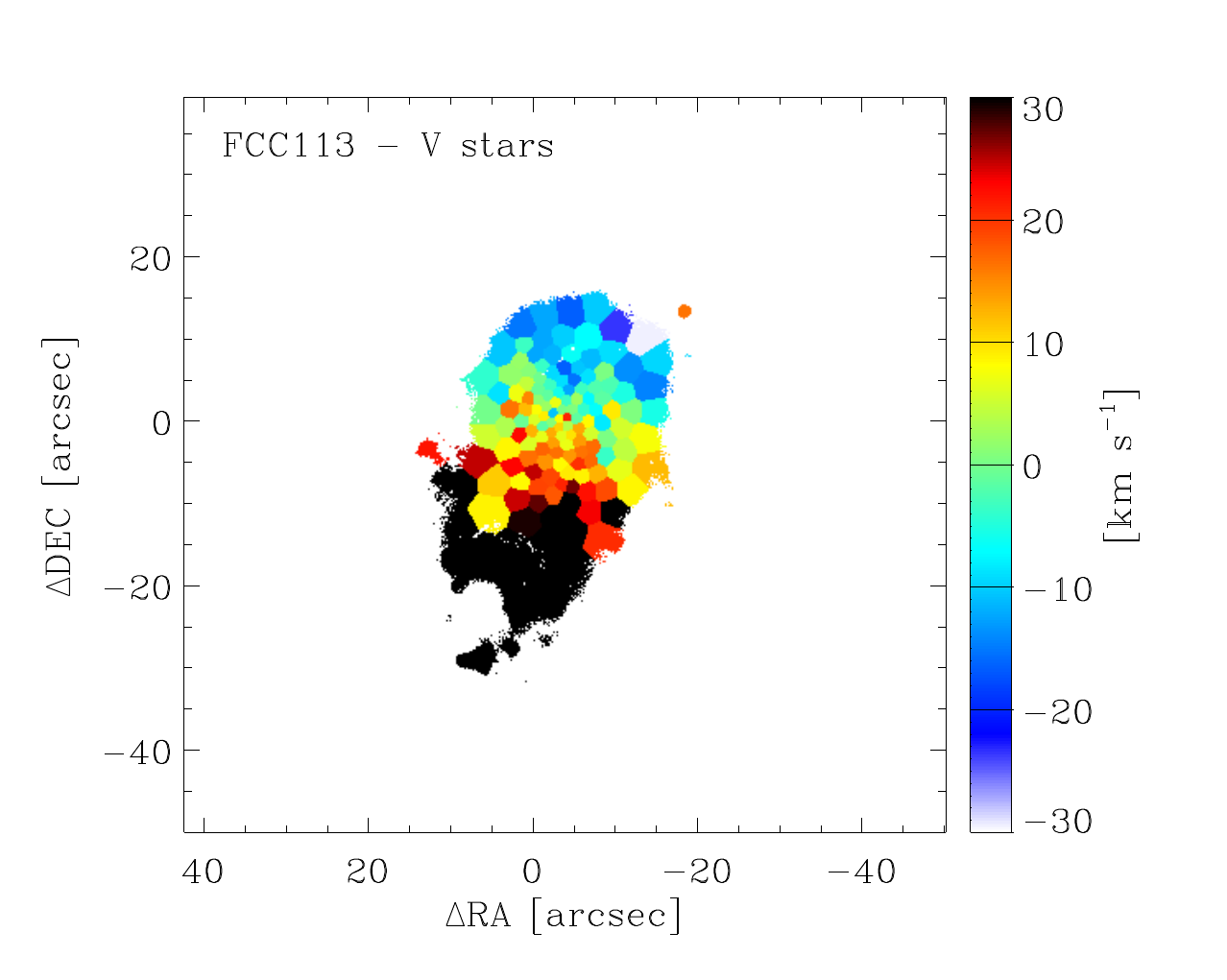} &
\includegraphics[width=6cm]{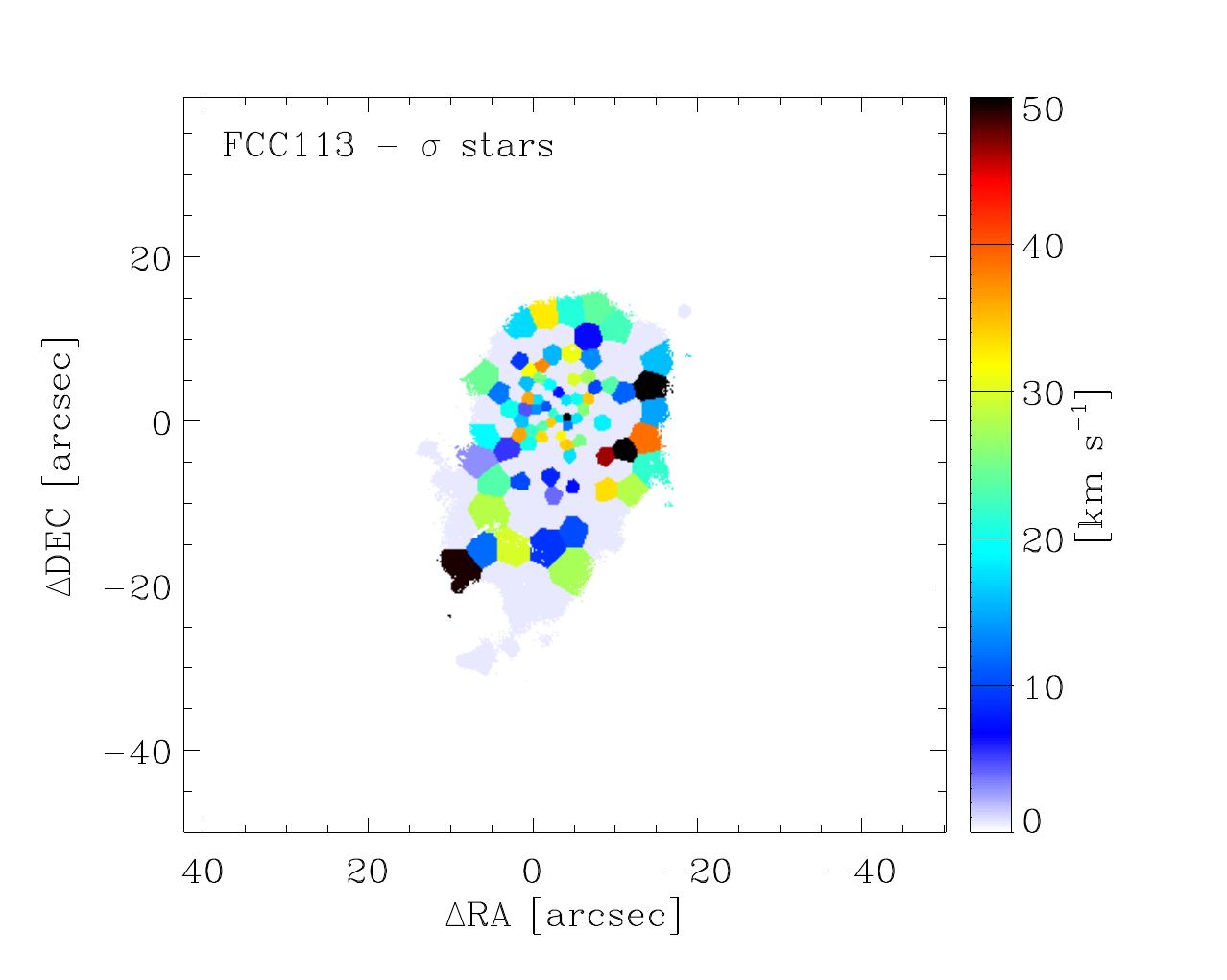} \\
\makebox[0pt][c]{\hspace{-0.65cm}\raisebox{0.2cm}{\includegraphics[width=4.5cm]{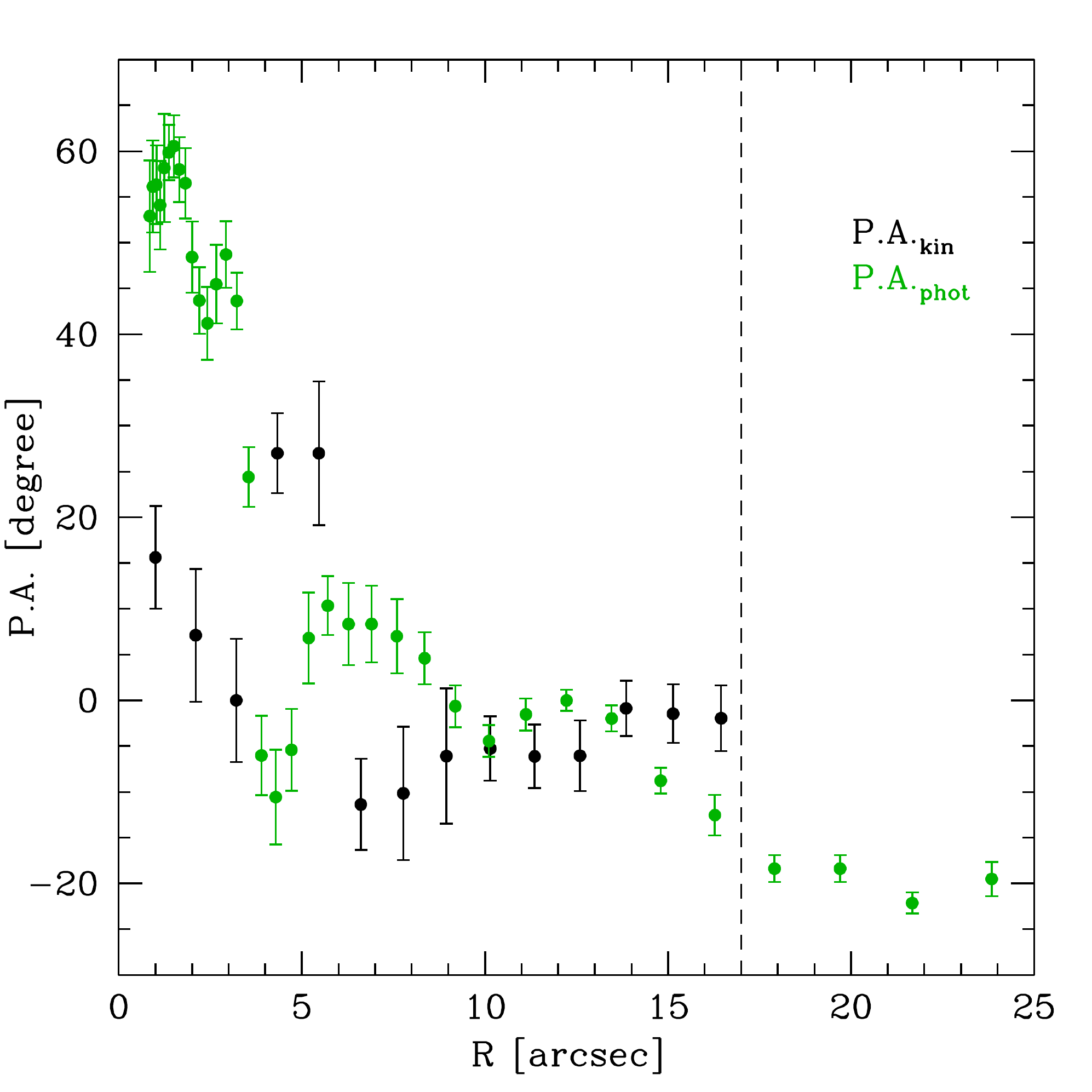}}} & & \\
\includegraphics[width=6cm]{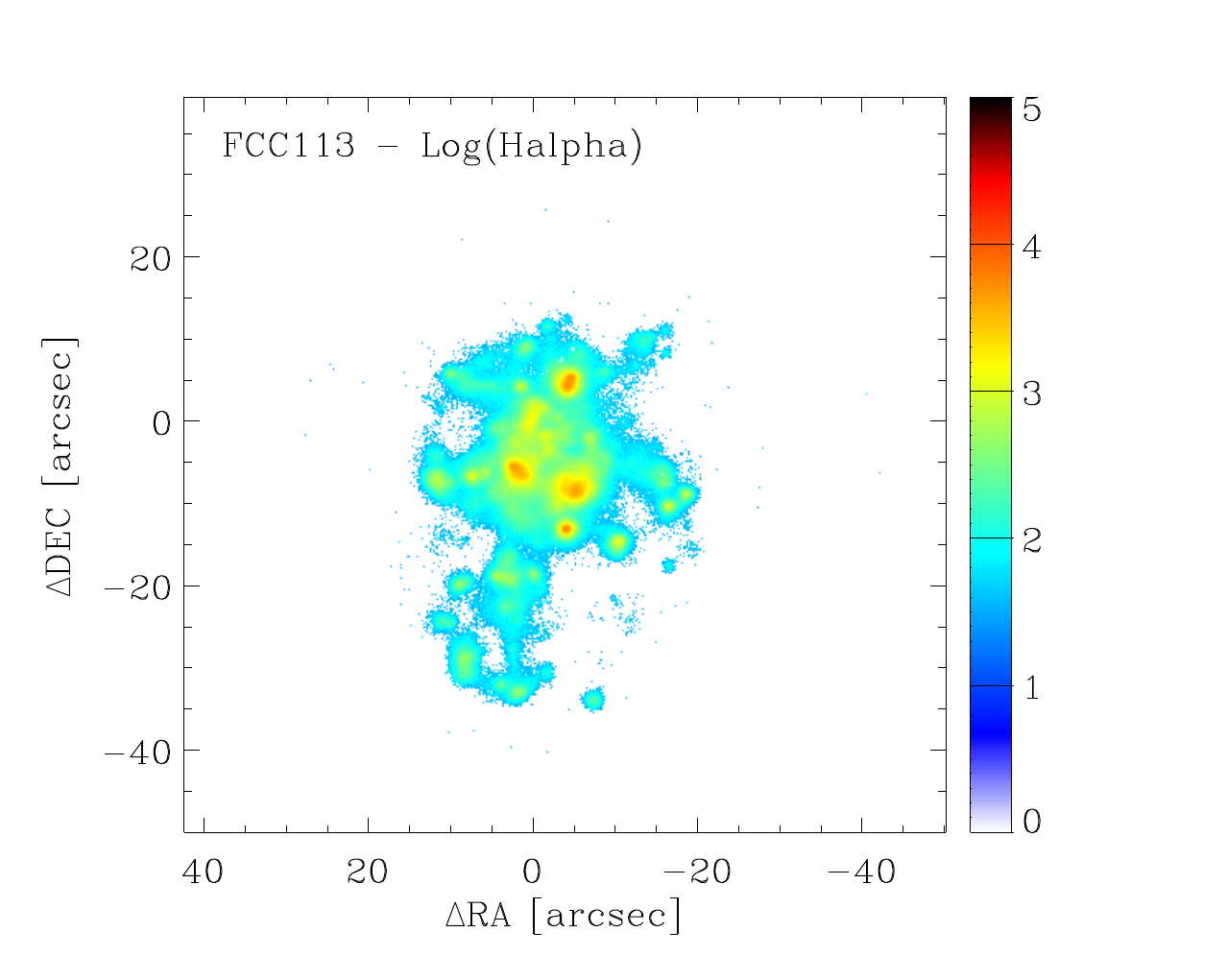} &
\includegraphics[width=6cm]{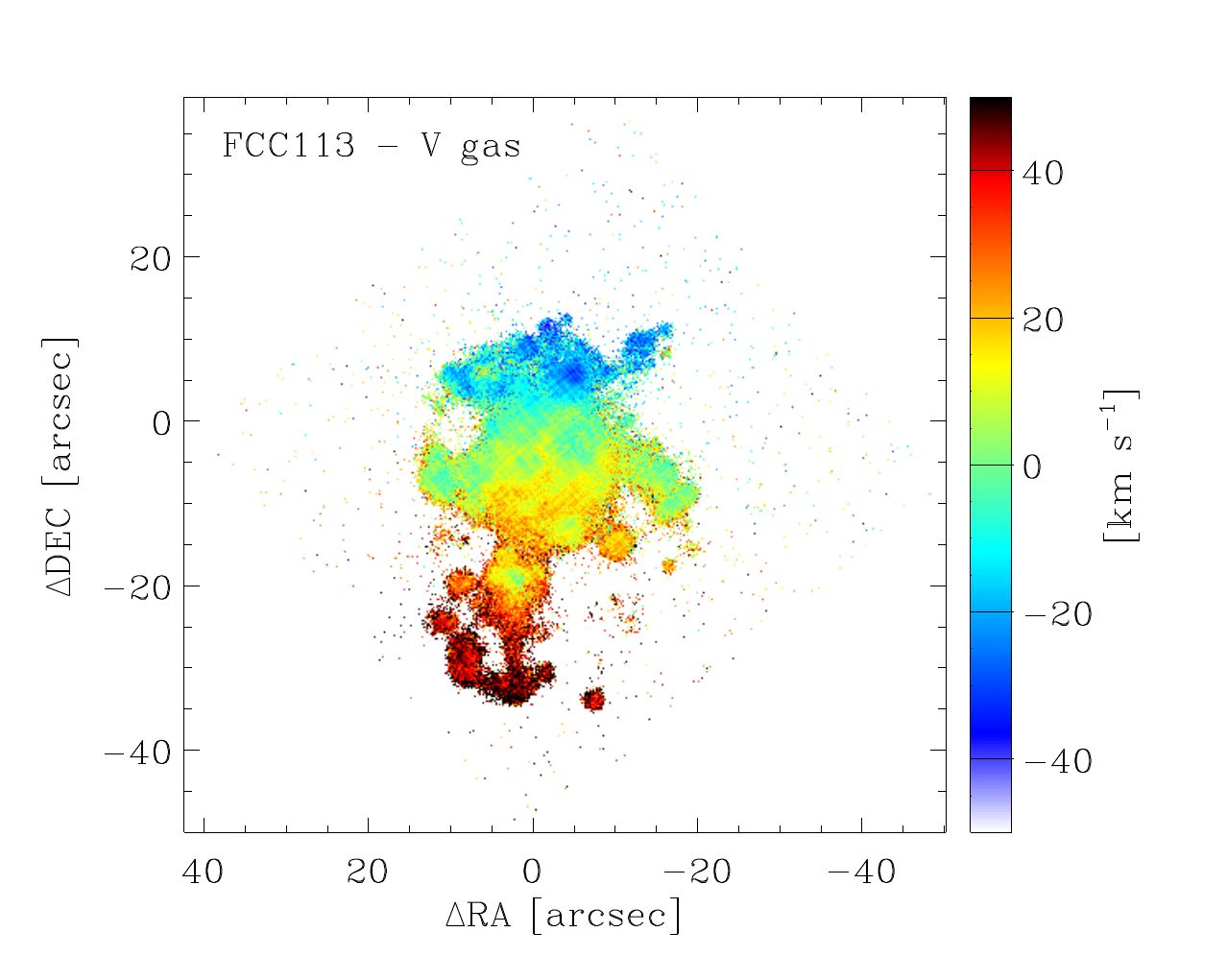} &
\includegraphics[width=6cm]{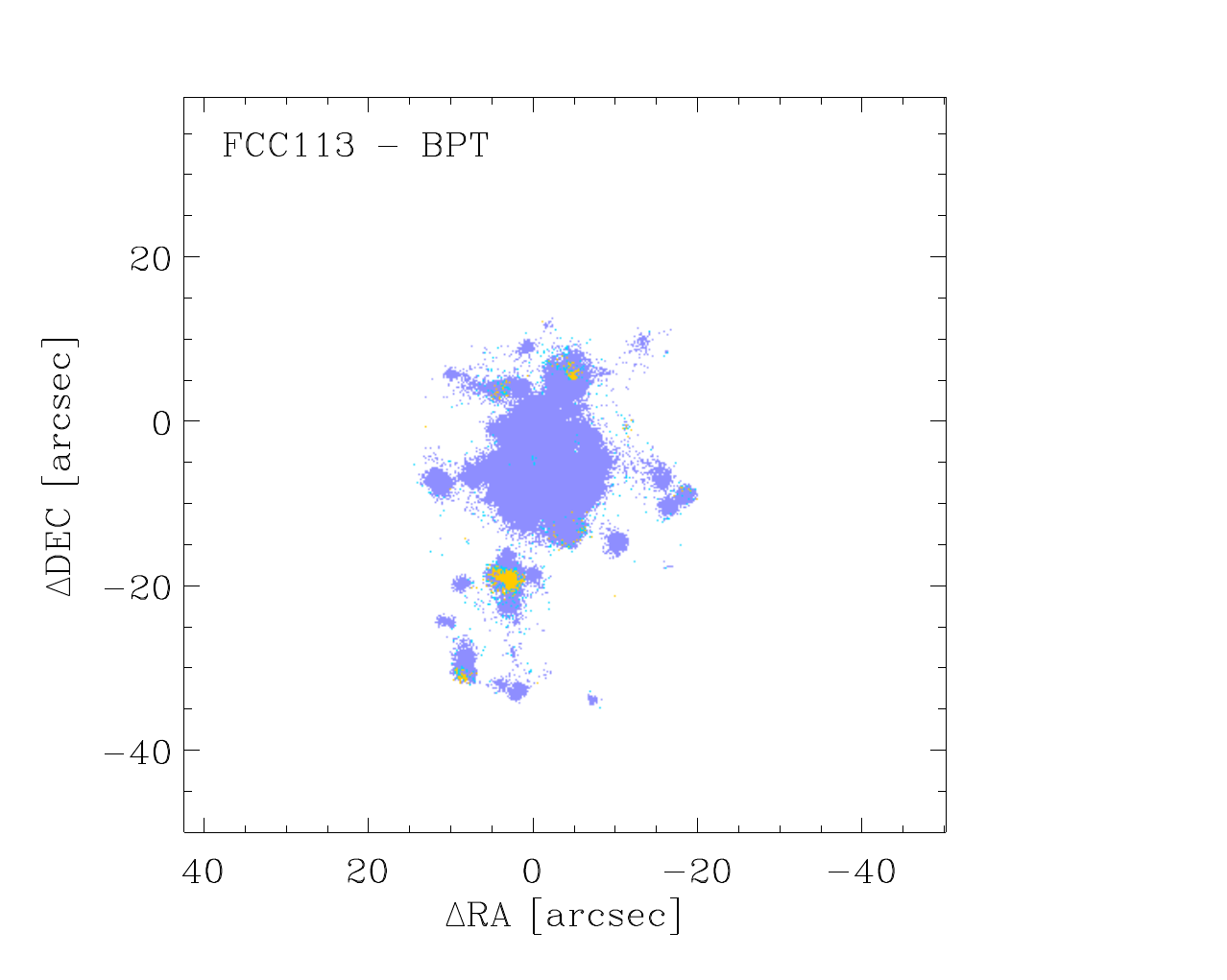} \\
\end{tabular}
\caption{Stellar and ionised-gas analysis for FCC~113.
{\it First row panels:\/} MUSE reconstructed image ({\it left\/}). The dotted  ellipse corresponds to the isophote at $\mu_B=25$~mag arcsec$^{-2}$.
Maps of the mean velocity ({\it middle\/}) and velocity dispersion ({\it right\/}) of the stellar LOSVD.
{\it Second row panel:\/} Radial profiles of the kinematic (black circles) and photometric (green circles) position angle. 
{\it Third row panels:\/} Map of the H$\alpha$ flux ({\it left\/}). Data are shown in logarithmic scale and in units of $10^{-20}$ erg~cm$^{-2}$~s$^{-1}$ where the $S/N$ of the line is sufficiently elevated to exclude a false positive detection. 
Map of the H$\alpha$ velocity ({\it middle\/}) with respect to the systemic velocity, as derived from the median gas-velocity value in the central regions of the galaxy.
 Map of the ionised-gas classification ({\it right\/}) into \ion{H}{ii} regions (light-blue), highly-ionised or AGN regions (yellow), LINER-like emission regions (red), and regions where the nebular emission is powered by some combination of massive-star radiation and other ionisation sources (green). }
\label{fig:FCC113map}
\end{figure*}

\begin{figure*}[t!]
\caption{Same as in Fig.~\ref{fig:FCC090map}, but for FCC~119.}
\label{fig:FCC119map}
\end{figure*}

\begin{figure*}[t!]
\caption{Same as in Fig.~\ref{fig:FCC083map}, but for FCC~143.}
\label{fig:FCC143map}
\end{figure*}

\begin{figure*}[t!]
\caption{Same as in Fig.~\ref{fig:FCC083map}, but for FCC~147.}
\label{fig:FCC147map}
\end{figure*}

\begin{figure*}[t!]
\caption{Same as in Fig.~\ref{fig:FCC083map}, but for FCC~148.}
\label{fig:FCC148map}
\end{figure*}

\begin{figure*}[t!]
\caption{Same as in Fig.~\ref{fig:FCC083map}, but for FCC~153.}
\label{fig:FCC153map}
\end{figure*}

\begin{figure*}[t!]
\caption{Same as in Fig.~\ref{fig:FCC083map}, but for FCC~161.}
\label{fig:FCC161map}
\end{figure*}

\begin{figure*}[t!]
\caption{Same as in Fig.~\ref{fig:FCC090map}, but for FCC~167.}
\label{fig:FCC167map}
\end{figure*}

\begin{figure*}[t!]
\caption{Same as in Fig.~\ref{fig:FCC083map}, but for FCC~170.}
\label{fig:FCC170map}
\end{figure*}

\begin{figure*}[t!]
\begin{tabular}{ccc}
\includegraphics[width=6cm]{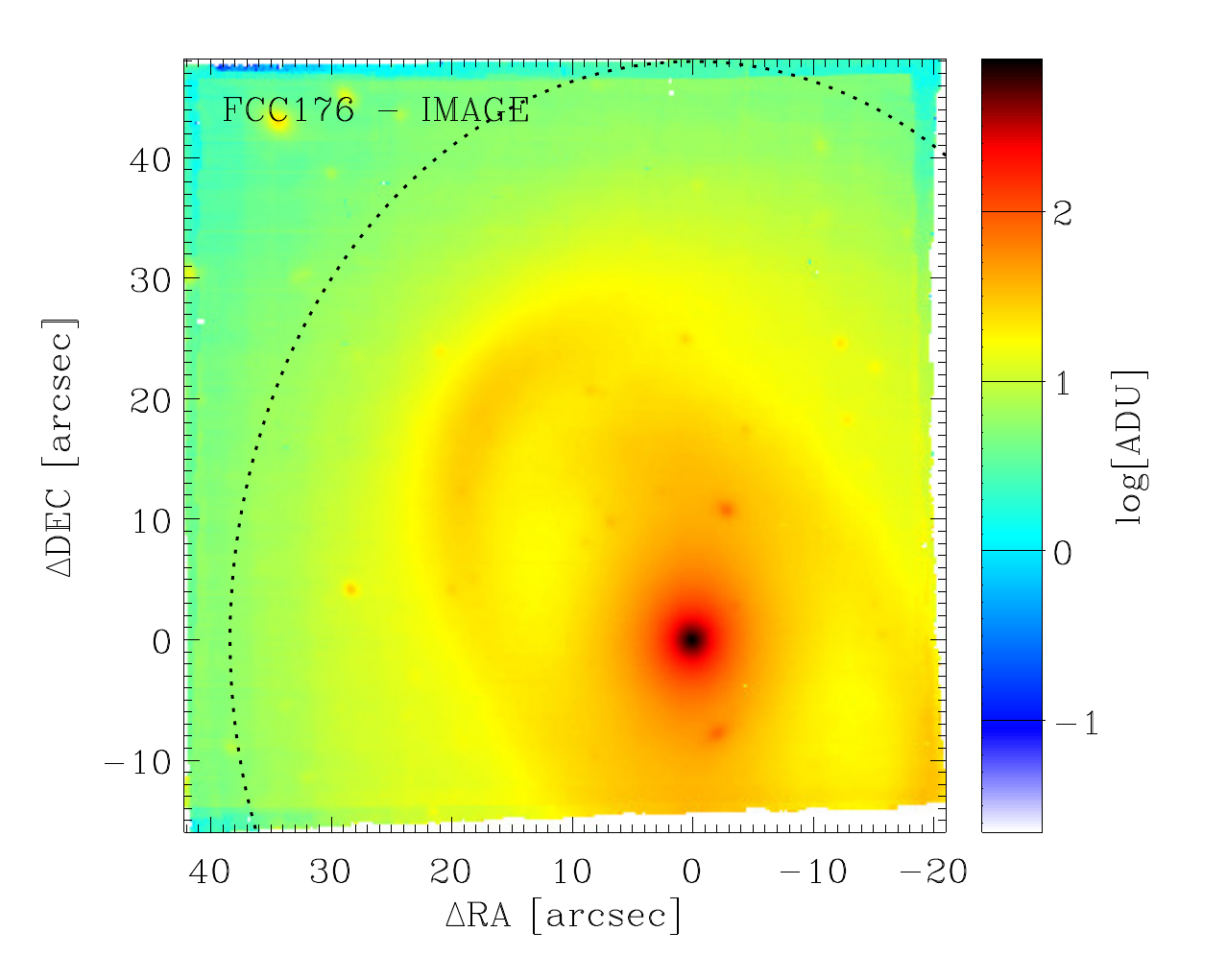} &
\includegraphics[width=6cm]{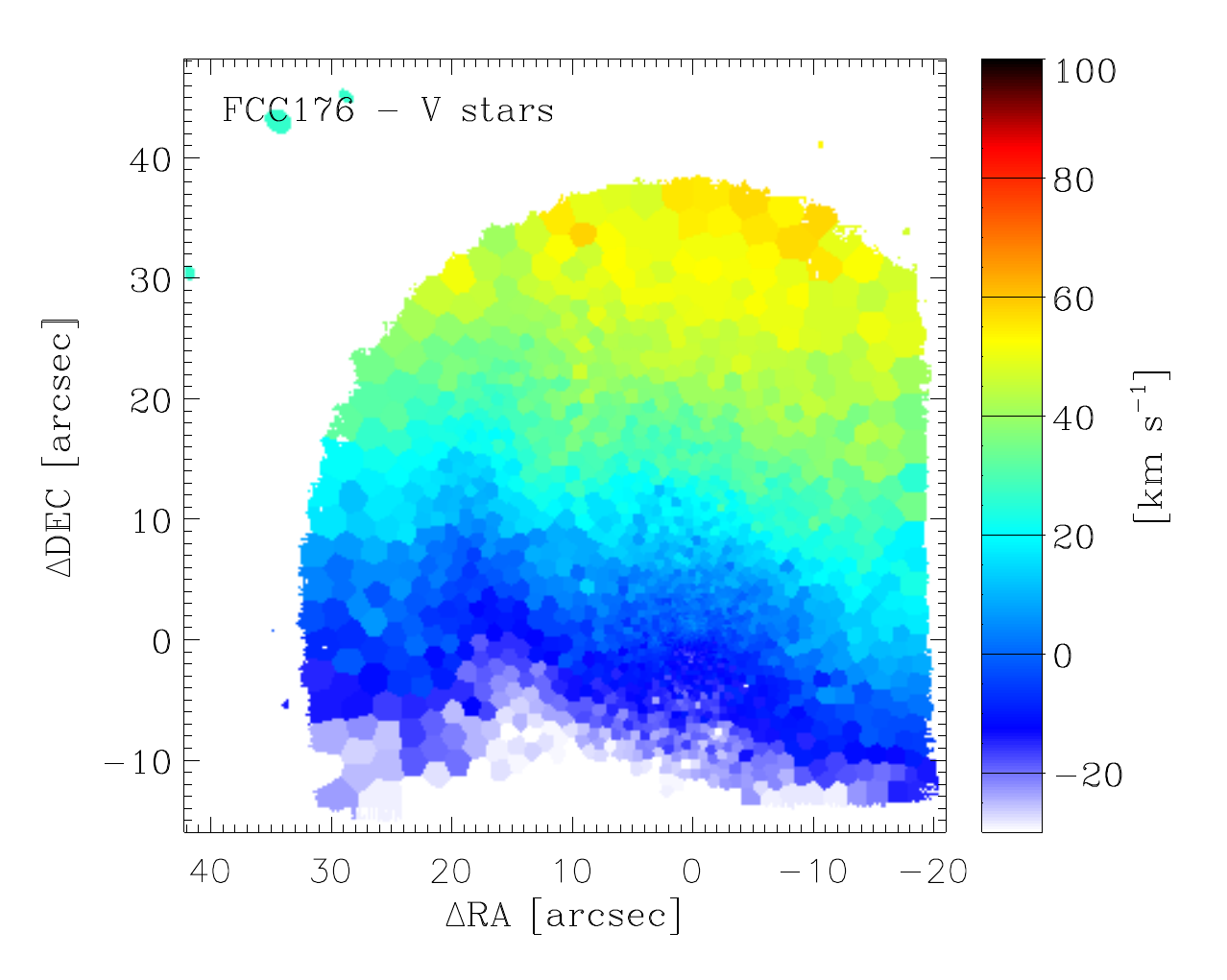} &
\includegraphics[width=6cm]{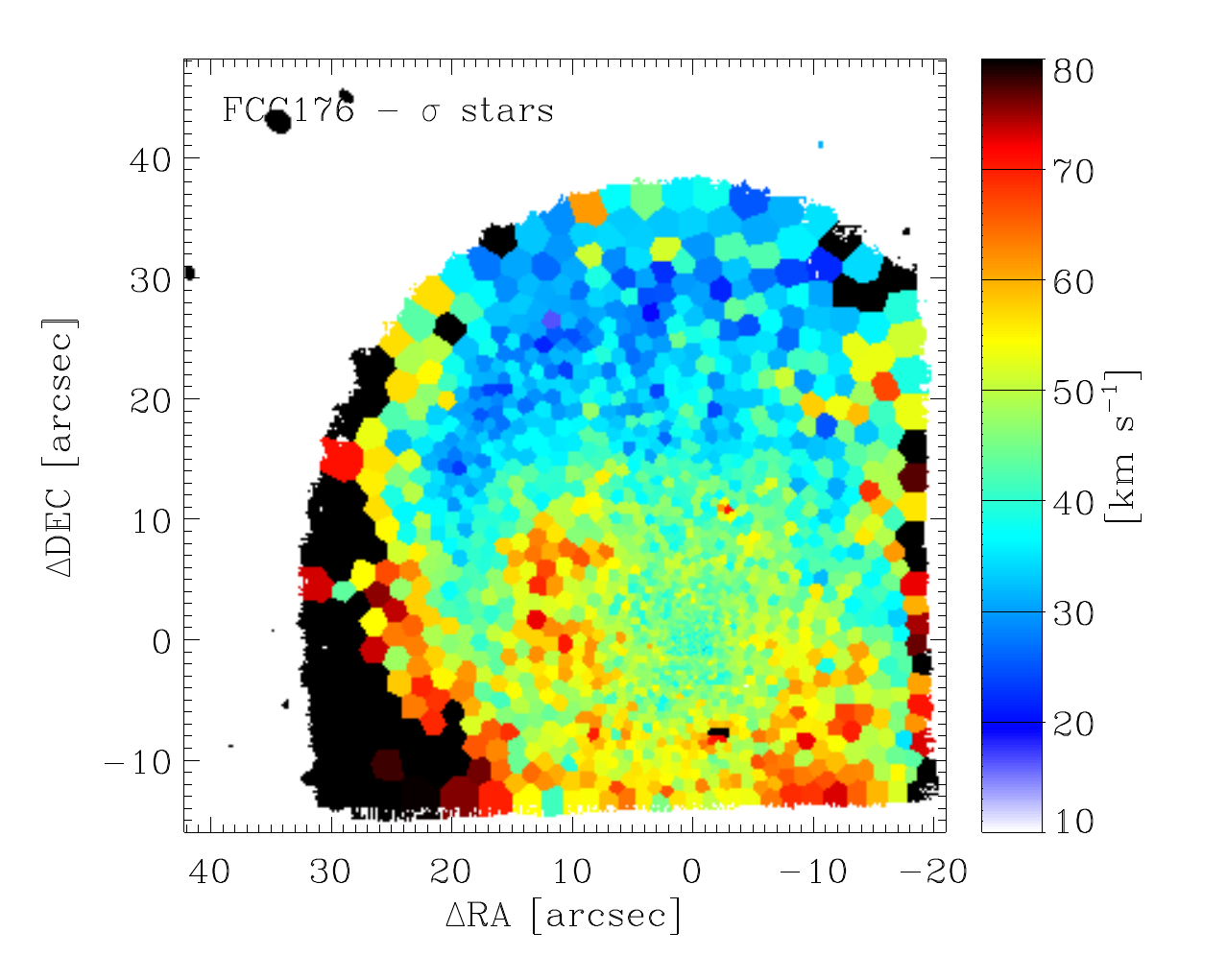} \\
\makebox[0pt][c]{\hspace{-0.55cm}\raisebox{0.25cm}{\includegraphics[width=4.6cm]{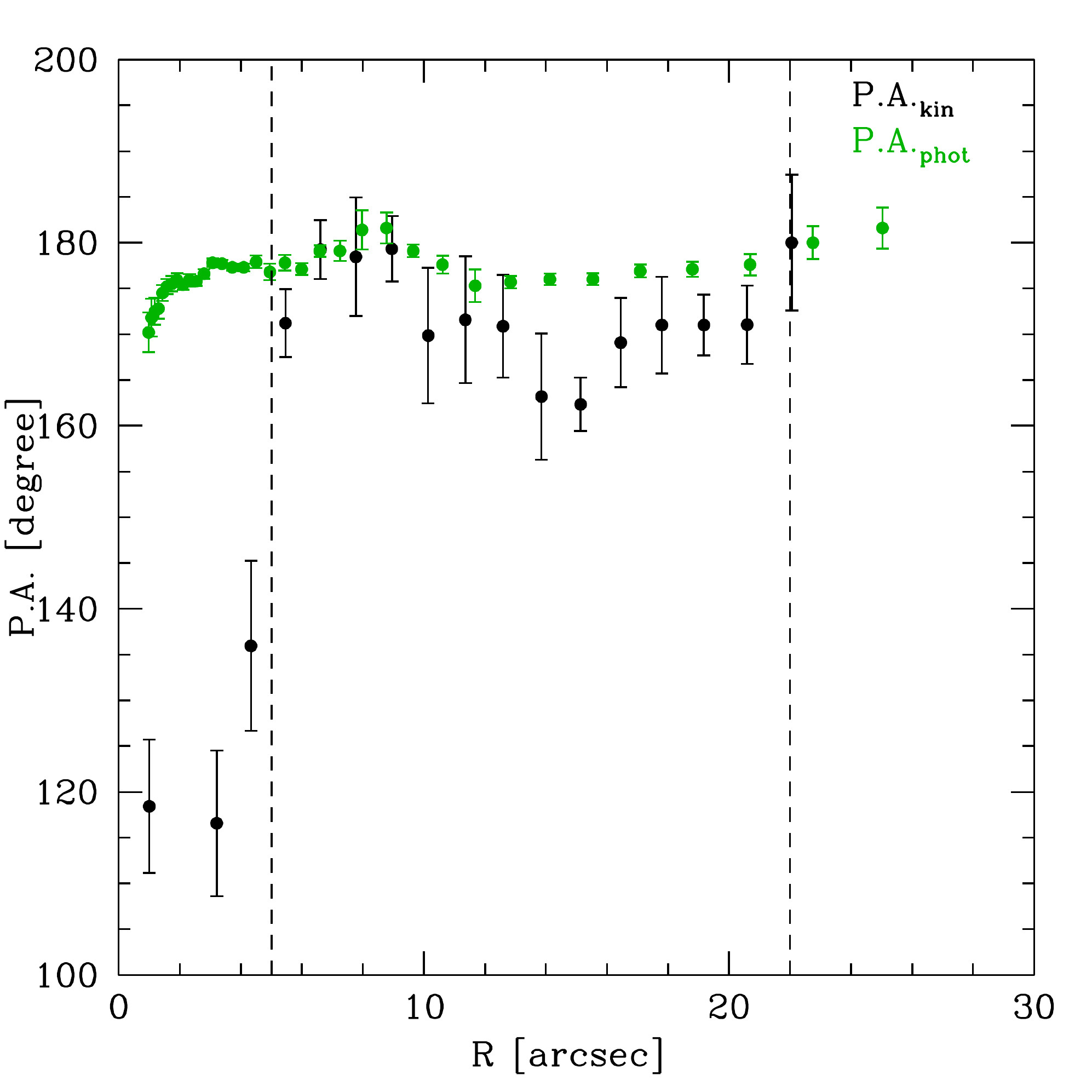}}} &
\includegraphics[width=6cm]{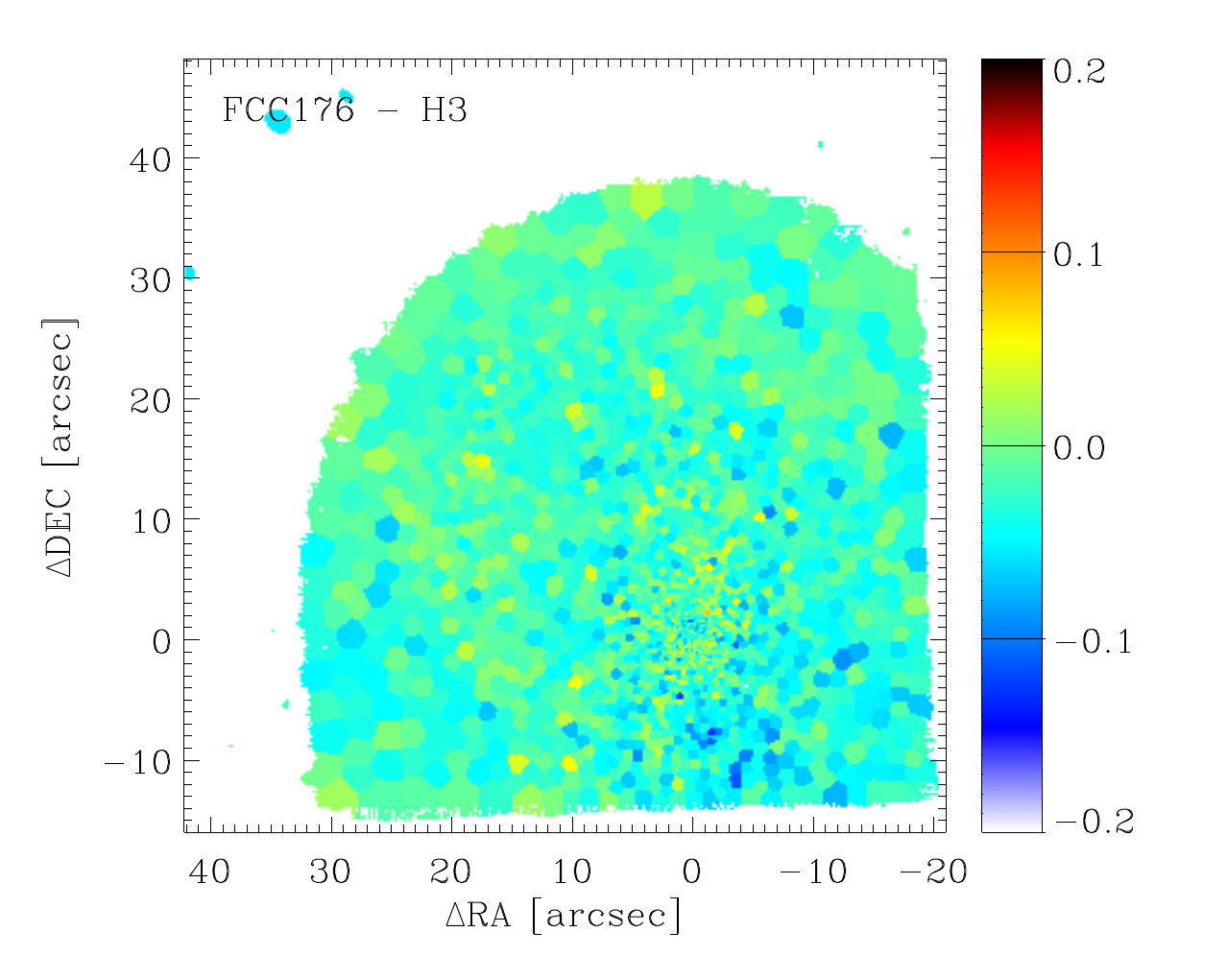} &
\includegraphics[width=6cm]{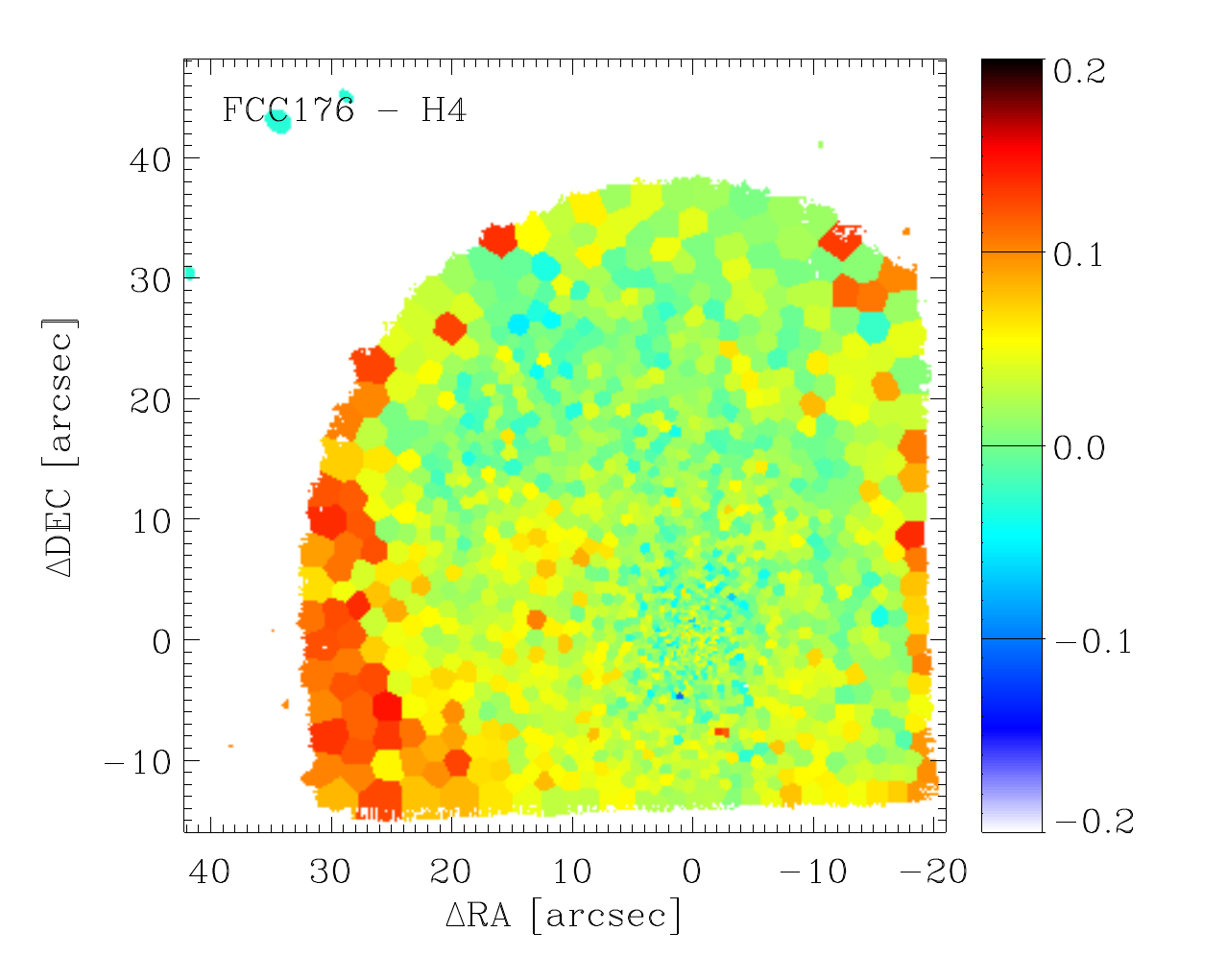} \\
\end{tabular}
\caption{Stellar analysis for FCC~176.
{\it First row panels:\/} MUSE reconstructed image ({\it left\/}). The dotted ellipse correspond to the isophotes at  
$\mu_B=25$~mag arcsec$^{-2}$. Maps of the mean velocity ({\it middle\/}) and velocity dispersion ({\it right\/}) of the stellar LOSVD.
{\it Second row panels:\/} Radial profiles of the kinematic (black circles) and photometric (green circles) position angle ({\it left\/}). The vertical dashed lines mark the radial range where the average position angles are computed. Maps of the third ({\it middle\/}) and fourth Gauss-Hermite coefficient ({\it right\/}) of the stellar LOSVD.}
\label{fig:FCC176map}
\end{figure*}

\begin{figure*}[t!]
\caption{Same as in Fig.~\ref{fig:FCC083map}, but for FCC~177.}
\label{fig:FCC177map}
\end{figure*}

\begin{figure*}[t!]
\begin{tabular}{ccc}
\makebox[0pt][c]{\hspace{0.2cm}\raisebox{0cm}{\includegraphics[width=6.4cm]{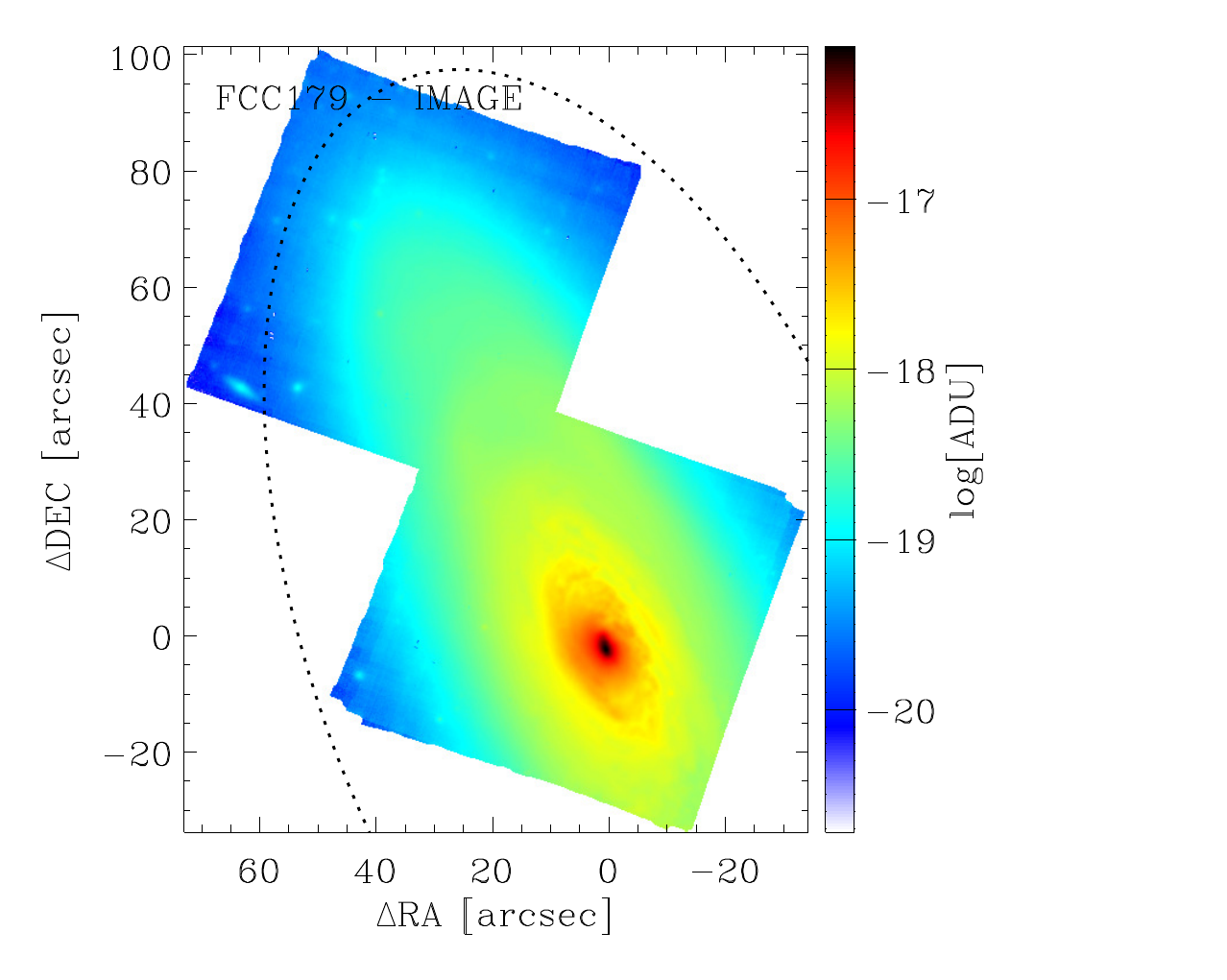}}} &
\includegraphics[width=6cm]{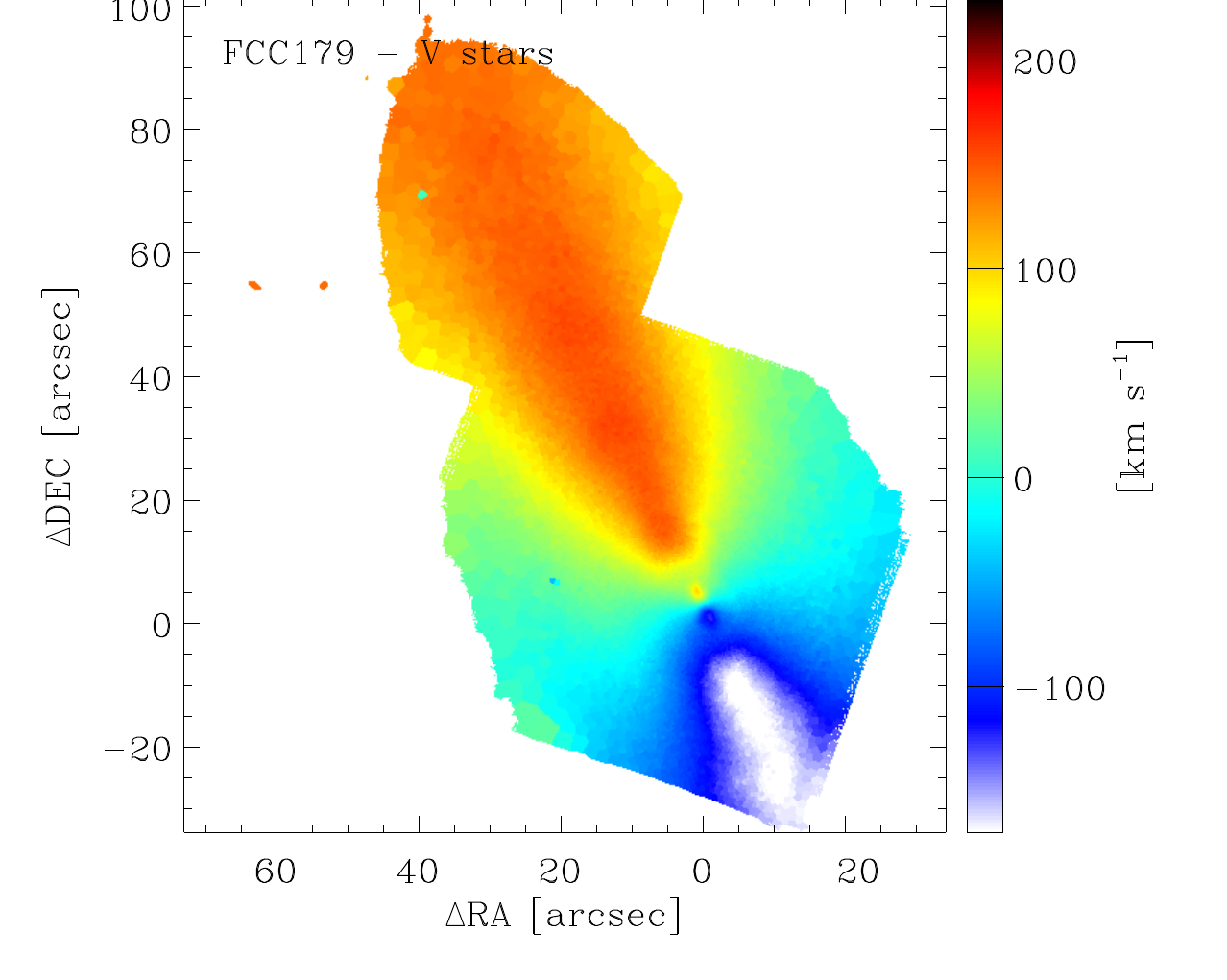} &
\includegraphics[width=6cm]{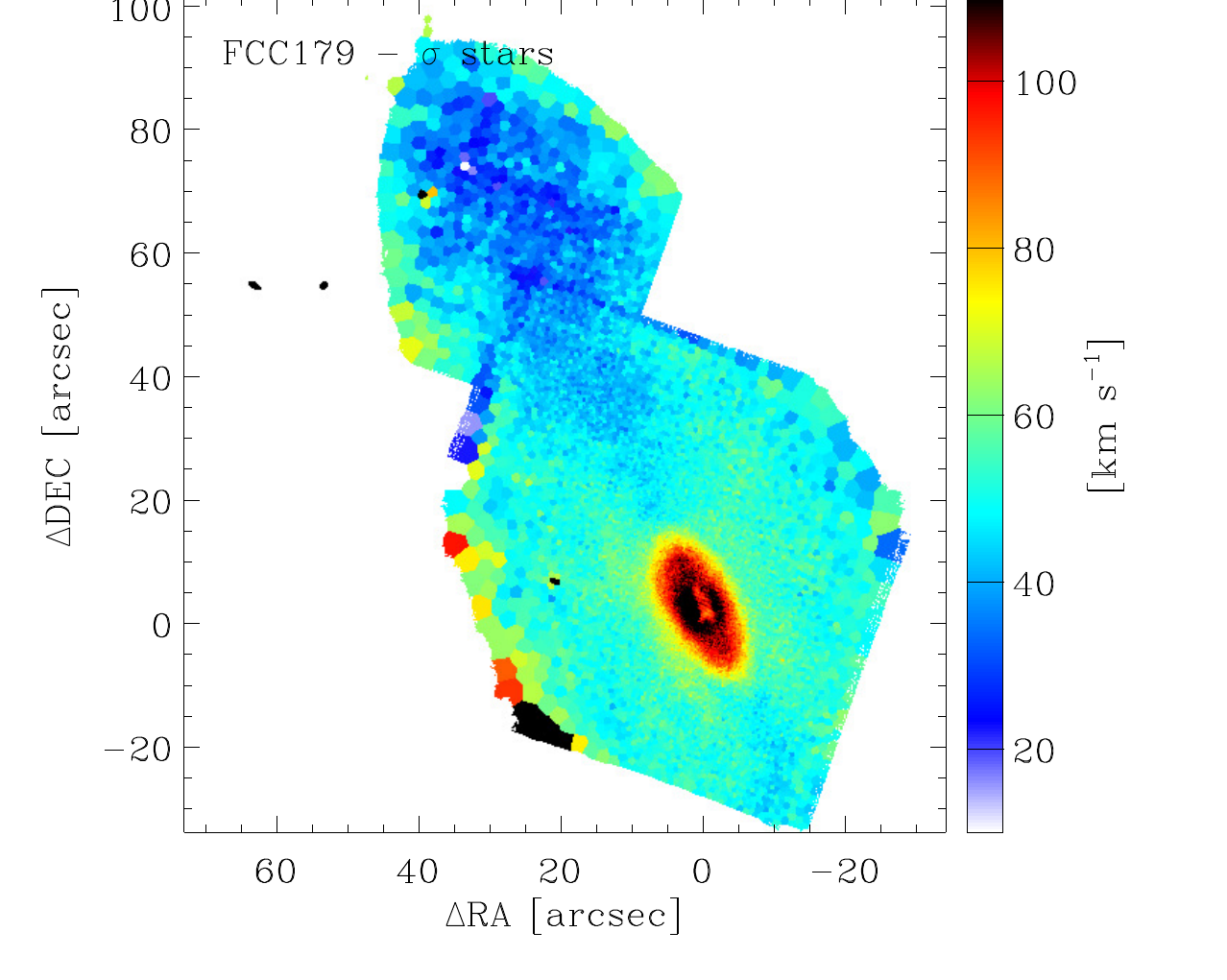} \\
\makebox[0pt][c]{\hspace{-0.5cm}\raisebox{0.3cm}{\includegraphics[width=4.7cm]{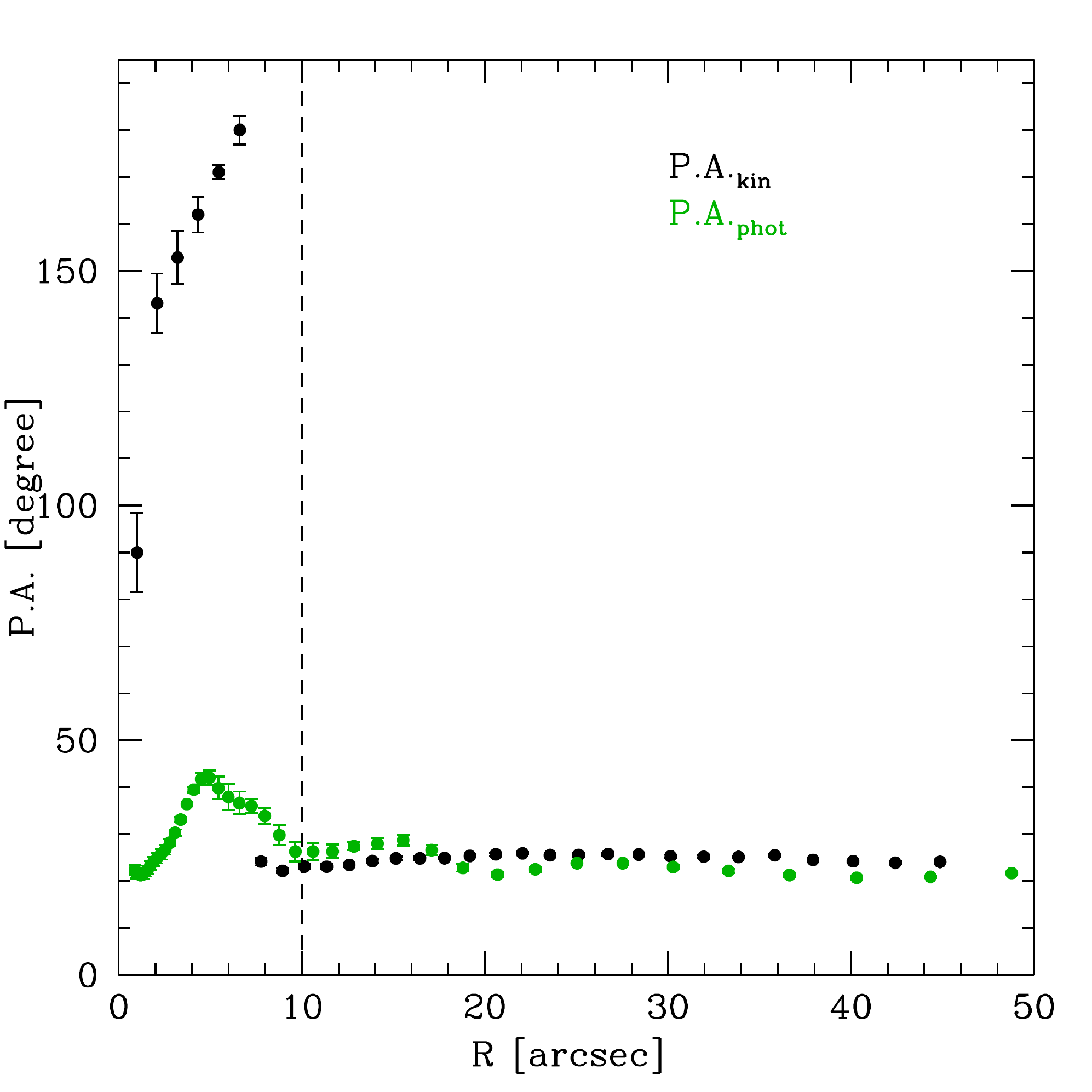}}} &
\includegraphics[width=6cm]{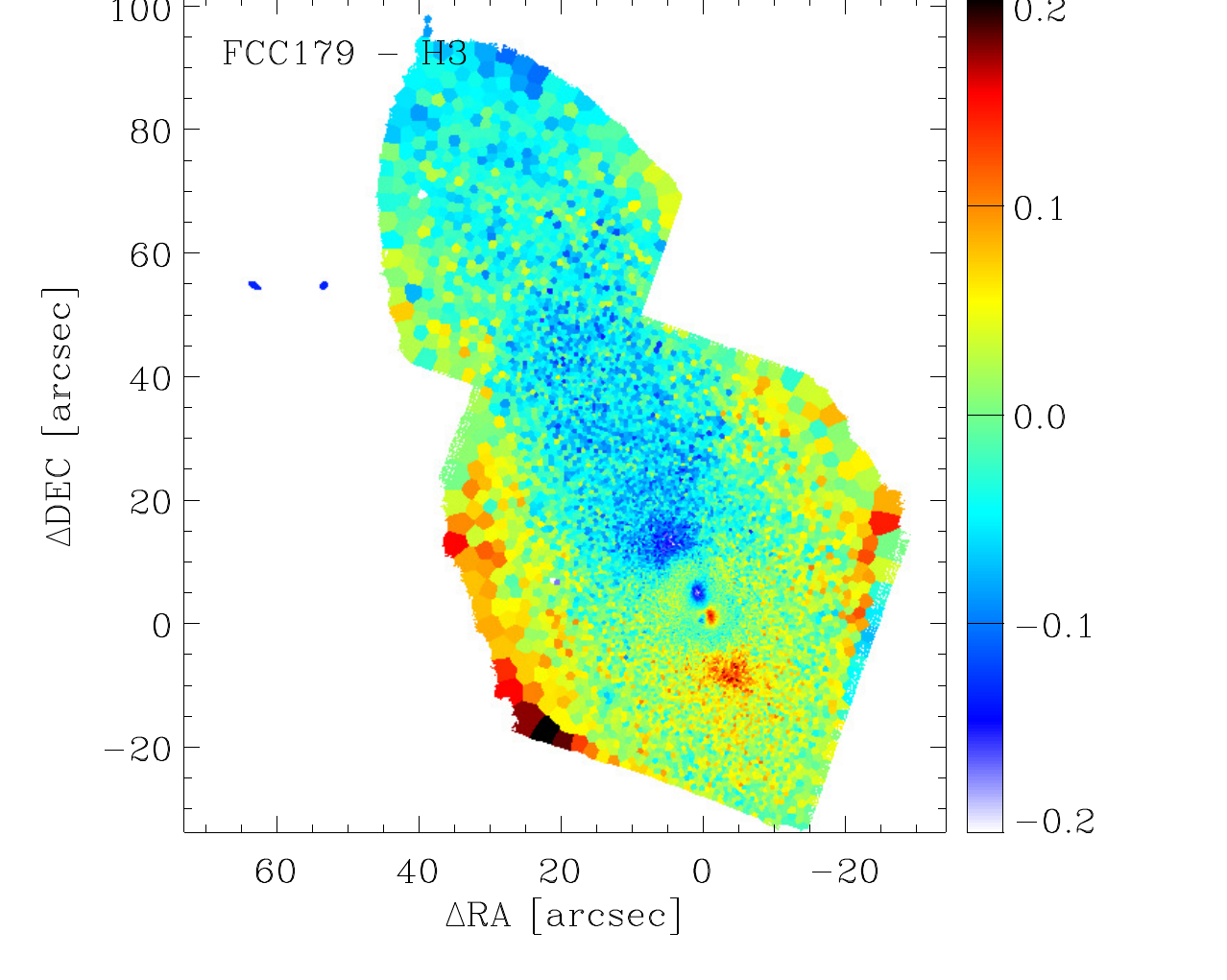} &
\includegraphics[width=6cm]{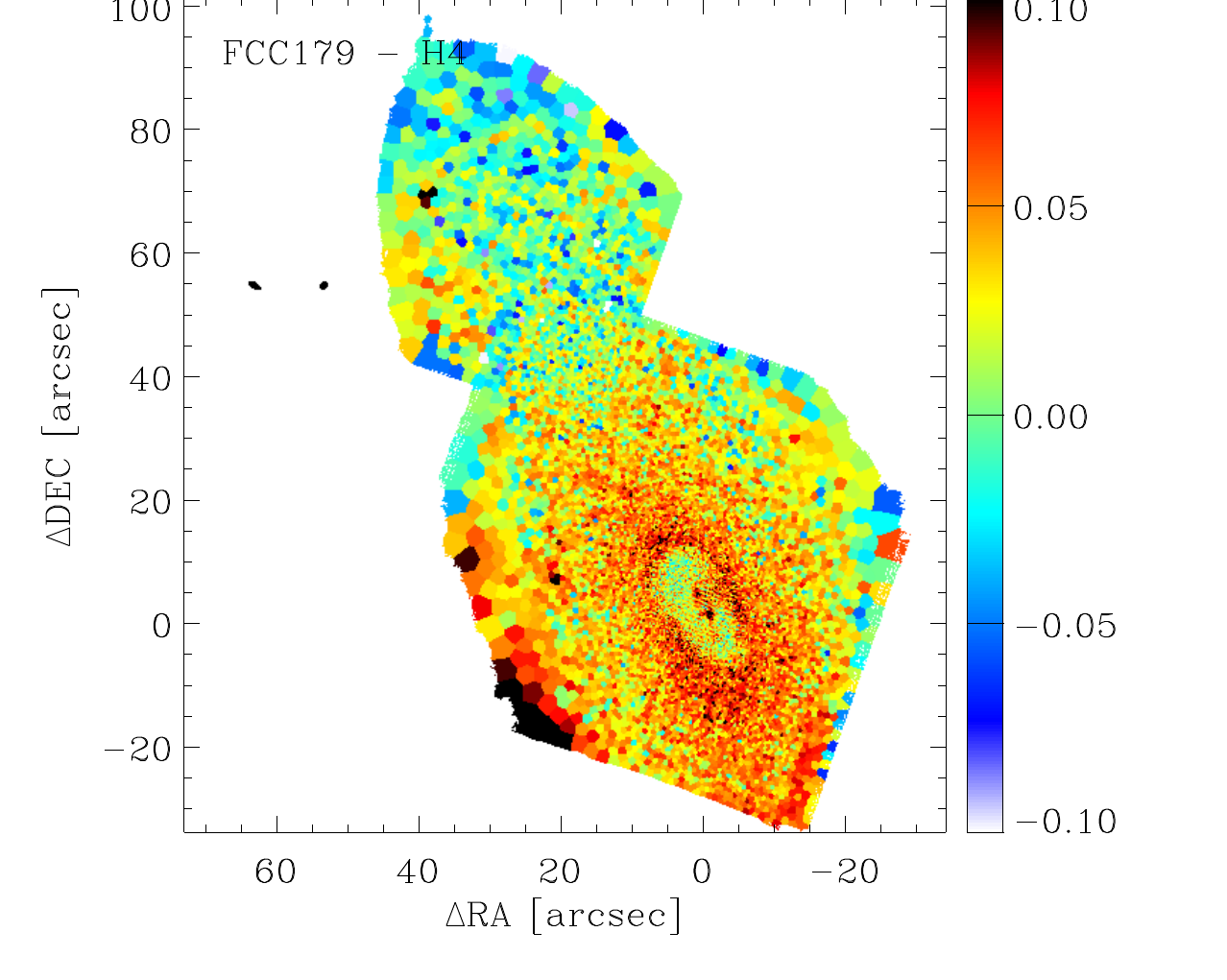} \\
\includegraphics[width=6cm]{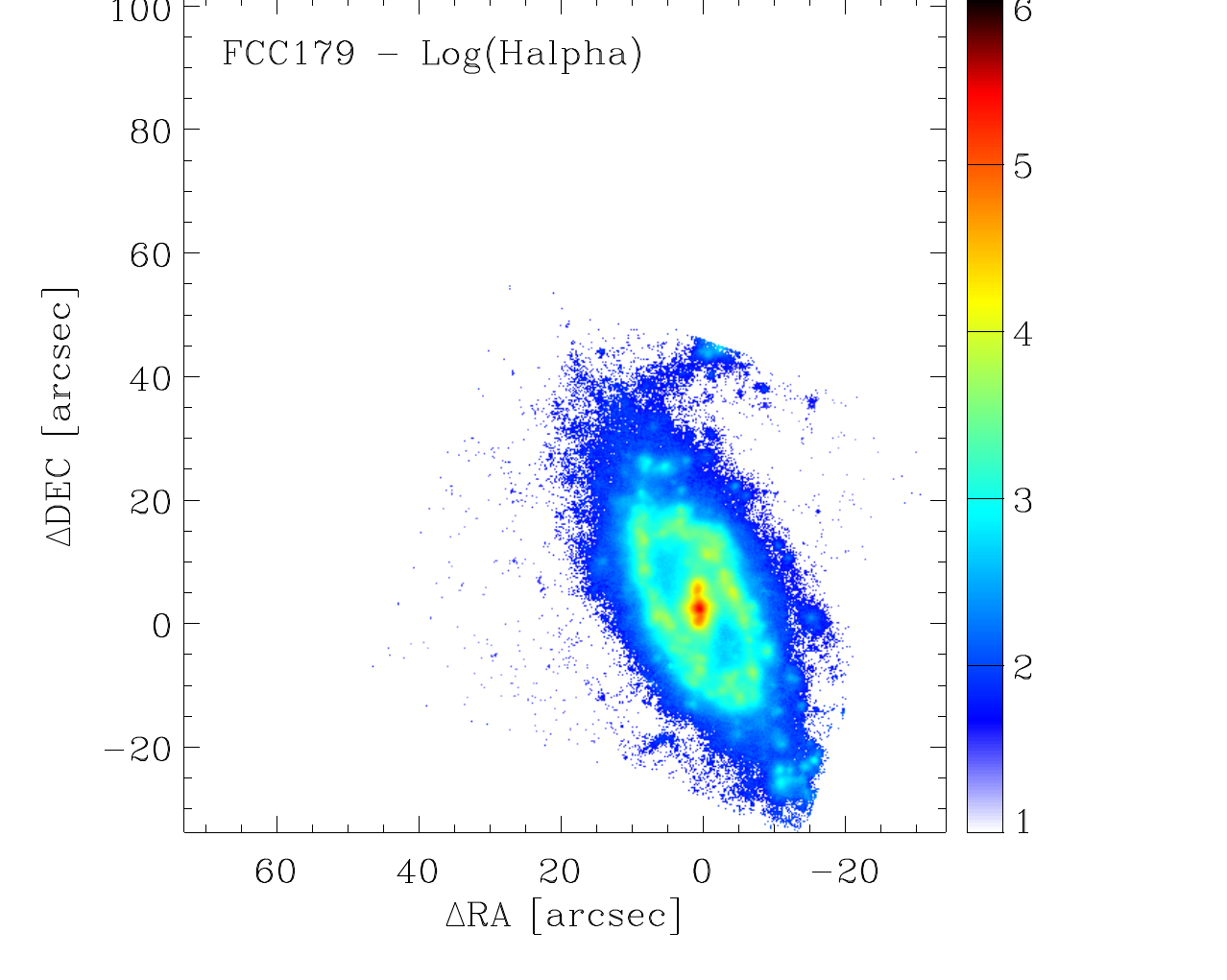} &
\includegraphics[width=6cm]{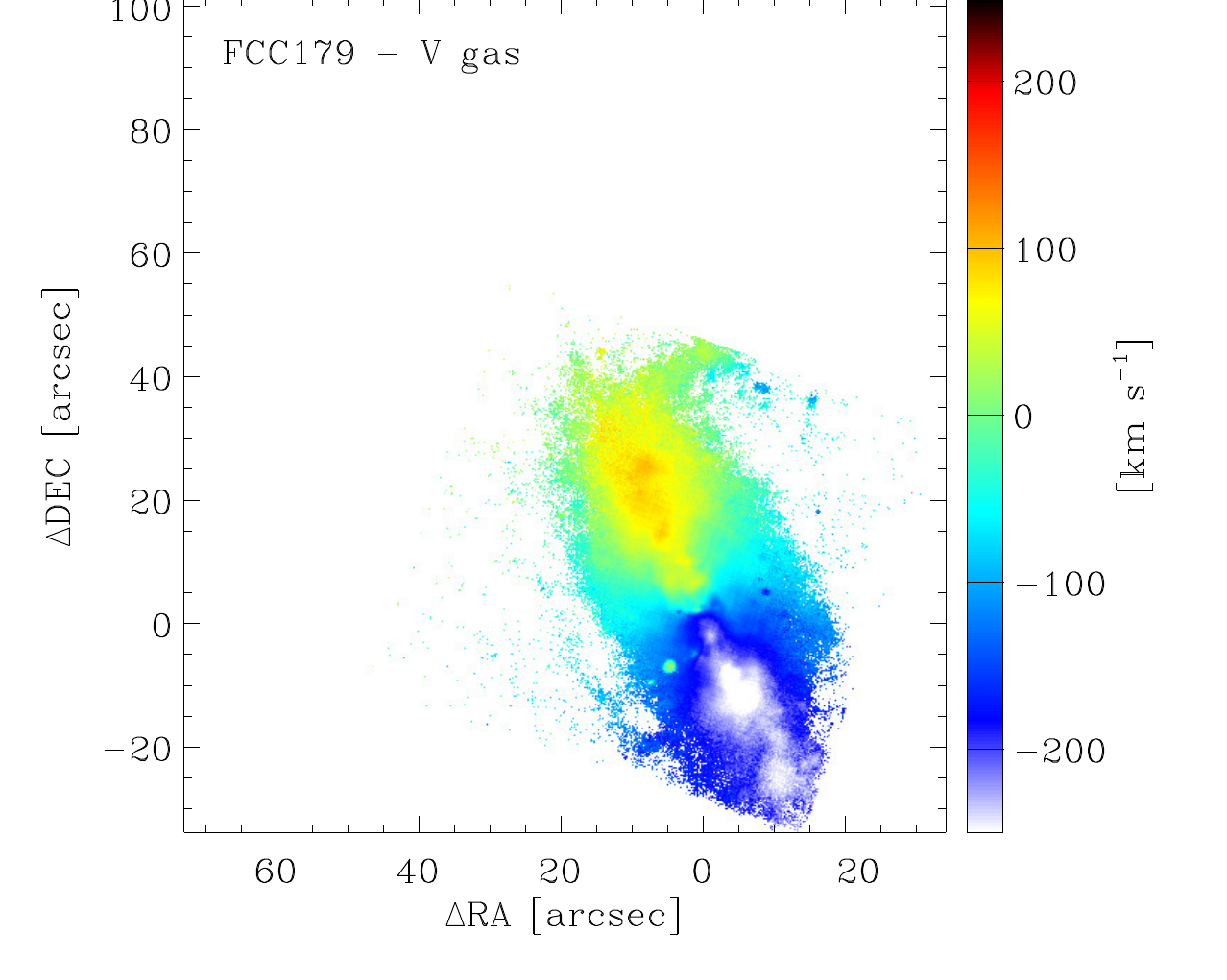} &
\includegraphics[width=6cm]{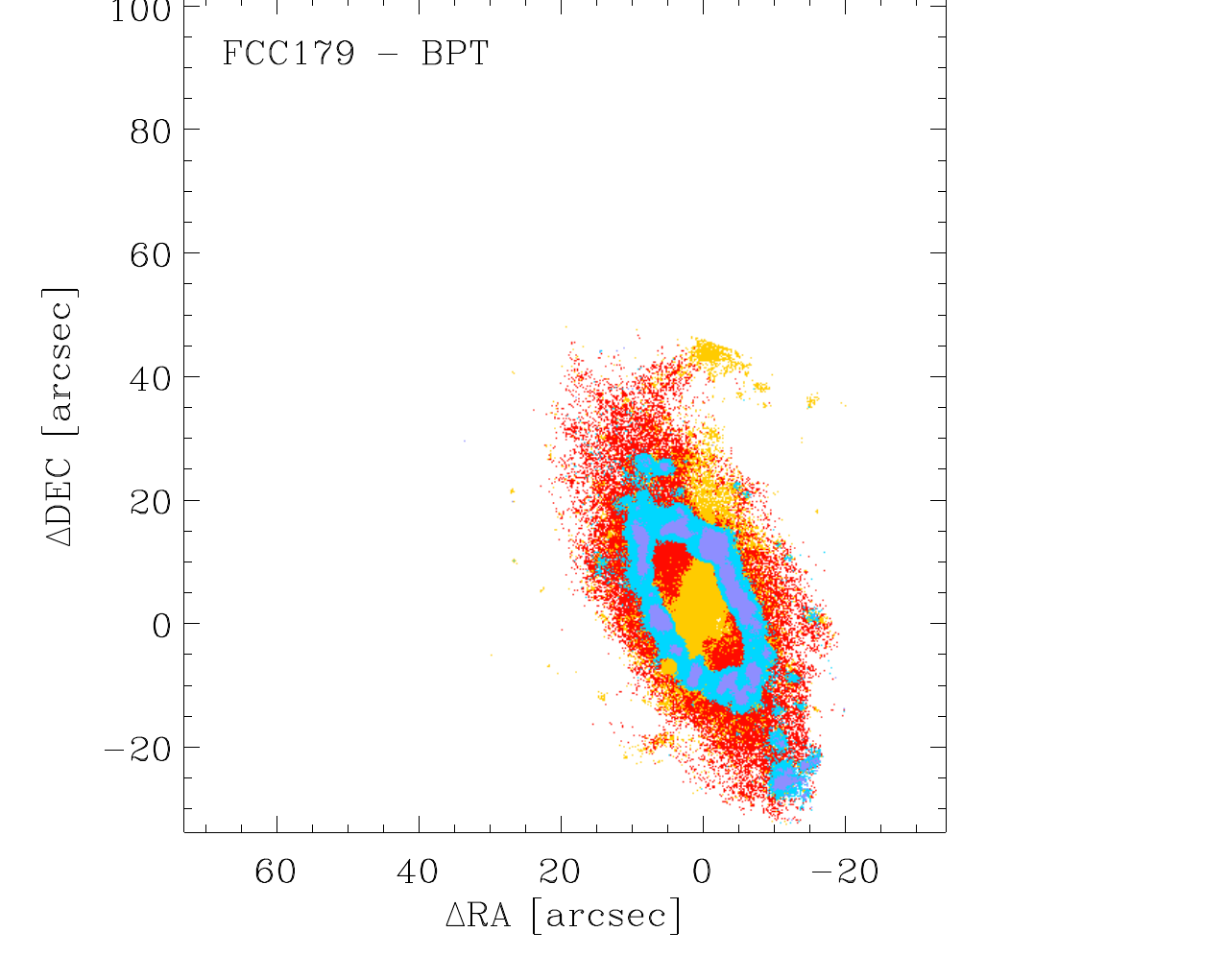} \\
\end{tabular}
\caption{Stellar and ionised-gas analysis for FCC~179.
{\it First row panels:\/} MUSE reconstructed image ({\it left\/}). The dotted ellipse corresponds to the isophote at  
$\mu_B=25$~mag arcsec$^{-2}$. Maps of the mean velocity ({\it middle\/}) and velocity dispersion ({\it right\/}) of the stellar LOSVD.
{\it Second row panels:\/} Radial profiles of the kinematic (black circles) and photometric (green circles) position angle ({\it left\/}). The vertical dashed lines mark the radial range where the average position angles are computed. Maps of the third ({\it middle\/}) and fourth Gauss-Hermite coefficient ({\it right\/}) of the stellar LOSVD.
{\it Third row panels:\/} Map of the ionised-gas classification ({\it right\/}) into \ion{H}{ii} regions (blue), highly-ionised or AGN regions (yellow), LINER-like emission regions (red), and regions where the nebular emission is powered by some combination of massive-star radiation and other ionisation sources (green).}
\label{fig:FCC179map}
\end{figure*}

\begin{figure*}[t!]
\caption{Same as in Fig.~\ref{fig:FCC083map}, but for FCC~182.}
\label{fig:FCC182map}
\end{figure*}

\begin{figure*}[t!]
\caption{Same as in Fig.~\ref{fig:FCC090map}, but for FCC~184.}
\label{fig:FCC184map}
\end{figure*}

\begin{figure*}[t!]
\caption{Same as in Fig.~\ref{fig:FCC083map}, but for FCC~190.}
\label{fig:FCC190map}
\end{figure*}

\begin{figure*}[t!]
\caption{Same as in Fig.~\ref{fig:FCC083map}, but for FCC~193.}
\label{fig:FCC193map}
\end{figure*}

\begin{figure*}[t!]
\caption{Same as in Fig.~\ref{fig:FCC090map}, but for FCC~219.}
\label{fig:FCC219map}
\end{figure*}

\begin{figure*}[t!]
\caption{Same as in Fig.~\ref{fig:FCC083map}, but for FCC~249.}
\label{fig:FCC249map}
\end{figure*}

\begin{figure*}[t!]
\caption{Same as in Fig.~\ref{fig:FCC083map}, but for FCC~255.}
\label{fig:FCC255map}
\end{figure*}

\begin{figure*}[t!]
\caption{Same as in Fig.~\ref{fig:FCC179map}, but for FCC~263.}
\label{fig:FCC263map}
\end{figure*}

\begin{figure*}[t!]
\caption{Same as in Fig.~\ref{fig:FCC083map}, but for FCC~276.}
\label{fig:FCC276map}
\end{figure*}

\begin{figure*}[t!]
\caption{Same as in Fig.~\ref{fig:FCC083map}, but for FCC~277.}
\label{fig:FCC277map}
\end{figure*}

\begin{figure*}[t!]
\caption{Same as in Fig.~\ref{fig:FCC113map}, but for FCC~285.}
\label{fig:FCC285map}
\end{figure*}

\begin{figure*}[t!]
\caption{Same as in Fig.~\ref{fig:FCC179map}, but for FCC~290.}
\label{fig:FCC290map}
\end{figure*}

\begin{figure*}[t!]
\caption{Same as in Fig.~\ref{fig:FCC083map}, but for FCC~301.}
\label{fig:FCC301map}
\end{figure*}

\begin{figure*}[t!]
\caption{Same as in Fig.~\ref{fig:FCC179map}, but for FCC~306.}
\label{fig:FCC306map}
\end{figure*}

\begin{figure*}[t!]
\caption{Same as in Fig.~\ref{fig:FCC179map}, but for FCC~308.}
\label{fig:FCC308map}
\end{figure*}

\begin{figure*}[t!]
\caption{Same as in Fig.~\ref{fig:FCC083map}, but for FCC~310.}
\label{fig:FCC310map}
\end{figure*}

\begin{figure*}[t!]
\caption{Same as in Fig.~\ref{fig:FCC179map}, but for FCC~312.}
\label{fig:FCC312map}
\end{figure*}

\clearpage

\section{Results on individual galaxies}\label{sec:description}

This section provides a brief description of the main properties of each F3D galaxy based on the kinematic and line-strength maps shown in Appendix~\ref{sec:kin_map}. The classification of the galaxies into ancient, intermediate, and recent infallers is based on Fig.~\ref{fig:PPS} and their  location in the cluster is shown in Fig.~\ref{fig:Xray}.

\smallskip      
    
\subsection{FCC~83 (NGC~1351) -- Fig.~\ref{fig:FCC083map}}

FCC~83 is located on the western side of the cluster at a distance $R_{\rm proj}=1.7^\circ$ from the cluster centre. It is among the brightest and more massive ETGs in the low-density region of the cluster and is an intermediate infaller. It shows a regular stellar velocity map with a maximum rotation velocity of $\sim 100$~km~s$^{-1}$. The stellar velocity dispersion peaks in the centre ($\sigma \sim 150$~km~s$^{-1}$) and decreases outwards ($\sigma \sim 60$~km~s$^{-1}$ at $R\sim 50$~arcsec). Except for the central regions ($R\leq5$~arcsec), $\rm PA_{\rm kin}$ is consistent with $\rm PA_{\rm phot}$. The Mg$b$ and Fe5015 line-strength indices follow the light distribution showing a central peak, while the H$\beta$ index is quite constant at all radii ($\sim2$~\AA) although it shows a small central decrease of $\sim0.2$~\AA.


\subsection{FCC~90 -- Fig.~\ref{fig:FCC090map}}

FCC~90 is a small and faint peculiar elliptical galaxy \citep{Ferguson1989} on the south-western side of the cluster at a distance $R_{\rm proj}=1.7^\circ$ from the centre. It is a recent infaller in the low-density region of the cluster. It is the only non-rotating galaxy of the F3D sample, so that $\rm PA_{\rm kin}$ is not well-determined. The stellar velocity dispersion shows a clear decrease of $\sim30$~km~s$^{-1}$ towards the galaxy centre ($R\la10$~arcsec). The values of the line-strength indices in the central parts are also different from those measured in the outskirts: H$\beta$ is larger (by $\sim2$~\AA) while Mg$b$ and Fe5015 are smaller (by $\sim1$~\AA). 
The ionised-gas emission is concentrated in the inner 10\arcsec\ and is entirely powered by star formation, consistent
with the very blue optical colors found in FDS ($g-r\sim 0.4$~mag). The ionised-gas shows modest rotation in harmony with the
molecular-gas \citep{Zabel2019}. Finally, a plume of gaseous material to the west shows distinct kinematics, suggesting a
past galaxy-galaxy interaction.


\subsection{FCC~113 (ESO~358-G015) -- Fig.~\ref{fig:FCC113map}}

FCC~113 is a late-type spiral on the north-western side of the cluster at a distance of $R_{\rm proj}=1.3^\circ$ from FCC~213 (NGC~1399). It is one of the intermediate infallers residing in the low-density region. It was classified as a star-forming dwarf galaxy by \citet{Drinkwater2001}. Its morphology resembles that of a lopsided spiral as confirmed by the smooth stellar velocity map. The maximum rotation velocity is $v_{\rm max}\sim50$~km~s$^{-1}$ at $R\sim20$~arcsec, while the stellar velocity dispersion is low ($\sigma\sim15$~km~s$^{-1}$) and constant at all radii. 
This galaxy shows extended ionised-gas emission powered by massive stars consistent with the previous findings from \citet{Drinkwater2001}, 
with several localised regions where star-formation is more intense. For one HII-region the measured ionisation level moves it outside the adopted BPT boundaries for star-formation, 
although there are no indications of other ionisation mechanisms. The ionised-gas
kinematics shows coherent, if not perfectly regular rotation in the same direction as the stars.



\subsection{FCC~119 -- Fig.~\ref{fig:FCC119map}}

FCC~119 is one the faintest ETG of the F3D sample. It is an intermediate infaller located in the low-density region on the northern side of the cluster at a distance of $R_{\rm proj}=2.1^\circ$ from the centre. The stellar kinematics is characterised by low rotation velocity ($v_{\rm max}\sim20$~km~s$^{-1}$ at $R\sim15$~arcsec) and velocity dispersion ($\sigma\sim20$~km~s$^{-1}$) which is constant at all radii. The light distribution from the reconstructed image shows irregular isophotes in the central region ($R\la10$~arcsec). The H$\beta$ line-strength index measured in the central parts of the galaxy is $\sim1$~\AA\ larger while the Mg$b$ and Fe5015 indices are both $\sim1$~\AA\ smaller compared to the outskirts. 
Ionised-gas emission is detected only in the galaxy centre, where it appears to rotate in the same direction as the stars and is, probably, mainly
powered by ongoing star formation.



\subsection{FCC~143 (NGC~1373) -- Fig.~\ref{fig:FCC143map}}

FCC~143 is one of the most compact and faint elliptical galaxies of the F3D sample. It is located on the western side of the cluster at $R_{\rm proj}=0.8^\circ$ from the centre and it is close in projection to FCC~147 (NGC~1374). According to \citet{Blakeslee2009}, the difference in distance between FCC~143 and FCC~147 is only $\sim0.3$ Mpc. Deep FDS images show a possible ongoing interaction between the two galaxies \citep{Iodice2018}. FCC~143 is an ancient infaller in the north-south clump of galaxies. The stellar kinematics reveals a central distinct component, which extends along the photometric major axis out to $R\sim20$~arcsec with a maximum rotation velocity $v_{\rm max}\sim40$~km~s$^{-1}$. Its stellar velocity dispersion is larger ($\sigma\sim80$~km~s$^{-1}$) than that in the surrounding regions ($\sigma\sim50$~km~s$^{-1}$). This feature is characterised by larger values of the Mg$b$ and Fe5015 line-strength indices than in the outskirts, while the H$\beta$ index is almost constant at all radii. The colour of this distinct component is redder ($g-i\sim1.2$~mag) than the outskirts ($g-i\sim0.7$~mag) in the FDS measurements.


\subsection{FCC~147 (NGC~1374) -- Fig.~\ref{fig:FCC147map}}

FCC~147 is one of the brightest elliptical galaxies of the sample. It is located on the western side of the cluster at $R_{\rm proj}=0.7^\circ$ from FCC~213 (NGC~1399). It is an ancient infaller belonging to the north-south galaxy clump. Its stellar kinematics is characterised by a small maximum rotation ($v_{\rm max}\sim50$~km~s$^{-1}$) and a high central velocity dispersion ($\sigma\sim200$~km~s$^{-1}$). The gradient measured in line-strength maps of the Mg$b$ and Fe5015 follows the gradient of velocity dispersion. The H$\beta$ index in the central parts is $\sim1$~\AA\ smaller than in the surrounding regions.


\subsection{FCC~148 (NGC~1375) -- Fig.~\ref{fig:FCC148map}}

FCC~148 is a lenticular galaxy with a prominent boxy bulge located on the western side of the cluster at $R_{\rm proj}=0.7^\circ$ from the centre. It is close in projection to FCC~147 (NGC~1374), but there are no signs of interaction between the two objects. This is consistent with FCC~148 being a recent infaller. The stellar velocity map resembles that of a typical disc with a maximum value of $v_{\rm max}\sim100$~km~s$^{-1}$ at $R\sim50$~arcsec. The stellar velocity dispersion is peaked in the centre ($\sigma\sim80$~km~s$^{-1}$) with a steep decrease outwards ($\sigma\sim20$~km~s$^{-1}$ at $R\sim30$~arcsec). At larger radii, it increases by $\sim20$~km~s$^{-1}$. The maps of Mg$b$ and Fe5015 also show different properties inside and outside $R\sim30$~arcsec. They have  larger values (by $\sim2$~\AA) in the central parts than in the outskirts. H$\beta$ is nearly constant at all radii ($\sim2$ \AA).

\subsection{FCC~153 (IC~1963) -- Fig.~\ref{fig:FCC153map}}

FCC~153 is an edge-on lenticular galaxy located in the low-density region on the northern side of the cluster at a distance of $R_{\rm proj}=1.2^\circ$ from the centre. 
It is one of the intermediate infallers in the north of the clump of galaxies. The analysis of the kinematics and properties of the stellar component of this object based on F3D data is presented and discussed in detail by \citetalias{Pinna2019a}.


\subsection{FCC~161 (NGC~1379) -- Fig.~\ref{fig:FCC161map}}

FCC~161 is one the brightest elliptical galaxies in the high-density region of the cluster. It is located at $R_{\rm proj}=0.5^\circ$ from FCC~213 (NGC~1399) on the western side of the cluster. It is one of the ancient infallers in the north-south clump of galaxies. Similarly to FCC~147, which is the other giant elliptical galaxy of this cluster region, also the stellar component of FCC~161 is characterised by a small maximum rotation ($v_{\rm max}\sim30$~km~s$^{-1}$) and a high velocity dispersion in the centre ($\sigma\sim150$~km~s$^{-1}$). As for FCC~147, the line-strength maps of Mg$b$ and Fe5015 are characterised by an increasing gradient toward the centre, which is consistent with the gradient measured in the stellar velocity dispersion. The H$\beta$ index in the central parts is $\sim1$~\AA\ smaller compared to the outskirts. The kinematics along the major axis is covered only by the central MUSE pointing, since the halo pointing was obtained at $90^\circ$ west of it. The main reason for this is to constrain the stellar populations in this particular region of the outskirts that shows a strong signature of an ongoing accretion from outside \citep{Iodice2018}. 


\subsection{FCC~167 (NGC~1380) -- Fig.~\ref{fig:FCC167map}}

FCC~167 is the brightest lenticular galaxy inside the virial radius of the Fornax cluster at $R_{\rm proj}=0.6^\circ$ from the centre. It is an ancient infaller in the north-south clump of galaxies on the western side of the cluster. This galaxy shows an extended rotationally-supported stellar disc with a large rotation velocity ($v_{\rm max}\sim300$~km~s$^{-1}$) and a high velocity dispersion in the centre ($\sigma\sim200$~km~s$^{-1}$), where there is also faint H$\alpha$ emission indicative of traces of star formation. 
\citet{Sarzi2018} chose FCC~167 as an illustrative example to describe the details of the F3D data reduction and analysis. \citet{Viaene2019} studied the dust and ionised-gas properties in the centre of FCC~167, and \citet{MartinNavarro2019} provided a detailed investigation of its stellar population properties and initial mass function.


\subsection{FCC~170 (NGC~1381) -- Fig.~\ref{fig:FCC170map}}

FCC~170 is an edge-on lenticular galaxy in the high-density region of the cluster at $R_{\rm proj}=0.4^\circ$ west of FCC~213 (NGC~1399). It belongs to the north-south clump of galaxies and is an ancient infaller. It is characterised by a boxy bulge and a thin disc, which becomes thicker at larger radii (\citealt{Iodice2018}; \citetalias{Pinna2019b}). The analysis of the kinematics and properties of the stellar component of this object based on F3D data is presented and extensively discussed by \citetalias{Pinna2019b}.


\subsection{FCC~176 (NGC~1369) -- Fig.~\ref{fig:FCC176map}}

FCC~176 is one of the most luminous LTGs of the F3D sample and is characterised by a prominent bar and an outer ring. It is located at $R_{\rm proj}=0.9^\circ$ from the centre on the south-western side of the cluster. It is an intermediate infaller in the transion between the high and low-density regions of the cluster where the X-ray emission is still present. Ram-pressure stripping could have acted to stop star formation once the galaxy entered into the cluster core. FCC~176 is indeed the only LTG of the F3D sample that does not show ionised-gas emission. The stellar velocity map shows a regular rotation with a maximum velocity $v_{\rm max}\sim120$~km~s$^{-1}$ at $R\sim30$~arcsec. The stellar velocity dispersion map is characterised by smaller values ($\sigma\sim30$~km~s$^{-1}$) for locations on the spiral arm on the north-western side of the galaxy and larger values along the bar ($\sigma\sim60$~km~s$^{-1}$). On the south-eastern side, the stellar velocity dispersion displays even higher values ($\sigma\geq80$~km~s$^{-1}$), which needs further investigation as the light from the bright foreground star on this galaxy side could have affected the spectra.

\subsection{FCC~177 (NGC~1380A) -- Fig.~\ref{fig:FCC177map}}

FCC~177 is an edge-on lenticular galaxy located in the high-density region on the northern side of the cluster at $R_{\rm proj}=0.8^\circ$. It is an ancient infaller in the north-south clump of galaxies. It is characterised by a small and bright bulge, an extended metal-rich and old thin disc, and a metal-poor and younger thick disc (\citealt{Iodice2018}, \citetalias{Pinna2019a}). The detailed analysis of the kinematics and properties of the stellar component of FCC~177 from F3D data is presented by \citetalias{Pinna2019a}.

\subsection{FCC~179 (NGC~1386) -- Fig.~\ref{fig:FCC179map}}

FCC~179 is a Seyfert 2 early-type spiral at $R_{\rm proj}=0.7^\circ$ from the cluster centre on the south-western side. Although it is close in projection to the cluster core and located in the high-density region, FCC~179 is one of the recent infallers of the cluster. This is consistent with the absence of star formation in the outer parts of the disc, probably quenched by ram pressure stripping, while it is still ongoing in the inner regions. This galaxy shows a prominent spiral structure for $R\la40$~arcsec which disappears outwards. 
The H$\alpha$ map reveals the presence of a circumnuclear ring where most of the central star-formation of FCC179 is found. 
Elsewhere in the disc other sources of ionisation are likely present, in particular
in the north-south direction where the impact of AGN activity would be
consistent with the central gas outflows in the same direction  \citep{Lena2015}.
The stellar disc is rotationally supported with $v_{\rm max}\sim250$~km~s$^{-1}$ at $R\sim90$~arcsec. The stellar velocity dispersion is large in the centre ($\sigma \sim150$~km~s$^{-1}$) but decreases steeply outwards.

\subsection{FCC~182 -- Fig.~\ref{fig:FCC182map}}

FCC~182 is a small galaxy at $R_{\rm proj}=0.3^\circ$ west of FCC~213 (NGC~1399). It is one of the ancient infallers in the north-south galaxy clump. Although it was classified as a barred galaxy by \citet{Ferguson1989}, the stellar velocity map is consistent with that of an ETG hosting a distinct component in the central $R\la10$~arcsec with a maximum rotation velocity $v_{\rm max }\sim15$~km~s$^{-1}$. The stellar velocity dispersion is small ($\sigma\sim40$~km~s$^{-1}$) and is almost constant at all radii. The line-strength maps show that the inner distinct component has larger values of Mg$b$ and Fe5015 (by $\sim 1$~\AA) than the outskirts, while H$\beta$ does not show significant variation inside the galaxy ($\sim2$ \AA). FDS imaging shows that this structure is characterised by redder colours than its surrounding regions \citep{Iodice2018}.

\subsection{FCC~184 (NGC~1387) -- Fig.~\ref{fig:FCC184map}}

FCC~184 is the bright elliptical galaxy located at $R_{\rm proj}=0.3^\circ$ west of FCC~213 (NGC~1399). Their distance 
differs by $\sim2$ Mpc according to \citet{Blakeslee2009}. Photometric and spectroscopic studies proved that FCC~184 is 
tidally interacting with FCC~213 with an ongoing stripping of the eastern part of the outer stellar envelope 
\citep{Dabrusco2016, Spiniello2018, Iodice2018}. FCC~184 is one of the ancient infallers located in the north-south galaxy 
clump. The MUSE pointings were aligned along $\rm PA_{\rm phot}\sim100^\circ$, while the stellar kinematics suggests that 
the rotation axis has $\rm PA_{\rm kin}\sim60^\circ$. FCC~184 has an inner bar, a prominent boxy bulge, and a small 
kinematically decoupled component. The stellar velocity dispersion is large inside the bulge 
($\sigma\sim200$~km~s$^{-1}$) with isocontours consistent with the bulge boxy isophotes. A nuclear dust-ring of 
$\sim6$~arcsec was found from 
near-infrared \citep{Laurikainen2006} and optical deep imaging \citep{Iodice2018}. This feature is also detected in the 
maps of stellar velocity dispersion (smaller $\sigma$ values) and H$\beta$ line-strength index (larger H$\beta$ values). 
The Mg$b$ and Fe5015 indices are higher in the inner bar compared to larger radii. The ionised-gas emission shows a 
central bright ring with prominent star formation and a spiral-like structure. 
The ionised-gas emission shows a central bright ring where most of the
gas has composite star-formation/AGN or LINER-like classification,
with fewer regions showing gas mainly powered by star formation. The
gas is counter-rotating with respect to the stars, which indicates an
external origin for such a structure.

\subsection{FCC~190 (NGC~1380B) -- Fig.~\ref{fig:FCC190map}}

FCC~190 is one of the four barred lenticular galaxies close to the cluster centre. It is located at $R_{\rm proj}=0.4^\circ$ on the north-western side of the cluster. It is one of the ancient infallers of the north-south clump within the high-density region of the cluster. This galaxy has a prominent inner and red ($g-i\sim1.1$~mag) with respect to the bluer outskirts. $\rm PA_{\rm kin}$ and $\rm PA_{\rm phot}$ differ by $\sim90^\circ$ in the region of the bar ($R\la40$~arcsec). The maximum rotation velocity in the stellar disc is $v_{\rm max}\sim100$~km~s$^{-1}$. The stellar velocity dispersion is larger along the bar ($\sigma\sim80$~km~s$^{-1}$) and decreases outwards ($\sigma\sim50$~km~s$^{-1}$). Along the bar, the Fe5015 line-strength index is higher ($\sim5$~\AA) than in the disc ($\sim4$~\AA). In the same region, the H$\beta$ and Mg$b$ indices do not show different values compared to the disc. 

\subsection{FCC~193 (NGC~1389) -- Fig.~\ref{fig:FCC193map}}

FCC~193 is one the most luminous barred lenticulars inside the virial radius 
of the Fornax cluster. It is located at $R_{\rm proj}=0.4^\circ$ from FCC~213 (NGC~1399) on the south-western cluster 
side. It is an ancient infaller which belongs to the north-south clump within the high-density region of the cluster. As in FCC~190, the presence of the bar for $R\la30$ arcsec is evident in the difference of $\sim20^\circ$ between $\rm 
PA_{\rm kin}$ and $\rm PA_{\rm phot}$. The stellar velocity map shows a nuclear disc-like structure ($R\la5$~arcsec) with 
a maximum rotation velocity $v_{\rm max}\sim100$~km~s$^{-1}$. This feature is aligned with the major axis of the outer disc 
and is characterised by a smaller velocity dispersion ($\sigma\sim90$~km~s$^{-1}$) with respect to the regions of the 
bar/outer disc. The maximum rotation velocity in the outer stellar disc is $v_{\rm max}\sim130$~km~s$^{-1}$ at 
$R\sim50$~arcsec, where the velocity dispersion decreases to $\sigma \sim40$~km~s$^{-1}$. The inner disc, bar, and outer 
disc have also different properties in the maps of the line-strength indices. The inner disc has larger values of Mg$b$ 
($\ga4$~\AA) and Fe5015 ($\ga5$~\AA) than the bar and outer disc, while it shows smaller values of H$\beta$ ($\sim1$~\AA). There is a remarkable gradient of the Mg$b$ and Fe5015 indices towards lower values  in the bar and outer disc.

\subsection{FCC~219 (NGC~1404) -- Fig.~\ref{fig:FCC219map}}

FCC~219 is the second brightest ETG inside the virial radius of  Fornax. It is $\sim1$~mag fainter than FCC~213 (NGC~1399) and located at a distance of $R_{\rm proj}=0.2^\circ$ from it on the south-eastern side of the cluster. FCC~219 is an ancient infaller settled in the cluster core. The stellar kinematic maps are regular and show large rotation ($v_{\rm max}\sim100$~km~s$^{-1}$ at $R\sim50$~arcsec) along the photometric major axis. The velocity dispersion profile peaks in the centre ($\sigma\sim250$~km~s$^{-1}$) and decreases outwards ($\sigma\sim90$~km~s$^{-1}$ at $R\sim50$~arcsec). There is the signature of a nuclear decoupled component ($R\la9$~arcsec). The line-strength maps of Mg$b$ and Fe5015 are peaked to the centre and decrease in the outskirts. A filament-like structure, along the north-south direction, was detected in the ionised-gas emission with an intensity peak on the south side of the galaxy. Deep FDS imaging shows a small and red feature that seems to be connected to a red tail on its eastern side \citep{Iodice2018}. The presence of ionised-gas emission suggests that this feature may be the remnant of a disrupted gas-rich small galaxy.

\subsection{FCC~249 (NGC~1419) -- Fig.~\ref{fig:FCC249map}}

FCC~249 is a compact low-luminosity non-rotating elliptical galaxy 
at a distance of $R_{\rm proj}=2.1^\circ$ from the centre, close to  the  virial radius of Fornax. It is an intermediate infaller in the southern low-density region of the cluster. The stellar kinematic maps reveal a nuclear ($R\la5$~arcsec) component with $v_{\rm max}\sim50$~km~s$^{-1}$. The stellar velocity dispersion peaks in the centre ($\sigma\sim110$~km~s$^{-1}$) and decreases outwards ($\sigma\sim80$~km~s$^{-1}$ at $R\sim20$~arcsec).  Similar behaviour is observed in Mg$b$ and Fe5015, which are larger in the centre ($\sim4$~\AA) and decrease outwards ($\sim2$ \AA).  H$\beta$ is nearly constant at all radii ($\sim2$~\AA).

\subsection{FCC~255 (ESO~358-G050) -- Fig.~\ref{fig:FCC255map}}

FCC~255 is a lenticular galaxy on the northern side of Fornax at $R_{\rm proj}=1.8^\circ$ from FCC~213 (NGC~1399). It is one of the intermediate infallers in the low-density region, with an inner thin disc and outer thicker envelope, also evident in the stellar kinematic and line-strength maps. The maximum rotation is $v_{\rm max}\sim50$~km~s$^{-1}$ at $R\sim30$~arcsec. The stellar velocity dispersion is small and constant ($\sigma\sim40$~km~s$^{-1}$) along the disc major axis and increases by $\sim20$~km~s$^{-1}$ along the minor axis. The thin disc has larger values of Mg$b$ and Fe5015 (by $\sim2$~\AA) and smaller values of H$\beta$ (by $\sim1$~\AA) compared to the thicker envelope.

\subsection{FCC~263 (ESO~358-G051) -- Fig.~\ref{fig:FCC263map}}

FCC~263 is a late-type barred spiral on the north-eastern side of the cluster at $R_{\rm proj}=0.8^\circ$ in the transition from high to low-density regions. It is a recent infaller with perturbed central isophotes. The stellar disc has a maximum rotation velocity $v_{\rm max}\sim90$~km~s$^{-1}$ at $R\sim30$~arcsec and small and constant velocity dispersion at all radii ($\sigma\sim20$~km~s$^{-1}$). 
The galaxy shows strong central H$\alpha$ emission powered by star-formation.  
The ionised-gas distribution and kinematics are not entirely regular, with emission extending towards the
north, consistent with the molecular gas data \citep{Zabel2019}, and suggests that either a tidal encounter or a
minor-merger occurred in the past.

\subsection{FCC~276 (NGC~1427) -- Fig.~\ref{fig:FCC276map}}

FCC~276 is the brightest elliptical galaxy on the eastern side of the cluster
at $R_{\rm proj}=0.8^\circ$ from FCC~213 (NGC~1399). It is the only ancient infaller on this side of the transition region from high to low density. This is one of the two slow rotators of the F3D sample. The stellar kinematic maps clearly show the presence of a distinct component in the centre ($R\la10$~arcsec), which has $v_{\rm max}\sim50$~km~s$^{-1}$ and $\sigma\sim180$~km~s$^{-1}$. In the region of the distinct component, the Mg$b$ and Fe5015 line-strength indices are larger ($\sim5$~\AA) than outside ($\sim3$~\AA) while H$\beta$ is nearly constant at all radii ($\sim2$~\AA). Differently from the other distinct components detected in the galaxies close to the cluster core, this feature has bluer colours \citep{Carollo1997, Iodice2018}.

\subsection{FCC~277 (NGC~1428) -- Fig.~\ref{fig:FCC277map}}

FCC~277 is a boxy lenticular galaxy located on the eastern side of the cluster at a distance of $R_{\rm proj}=0.8^\circ$ from FCC~213 (NGC~1399) in the transition from high to low-density regions. FCC~277 is $\sim2$ mag fainter than FCC~276, which is the other ETG in this area, and their distance differs by $\sim1$ Mpc \citep{Blakeslee2009}. Differently from FCC~276, FCC~277 is an intermediate infaller. The stellar kinematics is consistent with a bulge of $\sigma\sim90$~km~s$^{-1}$ for $R\la10$~arcsec and a disc of $v_{\rm max}\sim100$~km~s$^{-1}$ and $\sigma\sim50$~km~s$^{-1}$ at $R\sim30$~arcsec. A dip in the velocity dispersion is observed in the nucleus ($\sigma\sim60$~km~s$^{-1}$ for $R\la2$~arcsec). This feature has bluer colours than its surroundings \citep{Iodice2018}. The line-strength maps show that H$\beta$, Mg$b$, and Fe5015 are larger ($\sim4$~\AA) in a discy region of $\sim10$~arcsec along the major axis, whereas the values of all the indices are smaller outside.

\subsection{FCC~285 (NGC~1437A) -- Fig.~\ref{fig:FCC285map}}

FCC~285 is a very late-type spiral located on the south-eastern side at a distance of $R_{\rm proj}=1.2^\circ$ from the cluster centre. Despite its irregular morphology, the galaxy shows a stellar velocity map with a regular rotation pattern with $v_{\rm max}\sim20$~km~s$^{-1}$ at $R\sim20$~arcsec. The stellar velocity dispersion is extremely small ($\sigma\la5$~km~s$^{-1}$) at all radii. 
Intense H$\alpha$ emission is detected over the entire extent of the
galaxy. It mainly consists of HII-regions all along the spiral
structure. As in FCC113, the central parts of several of these
HII-regions show high level of ionisation that bring them outside the
domain of HII-regions in BPT diagrams. The ionised gas co-rotates with the stellar disc.
FCC~285 is one of the recent infallers in the low-density region of the cluster. This is consistent with the active star formation and presence of neutral hydrogen \citep{Schroder2001}.

\subsection{FCC~290 (NGC~1436) -- Fig.~\ref{fig:FCC290map}}

FCC~290 is a late-type spiral at $R_{\rm proj}=1.1^\circ$ from FCC~213 (NGC~1399) in the south-eastern low-density region of the cluster. It is one of the intermediate infallers in this area. Since the spiral arms are prominent inside $\sim2$~arcmin and the disc appears smooth and featureless outwards, this galaxy may evolve into a lenticular system \citep[see also][]{Raj2019}. The spiral arms cause perturbations in the central regions of the stellar kinematic maps. The disc has $v_{\rm max}\sim140$~km~s$^{-1}$ at $R\sim60$~arcsec and a very small velocity dispersion ($\sigma\la30$~km~s$^{-1}$). 
This galaxy shows a regular ionised-gas disc that co-rotates with the stars and with the molecular gas \citep{Zabel2019}. 
The gas distribution is characterised by the presence of
several HII-regions, generally powered by star-formation, except for the central, bulge-dominated region where the gas emission
is much fainter and takes on a LINER-like classification.

\subsection{FCC~301 (ESO 358-G059) -- Fig.~\ref{fig:FCC301map}}

FCC~301 is one of the faintest elliptical galaxies inside the virial radius of the Fornax cluster. It is located at a distance of $R_{\rm proj}=1.4^\circ$ from FCC~213 (NGC~1399) and is one of the recent infallers in the low-density region on the south-eastern side. As FCC~90 and FCC~147, which are the other two ETGs classified as recent infallers, also FCC~301 has different photometric and kinematic properties compared to the ETGs identified as ancient infallers. The central discy isophotes are consistent with the inner stellar disc detected in the maps of the velocity ($v_{\rm max}\sim60$~km~s$^{-1}$) and velocity dispersion ($\sigma\la40$~km~s$^{-1}$ at $R\sim10$~arcsec). At larger radii, the galaxy is characterised by a roundish envelope which is slowly rotating ($v_{\rm max}\sim30$~km~s$^{-1}$) and has a larger velocity dispersion ($\sigma\la80$~km~s$^{-1}$) with respect to the inner disc. The inner disc and outer envelope also have different properties in the line-strength maps. The inner disc shows larger values of Mg$b$ ($\sim3$~\AA) and Fe5015 ($\sim5$~\AA) compared to the envelope, whereas H$\beta$ remains almost constant ($\sim2$~\AA) throughout the galaxy.

\subsection{FCC~306 -- Fig.~\ref{fig:FCC306map}}

FCC~306 is the faintest and smallest galaxy inside the virial radius of the Fornax cluster. This extremely-late barred galaxy is located on the south-eastern side of the cluster at $R_{\rm proj}=1.7^\circ$ from the centre. It is one of the recent infallers populating the low-density region. The stellar velocity maps suggests a low rotation ($v_{\rm max}\sim20$~km~s$^{-1}$) along $\rm PA_{\rm kin}\sim-30^\circ$ whereas $PA_{\rm phot}\sim30^\circ$. Therefore, FCC~306 is classified as a prolate rotator. The stellar velocity dispersion is very small at all radii ($\sigma\la20$~km~s$^{-1}$). H$\alpha$ emission is detected in the inner 10 arcsec with several blobs on the north-western side of the galaxy.

\subsection{FCC~308 (NGC~1437B) -- Fig.~\ref{fig:FCC308map}}

FCC~308 is a late-type spiral which is close to FCC~306 at $R_{\rm proj}=1.8^\circ$ from FCC~213 (NGC~1399). It is one of the intermediate infallers located in the low-density region on the south-eastern side of the cluster. The central regions are perturbed by dust, which also affects the stellar kinematic maps. The disc has a maximum rotation velocity of $v_{\rm max}\sim50$~km~s$^{-1}$ at $R\sim50$~arcsec. The velocity dispersion is $\sim50$~km~s$^{-1}$ at all radii and tends to increase along the minor axis. 
Ionised-gas emission is detected throughout the central region, where star-formation is pervasive. The ionised-gas
distribution is characterise by the presence of several HII-regions associated to inner spiral structure.

\subsection{FCC~310 (NGC~1460) -- Fig.~\ref{fig:FCC310map}}

FCC~310 is the only barred lenticular galaxy on the eastern side of the cluster. It is located at a distance $R_{\rm proj}=2.0^\circ$ which corresponds to about the virial radius of the Fornax cluster. It is an intermediate infaller in the south-eastern low-density region of the cluster. The prominent peanut-shaped bar is evident in the reconstructed galaxy image and gives rise to the difference of $\sim40^\circ$ between $\rm PA_{\rm kin}$ and $\rm PA_{\rm phot}$. The underlying stellar disc has a maximum rotation velocity $v_{\rm max}\sim100$~km~s$^{-1}$ at $R\sim40$~arcsec. The stellar velocity dispersion is relatively large in the bulge and bar ($\sigma\la70$~km~s$^{-1}$) and  decreases outside ($\sigma\la40$~km~s$^{-1}$ at $R\sim40$~arcsec). As also observed in FCC~277 and FCC~301, which are the two other ETGs on the eastern side of the cluster, FCC~310 has a blue nucleus which causes a dip in the velocity dispersion ($\sigma\la50$~km~s$^{-1}$ at $R\la2$~arcsec). The bar structure clearly stands out in the line-strength maps of Mg$b$ ($\sim$3 dex) and Fe5015 ($\sim5$ dex). H$\beta$ is constant through the galaxy ($\sim2$ dex).

\subsection{FCC~312 (ESO~358-G063) -- Fig.~\ref{fig:FCC312map}}

FCC~312 is a late-type spiral and the most luminous LTG inside the virial radius of the Fornax cluster. It is located in the low-density region at distance $R_{\rm proj}=1.7^\circ$ from the centre of the cluster on its eastern side. Deep FDS imaging shows faint tidal tails warping out from the stellar disc \citep{Raj2019}, which has maximum rotation velocity of $v_{\rm max}\sim120$~km~s$^{-1}$ at $R\sim90$~arcsec. At the same distance the velocity dispersion is $\sim$30~km~s$^{-1}$. The disc is dominated by H$\alpha$ emission powered by star-formation,
although some highly-excited material protrudes out of the disc near, but not exactly along, the direction of the galaxy minor axis.
Star formation together with detected neutral hydrogen \citep{Schroder2001} and molecular gas \citep{Zabel2019} are consistent with this galaxy being a recent infaller into the cluster potential.

\end{appendix}

\end{document}